\documentclass[10pt]{article} 
\usepackage{amssymb,amsmath,amsfonts,theorem,subfigure} 
\usepackage{graphicx}
\usepackage{epic,curves}
\usepackage{multirow}
\usepackage{pstricks}
\usepackage{array}
\usepackage[usenames]{pstcol}
\usepackage[normalem]{ulem}

\newcommand{\lw}{\psset{linewidth=0.5pt}}
\newcommand{\unlw}{\psset{linewidth=1pt}}
\newrgbcolor{myc}{1 0 0}
\newrgbcolor{myc2}{0 0 1}
\newrgbcolor{myc3}{1 0 1}
\newrgbcolor{myc4}{0.588632 0.133333 1.}
\newrgbcolor{myc5}{0.0791028 0.666667 0}

\theorembodyfont{\itshape} 
\theoremheaderfont{\scshape}
\theoremstyle{plain}  
\newtheorem{Lemme}{Lemma}[section]
\newtheorem{Theoreme}[Lemme]{Theorem}
\newtheorem{Proposition}[Lemme]{Proposition}
\newtheorem{Corollaire}[Lemme]{Corollary}

\newtheorem{Hypotheses}{Hypotheses}[section]

\newcommand{\upto}{\textrm{UpTo}}
\newcommand{\im}{\textrm{im\,}}
\newcommand{\eptl}{\eptlnu_N}
\newcommand{\eptlnu}{\mathcal EPTL}
\newcommand{\uq}{U_q(sl_2)}
\newcommand{\cn}{(\mathbb C^2)^{\otimes N}}

\definecolor{forestgreen}{rgb}{0.13,0.54,0.13}

\begin{document}

%
%

\title{\bf Jordan cells of periodic loop models} 
\author{
{Alexi Morin-Duchesne}\footnote{\ttfamily a.morinduchesne{\char'100}uq.edu.au} \\
\it School of Mathematics and Physics\\ 
\it University of Queensland, St Lucia, Brisbane\\
\it Queensland 4072, Australia\\[10pt]
{Yvan Saint-Aubin}\footnote{\ttfamily yvan.saint-aubin{\char'100}umontreal.ca}\\
\it D\'epartement de math\'ematiques et de statistique\\ 
\it Universit\'e de Montr\'eal, C.P.\ 6128, succ.\ centre-ville, Montr\'eal\\ 
\it Qu\'ebec, Canada, H3C 3J7\\[10pt]}

\maketitle

%
%
 
\begin{abstract}
Jordan cells in transfer matrices of finite lattice models are a signature of the logarithmic character of the conformal field theories that appear in their thermodynamical limit. The transfer matrix of periodic loop models, $T_N$, is an element of the periodic Temperley-Lieb algebra $\eptl(\beta, \alpha)$, where $N$ is the number of sites on a section of the cylinder, and $\beta=-q-q^{-1} = 2 \cos \lambda$ and $\alpha$ the weights of contractible and non-contractible loops. The thermodynamic limit of $T_N$ is believed to describe a conformal field theory of central charge $c=1-6\lambda^2/(\pi(\lambda-\pi))$. The abstract element $T_N$ acts naturally on (a sum of) spaces $\tilde V_N^d$, similar to those upon which the standard modules of the (classical) Temperley-Lieb algebra act. These spaces known as {\em sectors} are labeled by the numbers of defects $d$ and depend on a {\em twist parameter} $v$ that keeps track of the winding of defects around the cylinder. Criteria are given for non-trivial Jordan cells of $T_N$ both between sectors with distinct defect numbers and within a given sector.

\medskip

\noindent Keywords: periodic Temperley-Lieb algebra, cylinder Temperley-Lieb algebra, affine Temperley-Lieb algebra, loop models, XXZ Hamiltonian, Jordan structure, indecomposable representations, standard modules. 

\end{abstract}

%
%

\tableofcontents

%
\section{Introduction} \label{sec:intro}
%

The hypergeometric differential equation describing the four-point correlation functions of a conformal field theory (CFT) appeared in the seminal work of Belavin, Polyakov and Zamolodchikov \cite{BPZ}. The possibility of logarithmic solutions when the theory contains two fields whose dimensions differ by an integer was there right from the start. But the algebraic origin of such logarithmic behavior was only fleshed out later by Gurarie \cite{Gurarie}. Through a straighforward analysis of the operator product expansion (OPE) of such fields, he concluded that, for their description, Virasoro representations would be needed in which the generator $L_0$ has Jordan cells. Examples of such reducible but indecomposable representations were constructed by Gaberdiel and Kausch \cite{GaberdielKausch} in 1996 and, the same year, Rohsiepe's thesis \cite{Rohsiepe} launched the systematic and rigorous study of some of these that he named staggered representations. More recently Kyt\"ol\"a and Ridout \cite{KytolaRidout} have classified the staggered modules where the generator $L_0$ has Jordan cells of rank at most two. (Their definition of staggered modules differs from Rohsiepe's.) The need for reducible but indecomposable representations is one of the main characteristics of logarithmic conformal field theory (LCFT), if not their defining one.

This algebraic study is clearly a fundamental step. The emergence of the Virasoro algebra from finite lattice models and the identification of the representation content of the continuum limit are the obvious next ones and remain outstanding open problems. A while ago, Koo and Saleur \cite{koo} proposed a program to build the Virasoro generators $L_n$ starting from lattice models. This program is now being fleshed out in a series of papers by Gainutdinov, Read and Saleur \cite{gainutdinov1,gainutdinov2} 
for the periodic $gl(1|1)$ spin chain. Like the present review, it uses the periodic Temperley-Lieb algebra, but the guiding principle there is to construct a well-defined local field theory.
This special number of Journal of Physics contains other articles reviewing the results achieved in this direction. 

We follow here another direction, maybe naive, but hopefully useful. We go back to finite lattice models and try to elucidate whether their natural ``evolution operators'' (their transfer matrices) have Jordan cells. If these transfer matrices are well-chosen and go in the limit, after proper scaling, to the generator $L_0$, then the representations of the Virasoro appearing in the continuum limit should be characterized by a non-diagonalizable generator $L_0$. Note that this approach may fail if Jordan cells are absent in these finite matrices but emerge in the limiting operators. Showing that these cells do exist and survive when the limit is taken (at least for a subsequence of lattice sizes) gives a strong support to this approach. The number of physical observables showing logarithmic behavior remains small. (See for example \cite{VasseurJacobsenSaleur, CardyZiff}.) A better understanding of Jordan cells in finite models might help reveal other genuinely logarithmic observables.

The problem of finding Jordan cells in matrices defined within representations of some associative algebras is somewhat awkward. Its aim is not to check the indecomposability of a given representation, but rather to establish the (non-)diagonalizability of one fixed element within this representation. In the context of representation theory, this problem is peripheral at best, and this might explain why it has been somewhat neglected. Fortunately one can, and we certainly did, profit from a large body of knowledge surrounding these lattice models, their algebraic formulation and the representation theory of the underlying algebras.

We choose to concentrate on loop models. Their relation with the Fortuin-Kasteleyn and the Potts models is known (\cite{BaxterKellandWu} or \cite{AMDSA} for a version closer to the present formulation). For specific boundary conditions, Pearce, Rasmussen and Zuber \cite{PRZ} showed that the transfer matrix they defined has Jordan blocks, at least in a few specific cases. In \cite{AMDSA} a general analysis of these transfer matrices for open boundary conditions led to a precise criterion on the closed loop weight $\beta=2\cos\lambda$ (the central charge is then $c=1-6\lambda^2/(\pi(\lambda-\pi))$) and the size of the lattice $N$ for such Jordan blocks to exist. This result will be recalled here as theorem \ref{thm:interstrip}. Its proof will not be reproduced, but it is important to recall that it uses the representation theory of the Temperley-Lieb algebra $TL_N(\beta=e^{i\lambda}+e^{-i\lambda})$ \cite{Martin, GoodmanWenzl, Wenzl, Westbury, DRYSA} and a special central element $F_N$ \cite{AMDSA, PRZ}.

We pursue the work started in \cite{AMDSA} and consider here the case of periodic loop models. Again a transfer matrix $T_N(\lambda,u)$ has been proposed and Jordan blocks have been observed for finite system sizes for $\lambda = \pi /2$ \cite{PRV}. The transfer matrix is an element of an abstract algebra, the periodic Temperley-Lieb algebra $\eptl$, whose structure and basic representation theory have already been worked out \cite{MartinSaleur, Lehrer, GreenFan, Greenseul, ErdmannGreen}. The presence of the Jordan cells in this element will depend of course on the representation in which it is realized. 

The two representations considered in this paper will be defined in subsection \ref{sec:EPTL}, but it is useful to describe them cursorily here. Both act on $N$-point link diagrams mixing {\em arches} linking positions pairwise and {\em defects} for unmatched positions. The representation $\rho$ acts on all $N$-link diagrams and allows for the number of defects to decrease. The second representation $\omega_d$ acts on $N$-links with a fixed number $d$ of defects and depends on a complex parameter $v$ measuring how the defects move under the action of the elements of the Temperley-Lieb algebra.

As it is often the case in CFT and LCFT, the difficulty of the periodic case, compared to that on the strip, calls for new techniques. One tool was introduced a long time ago by Martin and Saleur \cite{MartinSaleur}. It is an intertwiner between some representations of $\eptl$ that are appropriate for loop models and others related to the periodic XXZ spin chains. They used it to compute, among other things, the determinant of the Gram bilinear form. In \cite{AY:alpha} the intimate relation between this intertwiner and the bilinear form was explored further. 
It is interesting to note that it is for the XXZ models that Pasquier and Saleur \cite{PasquierSaleur} provided some of the first indications of non-diagonalizability of Hamiltonians, using in a crucial way the quantum algebra $\uq$. But in the periodic case, the XXZ Hamiltonians are hermitian and therefore diagonalizable, and the algebra $\uq$ does not commute with them. Still, both the periodic XXZ Hamiltonians and the algebra $\uq$ will play a central role in our analysis. Our main result are stated in theorem \ref{thm:inter} and corollary \ref{coro:leCoroFN} for the representation $\rho$ and in theorem \ref{thm:intra} for the $\omega_d$.

From its tie with Fortuin-Kasteleyn (FK) and Potts models, the relevance of the transfer matrix $T_N$ goes beyond the understanding of its Jordan structure. For periodic geometries like the cylinder and torus, properties of the underlying physical models can be computed from this formalism: In addition to partition functions, winding and homotopy properties of spin or FK clusters \cite{Pinson, Arguin} can also be considered. This requires keeping track, in the link representations, of how the contour of the clusters wrap around the cylinder or torus. One way to proceed was proposed by Richard and Jacobsen \cite{RichardJacobsen}. We propose here a different one. Besides the usual $q$ (or $\lambda$ in $q=-e^{i\lambda}$) that appears in the definition of the periodic Temperley-Lieb algebras $\eptl$, the representations of this algebra we shall use depend upon a further parameter $v$, the {\it twist parameter}. This parameter appears in both the link and XXZ representations of $\eptl$. Its use is somewhat non-trivial and requires generalizations 
of the common periodic XXZ Hamiltonians and the usual representations of $\uq$ on the tensor product $(\mathbb C^2)^{\otimes N}$ that both depend only on the usual parameter $q$. 

The paper is organized as follows. 
The next section recalls the main tools: the periodic Temperley-Lieb algebra and its representations, and the periodic loop transfer matrices and XXZ Hamiltonian. Section \ref{sec:BJentre} recalls the criterion for the presence of Jordan blocks for the transfer matrix on the strip and finds new ones for the periodic one. 
Section \ref{sec:tildeiNd} describes tools related to $\uq$ and the intertwiner $\tilde i_N^d$ that are needed in section \ref{sec:BJintra} to probe the Jordan structure of the loop Hamiltonian within sectors. It is in this last section that the harder case of the representation $\omega_d$ is attacked and that theorem \ref{thm:intra} providing our (partial) results for the existence of Jordan cells is proved. 
The technical computations are hidden in the appendices. The conclusion summarizes the result while discussing the strenghts of and possible extensions to the new method introduced in section \ref{sec:BJintra}.

%
\section{The periodic Temperley-Lieb algebra and the transfer matrix of loop models} \label{sec:TLP}
%

This section gathers definitions and results used in the following: it defines the periodic Temperley-Lieb algebra, the representations that will play a role in the paper and the transfer matrix of the periodic loop model to be considered. 

\subsection[The periodic Temperley-Lieb algebra]{The enlarged periodic Temperley-Lieb algebra $\eptl(\beta, \alpha)$}\label{sec:EPTL}

The periodic Temperley-Lieb algebra $\eptl(\beta,\alpha)$ (or affine Temperley-Lieb algebra) was introduced and studied by Martin and Saleur \cite{MartinSaleur}, Graham and Lehrer \cite{Lehrer}, and Green, Fan and Erdmann \cite{GreenFan, Greenseul, ErdmannGreen}. It is the associative algebra generated by $e_i, 1\le i\le N$, $\Omega, \Omega^{-1}$ and the unit $id$. They satisfy the relations
\begin{alignat}{3}\label{eq:TLPN}
e_i^2&= \beta e_i, &\qquad\qquad\qquad&& \Omega e_i \Omega^{-1} &=  e_{i-1}, \nonumber\\
e_ie_j&=e_je_i, \quad \text{\rm for }|i-j|>1,& &&\Omega \Omega^{-1} &=  \Omega^{-1} \Omega = id, \\ 
e_ie_{i\pm 1}e_i&=e_i,&&& (\Omega^{\pm 1} e_N)^{N-1} &= \Omega^{\pm N} (\Omega ^{\pm 1} e_N), \nonumber 
\end{alignat}
where the indices are understood modulo $N$ and taken in the range $1$ to $N$, as well as 
\begin{equation}
E \Omega^{\pm 1} E = \alpha E, \qquad \textrm{where\ }E = e_2e_4\dots e_{N-2}e_N
\label{eq:elementE}\end{equation}
if $N$ is even.
It is common to replace the parameter $\beta \in\mathbb C$ by another complex number $q\in\mathbb C^\times$ with $\beta=-q-q^{-1}$. Our main results will focus on the interval $\beta \in [-2,2]$. 
The algebra is infinite-dimensional as it contains the infinite subalgebra $\langle \Omega,\Omega^{-1}\rangle$. The dimension of the subalgebra generated by the $e_i$s only (without the $\Omega^{\pm 1}$) is also infinite as can be seen by considering the subalgebra generated by the single element $e_1e_2\dots e_N$.

Green and Fan \cite{GreenFan} showed that the abstract algebra $\langle id, e_i,\Omega^{\pm}\rangle / (\textrm{relations (\ref{eq:TLPN}) \textrm{and} (\ref{eq:elementE}))}$
is isomorphic to an algebra defined graphically. The isomorphism $\phi$ acts on the generators as
\vskip0.3cm
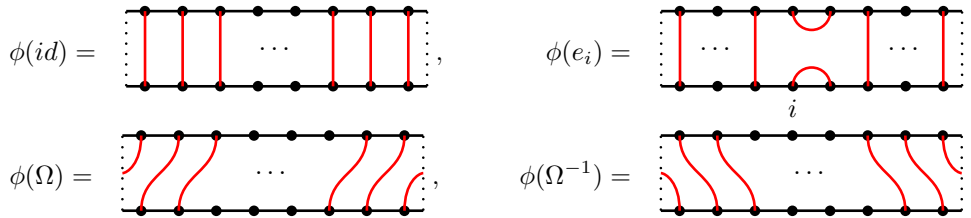
\begin{equation*} \phi(id) =
\psset{unit=0.5}
\begin{pspicture}(-0.1,-0.3)(8.5,0.8)
\psline[linewidth=1pt]{-}(0.5,1)(8.5,1)
\psline[linewidth=1pt,linestyle=dotted]{-}(0.5,1)(0.5,-1)
\psline[linewidth=1pt,linestyle=dotted]{-}(8.5,1)(8.5,-1)
\psline[linewidth=1pt]{-}(8.5,-1)(0.5,-1)
\psset{linewidth=1pt}
\psdots(1,1)(2,1)(3,1)(4,1)(5,1)(6,1)(7,1)(8,1)(1,-1)(2,-1)(3,-1)(4,-1)(5,-1)(6,-1)(7,-1)(8,-1)
\psset{linecolor=myc}
\psline(1,1)(1,-1)
\psline(2,1)(2,-1)
\rput(4.5,0){\dots}
\psline(3,1)(3,-1)
\psline(6,1)(6,-1)
\psline(7,1)(7,-1)
\psline(8,1)(8,-1)
\end{pspicture}\, \,, \qquad \qquad \hspace{-0.1cm}
\phi(e_i) =
\begin{pspicture}(-0.1,-0.3)(8.5,0.8)
\psline[linewidth=1pt]{-}(0.5,1)(8.5,1)
\psline[linewidth=1pt,linestyle=dotted]{-}(0.5,1)(0.5,-1)
\psline[linewidth=1pt,linestyle=dotted]{-}(8.5,1)(8.5,-1)
\psline[linewidth=1pt]{-}(8.5,-1)(0.5,-1)
\psset{linewidth=1pt}
\psdots(1,1)(2,1)(3,1)(4,1)(5,1)(6,1)(7,1)(8,1)(1,-1)(2,-1)(3,-1)(4,-1)(5,-1)(6,-1)(7,-1)(8,-1)
\psset{linecolor=myc}
\psarc(4.5,-1){0.5}{0}{180}\psarc(4.5,1){0.5}{180}{360}
\rput(2,0){\dots}\rput(7,0){\dots}
\rput(4,-1.6){$i$}
\psline(1,1)(1,-1)
\psline(3,1)(3,-1)
\psline(6,1)(6,-1)
\psline(8,1)(8,-1)
\end{pspicture}
\, \,, \vspace{0.8cm}
\end{equation*}
\begin{equation*} \phi(\Omega) =
\psset{unit=0.5}
\begin{pspicture}(-0.1,-0.3)(8.5,0.8)
\psline[linewidth=1pt]{-}(0.5,1)(8.5,1)
\psline[linewidth=1pt,linestyle=dotted]{-}(0.5,1)(0.5,-1)
\psline[linewidth=1pt,linestyle=dotted]{-}(8.5,1)(8.5,-1)
\psline[linewidth=1pt]{-}(8.5,-1)(0.5,-1)
\psset{linewidth=1pt}
\psdots(1,1)(2,1)(3,1)(4,1)(5,1)(6,1)(7,1)(8,1)(1,-1)(2,-1)(3,-1)(4,-1)(5,-1)(6,-1)(7,-1)(8,-1)
\psset{linecolor=myc}
\psbezier{-}(1,1)(1,0.25)(0.625,0)(0.5,0)
\psbezier{-}(2,1)(2,0)(1,0)(1,-1)
\psbezier{-}(3,1)(3,0)(2,0)(2,-1)
\psbezier{-}(8,1)(8,0)(7,0)(7,-1)
\psbezier{-}(7,1)(7,0)(6,0)(6,-1)
\psbezier{-}(8,-1)(8,-0.25)(8.375,0)(8.5,0)
\rput(4.5,0){\dots}
\end{pspicture}\, \,, \qquad \hspace{0.3cm}
\phi(\Omega^{-1}) =
\begin{pspicture}(-0.1,-0.3)(8.5,0.8)
\psline[linewidth=1pt]{-}(0.5,1)(8.5,1)
\psline[linewidth=1pt,linestyle=dotted]{-}(0.5,1)(0.5,-1)
\psline[linewidth=1pt,linestyle=dotted]{-}(8.5,1)(8.5,-1)
\psline[linewidth=1pt]{-}(8.5,-1)(0.5,-1)
\psset{linewidth=1pt}
\psdots(1,1)(2,1)(3,1)(4,1)(5,1)(6,1)(7,1)(8,1)(1,-1)(2,-1)(3,-1)(4,-1)(5,-1)(6,-1)(7,-1)(8,-1)
\psset{linecolor=myc}
\rput(4.5,0){\dots}
\psbezier{-}(1,-1)(1,-0.25)(0.625,0)(0.5,0)
\psbezier{-}(2,-1)(2,-0)(1,0)(1,1)
\psbezier{-}(3,-1)(3,-0)(2,0)(2,1)
\psbezier{-}(8,-1)(8,-0)(7,0)(7,1)
\psbezier{-}(7,-1)(7,-0)(6,0)(6,1)
\psbezier{-}(8,1)(8,0.25)(8.375,0)(8.5,0)
\end{pspicture}
\, \,. \vspace{0.5cm}
\end{equation*}
The graphical patterns above (called {\it connectivities}) are diagrams with $2N$ sites, distributed equally on the top and bottom, and non-intersecting curves connecting the sites pairwise. The patterns are to be understood as slices of a cylinder. The elements of the new algebra are linear combinations of connectivities. The product $ab$ of two generators $a$ and $b\in\eptl$ is represented by the vertical stacking of $\phi(a)$ and $\phi(b)$ with $\phi(b)$ being on top of $\phi(a)$. The result $\phi(ab)$ is then the connectivity where the $N$ dots on the bottom and the $N$ on the top are connected as they are in the stacked drawing. This connectivity is then weighted by a factor $\beta$ for each contractible loop and a factor $\alpha$ for each non-contractible one, that is, one that connects non-trivially the two boundary walls depicted by dotted lines on the drawings (hereafter referred to as the {\it imaginary boundary}). For example here is the product of two connectivities in $\eptlnu_8$:
\vspace{-0.1cm}
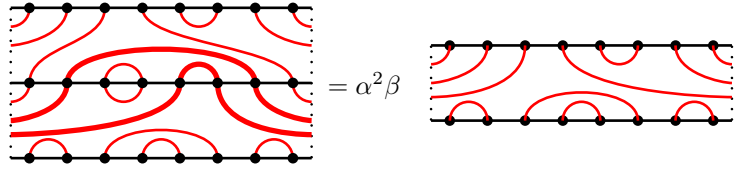
\begin{equation*}
\psset{unit=0.5}
\begin{pspicture}(-0.5,-1.3)(8.5,0.4)
\psline[linewidth=1pt]{-}(0.5,1)(8.5,1)
\psline[linewidth=1pt,linestyle=dotted]{-}(0.5,1)(0.5,-3)
\psline[linewidth=1pt,linestyle=dotted]{-}(8.5,1)(8.5,-3)
\psline[linewidth=1pt]{-}(8.5,-1)(0.5,-1)
\psline[linewidth=1pt]{-}(8.5,-3)(0.5,-3)
\psset{linewidth=1pt}
\psset{linecolor=myc}
\psarc(3.5,-1){0.5}{0}{180}
\psarc(0.5,1){0.5}{-90}{0}
\psarc(5.5,1){0.5}{180}{360}
\psarc(8.5,1){0.5}{180}{270}
\psbezier{-}(3,1)(3,0)(1,0)(1,-1)
\psbezier{-}(2,1)(2,0.25)(0.75,0)(0.5,0)
\psbezier{-}(4,1)(4,0)(8,0)(8,-1)
\psbezier{-}(7,1)(7,0.25)(8.25,0)(8.5,0)
\psarc(3.5,-1){-0.5}{0}{180}
\psarc(8.5,-1){0.5}{180}{270}
\psarc(0.5,-1){0.5}{-90}{0}
\psarc(1.5,-3){0.5}{0}{180}
\psarc(4.5,-3){0.5}{0}{180}
\psarc(7.5,-3){0.5}{0}{180}
\psbezier[linewidth=2pt]{-}(2,-1)(2,0.2)(7,0.2)(7,-1)
\psbezier[linewidth=2pt]{-}(2,-1)(2,-1.75)(0.75,-2)(0.5,-2)
\psbezier[linewidth=2pt]{-}(7,-1)(7,-1.75)(8.25,-2)(8.5,-2)
\psbezier{-}(3,-3)(3,-2)(6,-2)(6,-3)
\psarc[linewidth=2pt](5.5,-1){0.5}{0}{180}
\psbezier[linewidth=2pt]{-}(5,-1)(5,-2.25)(1,-2.375)(0.5,-2.375)
\psbezier[linewidth=2pt]{-}(6,-1)(6,-2.25)(8,-2.375)(8.5,-2.375)
\psset{linecolor=black}
\psdots(1,1)(2,1)(3,1)(4,1)(5,1)(6,1)(7,1)(8,1)(1,-1)(2,-1)(3,-1)(4,-1)(5,-1)(6,-1)(7,-1)(8,-1)(1,-3)(2,-3)(3,-3)(4,-3)(5,-3)(6,-3)(7,-3)(8,-3)
\end{pspicture}\, \, = \alpha^2 \beta 
\begin{pspicture}(-0.3,-2.3)(8.5,0.4)
\psline[linewidth=1pt,linestyle=dotted]{-}(0.5,-1)(0.5,-3)
\psline[linewidth=1pt,linestyle=dotted]{-}(8.5,-1)(8.5,-3)
\psline[linewidth=1pt]{-}(8.5,-1)(0.5,-1)
\psline[linewidth=1pt]{-}(8.5,-3)(0.5,-3)
\psset{linewidth=1pt}
\psdots(1,-1)(2,-1)(3,-1)(4,-1)(5,-1)(6,-1)(7,-1)(8,-1)(1,-3)(2,-3)(3,-3)(4,-3)(5,-3)(6,-3)(7,-3)(8,-3)
\psset{linecolor=myc}
\psarc(5.5,-1){-0.5}{0}{180}
\psarc(8.5,-1){0.5}{180}{270}
\psarc(0.5,-1){0.5}{-90}{0}
\psarc(1.5,-3){0.5}{0}{180}
\psarc(4.5,-3){0.5}{0}{180}
\psarc(7.5,-3){0.5}{0}{180}
\psbezier{-}(2,-1)(2,-1.75)(0.75,-2)(0.5,-2)
\psbezier{-}(7,-1)(7,-1.75)(8.25,-2)(8.5,-2)
\psbezier{-}(3,-3)(3,-2)(6,-2)(6,-3)
\psbezier{-}(3,-1)(3,-2.25)(1,-2.375)(0.5,-2.375)
\psbezier{-}(4,-1)(4,-2.25)(8,-2.375)(8.5,-2.375)
\end{pspicture}\, \, .
\end{equation*} 
\vskip0.9cm
\noindent In this product there is a single contractible loop and two non-contractible ones drawn with a thicker stroke. See \cite{AY:alpha} for more details and examples. 

We shall use the abstract and the graphical descriptions of $\eptl(\beta, \alpha)$ interchangeably. The object under study, the loop transfer matrix and the XXZ Hamiltonian, are defined as matrices representing an abstract element of $\eptl$ in certain representations to be introduced now. We now describe these representations. Hereafter $N\ge 2$ is a fixed integer. 

The first representations act on vector spaces $\tilde V_N$ and $\tilde V_N^d$ defined through their respective
bases $\tilde B_N$ and $\tilde B_N^d$. The elements of these bases are called {\em link states}. 
They are depicted by $N$ dots on a horizontal line, each of these being joined either to a (single) other one or to an imaginary point at infinity. 
The curves showing pairings may not cross and are hereafter referred to as half-arcs or bubbles. The set of all possible link states is denoted by $\tilde B_N$. The points joined to infinity are called {\em defects} and their number is denoted in this paper by $d, d'$ or $e$. The number of defects has the parity of $N$.
 The spaces $\tilde V_N^d$ are spanned by the bases $\tilde B_N^d$ of link states with $d$ defects. Here are the bases with $N=4$:
\begin{equation}
\psset{unit=0.6}
 \qquad  \tilde B_4^0 = \big\{
 \begin{pspicture}(-0.3,0)(1.8,0.4)
\psdots(0,0)(0.5,0)(1,0)(1.5,0)
\psset{linewidth=1pt}
\psset{linecolor=myc2}
\psarc{-}(0.25,0){0.25}{0}{180}
\psarc{-}(1.25,0){0.25}{0}{180}
\end{pspicture},
 \begin{pspicture}(-0.3,0)(1.8,0.64)
\psdots(0,0)(0.5,0)(1,0)(1.5,0)
\psset{linewidth=1pt}
\psset{linecolor=myc2}
\psarc{-}(0.75,0){0.25}{0}{180}
\psbezier{-}(0,0)(0,0.75)(1.5,0.75)(1.5,0)
\end{pspicture},
 \begin{pspicture}(-0.3,0)(1.8,0.4)
\psdots(0,0)(0.5,0)(1,0)(1.5,0)
\psset{linewidth=1pt}
\psset{linecolor=myc2}
\psbezier{-}(0.5,0)(0.5,0.65)(1.5,0.65)(1.75,0.4)
\psbezier{-}(0,0)(0,0.275)(-0.125,0.4)(-0.25,0.4)
\psarc{-}(1.25,0){0.25}{0}{180}
\end{pspicture},
  \begin{pspicture}(-0.3,0)(1.8,0.4)
\psdots(0,0)(0.5,0)(1,0)(1.5,0)
\psset{linewidth=1pt}
\psset{linecolor=myc2}
\psarc{-}(-0.25,0){0.25}{0}{90}
\psarc{-}(1.75,0){0.25}{90}{180}
\psarc{-}(0.75,0){0.25}{0}{180}
\end{pspicture},
\begin{pspicture}(-0.3,0)(1.8,0.4)
\psdots(0,0)(0.5,0)(1,0)(1.5,0)
\psset{linewidth=1pt}
\psset{linecolor=myc2}
\psbezier{-}(1,0)(1,0.65)(0,0.65)(-0.25,0.4)
\psbezier{-}(1.5,0)(1.5,0.275)(1.75,0.4)(1.75,0.4)
\psarc{-}(0.25,0){0.25}{0}{180}
\end{pspicture},
   \begin{pspicture}(-0.3,0)(1.8,0.4)
\psdots(0,0)(0.5,0)(1,0)(1.5,0)
\psset{linewidth=1pt}
\psset{linecolor=myc2}
\psarc{-}(-0.25,0){0.25}{0}{90}
\psarc{-}(1.75,0){0.25}{90}{180}
\psbezier{-}(0.5,0)(0.5,0.5)(0,0.6)(-0.25,0.6)
\psbezier{-}(1,0)(1,0.5)(1.5,0.6)(1.75,0.6)
\end{pspicture}
 \big\} ,
\label{eq:linkbasis40}
\end{equation}
\begin{equation*}
\psset{unit=0.6}
 \qquad  \tilde B_4^2 = \big\{
 \begin{pspicture}(-0.3,0)(1.8,0.4)
\psdots(0,0)(0.5,0)(1,0)(1.5,0)
\psset{linewidth=1pt}
\psset{linecolor=myc2}
\psarc{-}(0.25,0){0.25}{0}{180}
\psline{-}(1,0)(1,0.75)
\psline{-}(1.5,0)(1.5,0.75)
\end{pspicture},
 \begin{pspicture}(-0.3,0)(1.8,0.4)
\psdots(0,0)(0.5,0)(1,0)(1.5,0)
\psset{linewidth=1pt}
\psset{linecolor=myc2}
\psarc{-}(0.75,0){0.25}{0}{180}
\psline{-}(0,0)(0,0.75)
\psline{-}(1.5,0)(1.5,0.75)
\end{pspicture},
 \begin{pspicture}(-0.3,0)(1.8,0.4)
\psdots(0,0)(0.5,0)(1,0)(1.5,0)
\psset{linewidth=1pt}
\psset{linecolor=myc2}
\psarc{-}(1.25,0){0.25}{0}{180}
\psline{-}(0,0)(0,0.75)
\psline{-}(0.5,0)(0.5,0.75)
\end{pspicture},
\begin{pspicture}(-0.3,0)(1.8,0.4)
\psdots(0,0)(0.5,0)(1,0)(1.5,0)
\psset{linewidth=1pt}
\psset{linecolor=myc2}
\psarc{-}(-0.25,0){0.25}{0}{90}
\psarc{-}(1.75,0){0.25}{90}{180}
\psline{-}(1,0)(1,0.75)
\psline{-}(0.5,0)(0.5,0.75)
\end{pspicture}
\big\} , \qquad  \tilde B_4^4 = \big\{
 \begin{pspicture}(-0.3,0)(1.8,0.4)
\psdots(0,0)(0.5,0)(1,0)(1.5,0)
\psset{linewidth=1pt}
\psset{linecolor=myc2}
\psline{-}(0,0)(0,0.75)
\psline{-}(0.5,0)(0.5,0.75)
\psline{-}(1,0)(1,0.75)
\psline{-}(1.5,0)(1.5,0.75)
\end{pspicture}\big\} 
\end{equation*}
and $\tilde B_4=\tilde B_4^0\cup\tilde B_4^2\cup\tilde B_4^4$. (The dots connected to infinity appear as vertical segments.) The dimensions of the vector spaces just defined are $\dim \tilde V_N^d=\left(\begin{smallmatrix}N\\ (N-d)/2\end{smallmatrix}\right)$ and $\dim\tilde V_N=\sum_{d = N\textrm{\,mod\,}2}\dim \tilde V_N^d$.

The first representation to be described,
\begin{equation}
\rho:\eptl(\beta,\alpha)\longrightarrow \textrm{End\,}(\tilde V_N)\label{eq:reprRho},
\end{equation}
will be used in section \ref{sec:BJentre}. The action of a generator $a\in\eptl$ on a link state $w\in\tilde B_N$ is computed by simply drawing the link state $w$ on top of $\phi(a)$ and reading how the bottom $N$ points of $\phi(a)$ are connected, pairwise or to infinity. As for the product in $\eptl$, contractible loops and non-contractible ones are weighted by $\beta$ and $\alpha$ respectively. 
Note that, in the representation $\rho$, no ``memory'' remains on how defects are twisted by the action of the algebra. For example $\Omega$ acts as the identity on the unique basis element of $\tilde B_4^4$. Similarly the first and fourth elements of $\tilde B_4^0$ are exchanged under $\Omega$ and the remaining four elements of this basis form a four-element orbit of $\Omega$. It is easy to see that $\rho(\Omega^N)$ is the identity. 

The second representation
\begin{equation}
\omega_{d,v}:\eptl(\beta,\alpha)\longrightarrow \textrm{End\,}(\tilde V_N^d)\label{eq:reprSigma} 
\end{equation}
will be used in section \ref{sec:BJintra}. 
It depends on a parameter $v\in\mathbb C^\times$. 
Again the action is defined graphically as before, with one further condition and one more weight. The condition is that, if the resulting link state has less than $d$ defects, the result is set to zero. The further weight is a multiplicative factor $v^\Delta$ for each defect, where $\Delta$ is the distance the defect has traveled toward the left, that is, its position in the original state $w$ minus its new one in the resulting $\phi(a)w$. (The distance between consecutive points in link states is one unity.) The constant $v$ will be called the {\em twist parameter} and we will often omit it in writing $\omega_d$ instead of $\omega_{d,v}$. Examples of the action of $\omega_d$ are found in \cite{AY:alpha}. 
Still it is useful to note that, contrarily to $\rho$, some ``memory'' of the twisting of defects is kept under the action of $\omega_d$. For example, $\Omega$ acts on $\tilde V_4^4$ as $v^4\cdot id$ and, more generally, $\Omega^N$ acts as $v^{Nd}\cdot id$ on $\tilde V_N^d$.

The third and last representation of the Temperley-Lieb algebra needed here is one related to the description of the XXZ spin chain:
\begin{equation}
\tau:\eptl(\beta,\alpha)\longrightarrow \textrm{End\,}\cn.
\label{eq:reprTau}
\end{equation}
It also depends on a parameter $v\in\mathbb C^\times$. (The choice of ``$v$'', the same letter as in \eqref{eq:reprSigma}, will be justified by proposition \ref{thm:leszerosdei}.) 
To define it, the usual notation is used:
$$\sigma_j^a=\underbrace{id_2\otimes \dots\otimes id_2}_{j-1} \otimes \, \sigma^a\otimes \underbrace{id_2\otimes\dots\otimes id_2}_{N-j}$$ for $1\le j\le N$ and $a\in\{x,y,z,+,-\}$, and with $\sigma^a_{N+1} \equiv \sigma^a_{1}$. The tensor product contains $N$ two-by-two matrices and the Pauli matrix $\sigma^a$ is the $j$-th factor in this product. The matrices $\bar e_j\equiv\tau(e_j)\in \textrm{End } \cn$
are, in the usual spin basis $\big\{|x_1x_2\dots x_N\rangle,\ x_i\in\{+1,-1\}\big\}$,
\begin{align} \bar e_j &= \frac12 \Big( \frac{v^2+v^{-2}}2 (\sigma_j^x \sigma_{j+1}^x + \sigma_j^y \sigma_{j+1}^y) +  \frac{v^2-v^{-2}}{2i} (\sigma_j^x \sigma_{j+1}^y - \sigma_j^y \sigma_{j+1}^x) \nonumber 
\\ & \hspace{5cm} + \frac{q+q^{-1}}2 (\sigma_j^z \sigma_{j+1}^z - id) - \frac{q-q^{-1}}2 (\sigma_j^z - \sigma_{j+1}^z) \Big) \nonumber \\
&= v^{-2}\sigma^-_j\sigma^+_{j+1} +v^2 \sigma^+_j\sigma^-_{j+1} + (q+q^{-1}) \sigma^+_j \sigma^-_j\sigma^+_{j+1}\sigma^-_{j+1} - q \sigma^+_j \sigma^-_j - q^{-1}\sigma^+_{j+1}\sigma^-_{j+1} \label{eq:ebar} \\
  &=  \underbrace{id_2 \otimes id_2 \otimes \dots \otimes id_2}_{j-1} \otimes \, \bar{e} \otimes \underbrace{id_2 \otimes id_2 \otimes \dots \otimes id_2}_{N-j-1} \nonumber
\end{align}
with
$$\bar{e} = \begin{pmatrix} 
0 & 0 & 0 & 0 \\
0 & -q & v^2 & 0 \\
0 & v^{-2}  & -q^{-1} & 0 \\
0 & 0 & 0 & 0
\end{pmatrix},$$
where the allowed values for $j$ are from $1$ to $N$ for the first two forms and from $1$ to $N-1$ for the last. 

It is clear from the second form that each $\bar e_j$ commutes with $S^z=\frac12\sum_{1\le i\le N}\sigma_i^z$. The matrices $\bar e_j$ are not hermitian. But, if $q$ and $v$ are on the unit circle, the first three terms of the first form in \eqref{eq:ebar} are clearly hermitian. Only the term $-\frac12(q-q^{-1})(\sigma^z_j-\sigma^z_{j+1})$ is not. Finally one can verify that these matrices satisfy the three relations in the left column of \eqref{eq:TLPN}, with $\bar{e}_{N+1} \equiv \bar{e}_1$ and $\beta = -q-q^{-1}$. 

The left and right translations around the cylinder are denoted by $t$ and $t^{-1}$: 
\begin{align*}t \, |x_1 x_2 \dots x_{N} \rangle &=  |x_2 x_3 \dots x_{N} x_1 \rangle \\
t^{-1} |x_1 x_2 \dots x_{N} \rangle &=  |x_{N} x_1 x_2 \dots x_{N-1} \rangle
\end{align*}
and they satisfy $t^{\pm 1} \sigma_j^a = \sigma^a_{j\mp 1} t^{\pm 1}$. Then the matrices representing $\Omega^{\pm1}$ are
\begin{equation} 
\bar \Omega^{\pm 1} \equiv \tau(\Omega^{\pm 1})= v^{\pm 2S^z} t^{\pm 1}.
\label{eq:Omegabar}
\end{equation}
Because $t \bar e_j = \bar e_{j-1} t$ and $[v^{2S^z}, \bar e_j] = 0$, 
two of the equations in (\ref{eq:TLPN}) involving $\Omega$, namely 
$\bar \Omega \bar e_j \bar \Omega^{-1}= \bar e_{j-1}$ and $\bar \Omega \bar \Omega^{-1} = \bar \Omega^{-1} \bar \Omega = id$, are both satisfied. Similarly $\bar \Omega \bar e_N \bar \Omega^{-1}= \bar e_{1}$ is also satisfied and allows us to take the indices of the generators modulo $N$. The other defining relations,
\begin{equation} (\bar \Omega^{\pm 1} \bar e_N)^{N-1} = \bar \Omega^{\pm N} (\bar \Omega ^{\pm 1} \bar  e_N) \qquad \textrm{and}\qquad \bar E \bar \Omega^{\pm 1} \bar E = \alpha \bar E,
\label{eq:nontrivial}\end{equation}
can be shown to be satisfied if $\alpha$ is taken to be $v^N+v^{-N}$ (see \cite{AY:alpha}). Thus $v$ comes into play both as a twist parameter and in the weight of non-contractible loops wrapping around the cylinder. 
Later on, the (usual) bilinear form on $(\mathbb C^2)^{\otimes N}$ will be used: $\langle x_1x_2\dots x_N|y_1y_2\dots y_N\rangle=\prod_i\delta_{x_iy_i}$. This bilinear form will be used in the following only for $q$ and $v$ on the unit circle. For a state $| \chi \rangle$ that depends on $q$ and $v$, the state $\langle \chi|$ is defined as $(|\chi \rangle^T)^*$ where $q$ and $v$ are then respectively changed for $q^{-1}$ and $v^{-1}$. 

We shall omit the label $\rho$, $\omega_d$ or $\tau$ on the generators when the context makes it clear which representation is being used.

\subsection{The loop transfer matrix and the XXZ Hamiltonian}\label{sec:transferMatrix}

The ``evolution'' of the states in the loop models and the XXZ spin chain is defined through a transfer matrix for the first and a Hamiltonian for the second, and both are realizations within a representation of an element of the abstract algebra $\eptl$. This paragraph introduces both.

The loop transfer matrix $T_N(\lambda, \nu)$ is an element of $\mathcal E TLP_N(\beta,\alpha)$ defined by 
\vspace{-0.25cm}

\begin{equation} T_N(\lambda,\nu) = \quad
\psset{unit=0.8}
\psset{linewidth=1pt}
\overbrace{
\begin{pspicture}(-0,0.375)(5,1.2)
\psdots(0.5,0)(1.5,0)(4.5,0)
\psdots(0.5,1)(1.5,1)(4.5,1)
\lw
\psline{-}(-0.15,0.5)(0.0,0.5)
\psline{-}(5.15,0.5)(5.0,0.5)
\unlw
\psline{-}(0,0)(1,0)(1,1)(0,1)(0,0)\psarc[linewidth=0.5pt]{-}(0,0){0.25}{0}{90}\rput(0.5,0.5){$\nu$}
\psline{-}(1,0)(2,0)(2,1)(1,1)(1,0)\psarc[linewidth=0.5pt]{-}(1,0){0.25}{0}{90}\rput(1.5,0.5){$\nu$}
\psline{-}(4,0)(5,0)(5,1)(4,1)(4,0)\psarc[linewidth=0.5pt]{-}(4,0){0.25}{0}{90}\rput(4.5,0.5){$\nu$}
\psline{-}(2,0)(2.5,0)\psline[linestyle=dashed,dash=2pt 2pt]{-}(2.5,0)(3.5,0)\psline{-}(3.5,0)(4,0)
\psline{-}(2,1)(2.5,1)\psline[linestyle=dashed,dash=2pt 2pt]{-}(2.5,1)(3.5,1)\psline{-}(3.5,1)(4,1)
\psset{linecolor=myc}\unlw
\end{pspicture}}^N
\label{eq:tu}
\end{equation} \vspace{-0.2cm}

\noindent where each box represents the sum
\begin{equation*}
\psset{unit=0.8}
\psset{linewidth=1pt}
\begin{pspicture}(-0.5,-0.1)(0.5,0.5)
\psline{-}(-0.5,-0.5)(0.5,-0.5)(0.5,0.5)(-0.5,0.5)(-0.5,-0.5)
\psarc[linewidth=0.5pt]{-}(-0.5,-0.5){0.25}{0}{90}
\rput(0,0){$\nu$}
\end{pspicture}\ =\ \sin(\lambda-\nu)\ \ 
\begin{pspicture}(-0.5,-0.1)(0.5,0.5)
\psline{-}(-0.5,-0.5)(0.5,-0.5)(0.5,0.5)(-0.5,0.5)(-0.5,-0.5)
\psset{linecolor=myc}
\psarc{-}(0.5,-0.5){0.5}{90}{180}
\psarc{-}(-0.5,0.5){0.5}{270}{360}
\end{pspicture}\ +\ \sin \nu\ \ 
\begin{pspicture}(-0.5,-0.1)(0.5,0.5)
\psline{-}(-0.5,-0.5)(0.5,-0.5)(0.5,0.5)(-0.5,0.5)(-0.5,-0.5)
\psset{linecolor=myc}
\psarc{-}(-0.5,-0.5){0.5}{0}{90}
\psarc{-}(0.5,0.5){0.5}{180}{270}
\end{pspicture}
\ \ =\ \ 
\begin{pspicture}(-0.5,-0.1)(0.5,0.5)
\psline{-}(-0.5,-0.5)(0.5,-0.5)(0.5,0.5)(-0.5,0.5)(-0.5,-0.5)
\psarc[linewidth=0.5pt]{-}(0.5,-0.5){0.25}{90}{180}
\rput(0,0){\small$\lambda\!-\!\nu$}
\end{pspicture}\, \, .
\end{equation*} \vspace{-0.3cm}

\noindent Here, $\lambda$ labels the model and defines $\beta$ through $\beta=-q-q^{-1}$ with $q=-e^{i\lambda}$ (and therefore $\beta = 2 \cos \lambda$), $\nu$ is the anisotropy and the leftmost and rightmost boxes of $T_N$ are connected, that is, their edges are identified. Note the single tile above or the linear combination of the two with the quarter-circles are neither a generator nor an element of $\eptl$. But the diagrammatic juxtaposition of $N$ of them is. The explicit expansion of the $2^N$ terms encapsulated in the sums represented by the tiles is tedious, but the case $N=2$ provides a simple example:
$$T_2(\lambda, \nu)=\Omega\sin^2(\lambda-\nu)+(e_1\Omega+\Omega e_1)\sin\nu\sin(\lambda-\nu)+\Omega^{-1}\sin^2\nu.$$
We use the accepted word ``matrix'', but clearly $T_N$ is a particular element of the abstract algebra $\eptl$.

The definition \eqref{eq:tu} appeared in \cite{PRV} for the specific case $\beta = 0$ $(\lambda = \pi/2)$. We should also mention that transfer matrices on periodic lattices have been considered earlier, for example in \cite{RichardJacobsen2005}, where it is built from tiles tilted $45^\circ$, and in \cite{RichardJacobsen}, where the partition function for the Potts model is constructed for generic lattices, with the definition of the transfer matrix depending on the choice of lattice.

The loop transfer matrix, or simply {\em transfer matrix}, is related to the Fortuin-Kasteleyn description of two-dimensional lattice models and has many crucial mathematical properties. (Several of the following properties were proved in a general context in \cite{BPOB}. Proofs and discussion of these properties in a context similar to the present one can be found in \cite{PRZ, PRV}. The tie with lattice models is found in \cite{BaxterKellandWu} or, for a presentation similar to the one here, in \cite{AMDSA} for example.) Here are some of its properties:
\begin{enumerate}
\item[{(i)}] It forms a commuting family:
$[T_N(\lambda,\nu_1),T_N(\lambda,\nu_2)]=0$ for all $\nu_1$ and $\nu_2$.
\item[{(ii)}] It satisfies a crossing-reflection symmetry: $T_N(\lambda, \lambda-\nu) = R^{-1} T_N(\lambda, \nu) R$ where $R$ is the left-right reflection: $e_i = R^{-1} e_{N-i} R$.
\item[{(iii)}] It is invariant under translation: $[T_N(\lambda, \nu),\Omega] =0$.
\item[{(iv)}] Its expansion around $\nu=0$ is $$T_N(\lambda,\nu) \simeq \Omega \sin^N\lambda \, [(1-\nu N \cot \lambda)id + \nu \mathcal H/\sin\lambda] + \mathcal O(\nu^2)$$ where 
\begin{equation}\mathcal H = \sum_{i=1}^{N}e_i\label{eq:hamiltonienBoucle}
\end{equation} 
will be called the Hamiltonian for loop models.
\end{enumerate}

The transfer matrix $T_N(\lambda,\nu)$ and the linear term $\mathcal H$ are elements of $\eptl$. We will show (theorems \ref{thm:inter} and \ref{thm:intra}) that, in the representations $\rho$ and $\omega_d$, these elements have non-trivial Jordan cells.

The Hamiltonian of the XXZ spin chain $H:\cn\rightarrow \cn$ is simply $H = \tau(\mathcal H) = \sum_{1\le i\le N}\bar e_i$ where the $\bar e_i$s are given in \eqref{eq:ebar}. It depends explicitly on both parameters $q$ and $v$. If both are on the unit circle, the Hamiltonian $H =H(q,v)$ is hermitian, since the first three terms of \eqref{eq:ebar} are hermitian and the sum $\sum_j (\sigma^z_j-\sigma^z_{j+1})$ then vanishes. The usual XXZ model corresponds to the case $v^2=1$ (for the case with boundary see for example \cite{PasquierSaleur} and also \cite{AMDtoutseul} where the interplay between loop models and XXZ Hamiltonian has been exploited). 

Note that the periodic XXZ Hamiltonian studied in \cite{PasquierSaleur} also depends upon a twist parameter (named $e^{i \varphi}$ therein) which only enters the definition of the last generator $\bar e_N$. We note 
that a similarity transformation $\mathcal O \bar e_i \mathcal O^{-1}$, with $\mathcal O = v^{\sum_{j=1}^N j \sigma^z_j}$, maps our generators to theirs if $e^{i\varphi} = v^{2N}$.

%
\section[Jordan blocks between sectors]{Jordan blocks between sectors of loop models}\label{sec:BJentre}
%

{\it Throughout this section, the action of the periodic Temperley-Lieb algebra on the space $\tilde V_N$ spanned by all link states is that of the representation $\rho$. The following section is independent of the present one. The following notation will be used extensively:
$$C_n=\cos n\Lambda, \qquad S_n=\sin n\Lambda, \qquad \textrm{with} \qquad \Lambda=\pi-\lambda = \frac{a \pi}b.$$
(Note that the parameters $a$ and $b$ play the respective roles of $p$ and $p'$ in the definition of the logarithmic models $\mathcal{LM}(p,p')$ \cite{PRZ,PRV}.)
}

\noindent In a previous work \cite{AMDSA}, we studied loop models on the strip.
The algebraic structure is then the (usual) Temperley-Lieb algebra $TL_N(\beta)$ and the double-row transfer matrix $D_N(\lambda, u)$ an element of this algebra defined as \vspace{-0.3cm}

\begin{equation} D_N(\lambda,\nu) = \quad \ \
\psset{unit=0.8}
\psset{linewidth=1pt}
\overbrace{
\begin{pspicture}(-0.0,-0.175)(5,1.2)
\psdots(0.5,-1)(1.5,-1)(4.5,-1)
\psdots(0.5,1)(1.5,1)(4.5,1)
\lw
\unlw
\psline{-}(0,0)(1,0)(1,-1)(0,-1)(0,0)\psarc[linewidth=0.5pt]{-}(0,-1){0.25}{0}{90}\rput(0.5,-0.5){$\nu$}
\psline{-}(1,0)(2,0)(2,-1)(1,-1)(1,0)\psarc[linewidth=0.5pt]{-}(1,-1){0.25}{0}{90}\rput(1.5,-0.5){$\nu$}
\psline{-}(4,0)(5,0)(5,-1)(4,-1)(4,0)\psarc[linewidth=0.5pt]{-}(4,-1){0.25}{0}{90}\rput(4.5,-0.5){$\nu$}
\psline{-}(0,0)(1,0)(1,1)(0,1)(0,0)\psarc[linewidth=0.5pt]{-}(0,0){0.25}{0}{90}\rput(0.5,0.5){\small$\lambda\!-\!\nu$}
\psline{-}(1,0)(2,0)(2,1)(1,1)(1,0)\psarc[linewidth=0.5pt]{-}(1,0){0.25}{0}{90}\rput(1.5,0.5){\small$\lambda\!-\!\nu$}
\psline{-}(4,0)(5,0)(5,1)(4,1)(4,0)\psarc[linewidth=0.5pt]{-}(4,0){0.25}{0}{90}\rput(4.5,0.5){\small$\lambda\!-\!\nu$}
\psline{-}(2,0)(2.5,0)\psline[linestyle=dashed,dash=2pt 2pt]{-}(2.5,0)(3.5,0)\psline{-}(3.5,0)(4,0)
\psline{-}(2,1)(2.5,1)\psline[linestyle=dashed,dash=2pt 2pt]{-}(2.5,1)(3.5,1)\psline{-}(3.5,1)(4,1)
\psline{-}(2,-1)(2.5,-1)\psline[linestyle=dashed,dash=2pt 2pt]{-}(2.5,-1)(3.5,-1)\psline{-}(3.5,-1)(4,-1)
\psset{linecolor=myc}\unlw
\psarc(0,0){0.5}{90}{270}
\psarc(5,0){-0.5}{90}{270}
\end{pspicture}}^N \quad \ \ .
\label{eq:du}
\end{equation} \vspace{0.2cm}

\noindent It has many interesting properties that were investigated in \cite{PRZ}. Other definitions of transfer matrices on the strip (also $\in TL_N(\beta)$) were studied in \cite{JacobsenSaleur2008,JacobsenSaleur2008blobs,DubailJS2009}: The transfer matrices studied there are built from tiles that are tilted $45^\circ$ compared to those in \eqref{eq:du} and are examined with a larger class of boundary conditions, as elements of the $1$- or $2$-boundary Temperley-Lieb algebras. The results presented here are tied to definition \eqref{eq:du} only.

The double-row transfer matrix $D_N(\lambda, \nu)$ acts on a link module spanned by $B_N$, the subset of $\tilde B_N$ of link states with no arches straddling the imaginary boundary. The subset $B_N^d\subset B_N$ contains the states with $d$ defects and $V_N^d$ is the subspace spanned by $B_N^d$. Note that the action of the representation $\rho$ restricted to the subspace $V_N=\textrm{span }B_N$ defines a representation of $TL_N\subset \eptl$. The restriction of the matrix $\rho(c)$ to $V_N^d$ is noted by $\rho(c)|_{d}$; similarly $\rho(c)|_{[d',d]}$ is the restriction to $\oplus_{d'\le e\le d} V_N^e$, and the same notation will be used in the periodic case, when $V_N^d$ is replaced by $\tilde V_N^d$. Although a proof is still missing, it is believed that the restriction to a given $V_N^d$ of $\rho(D_N(\lambda, u))$ does not have Jordan blocks. Jordan blocks were shown \cite{AMDSA} to exist between $V_N^d$ and $V_N^{d'}$ for specific values of $\beta$, $N$ and pairs $d, d'$:

\noindent \begin{Theoreme}\label{thm:interstrip} 
Let $\Lambda=\pi-\lambda=\pi a/b$ and $a,b$ coprime integers. Then Jordan blocks in $\rho\big(D_N(\lambda, 
\nu)\big)_{[d',d]}$ occur as follows: 
\smallskip

\noindent\begin{tabular}{llc}
(i)   $a$ even: & no Jordan cells, \vspace{0.1cm}\\
(ii)  $a$ odd: & rank 2 Jordan blocks & if $|d-d'|<2b \quad \textrm{and} \quad {\textstyle{\frac12}}(d+d')\equiv b-1\hspace{-0.10cm}\mod\hspace{-0.02 cm} 2b.$
\end{tabular}
\smallskip

\noindent The existence of Jordan blocks is independent of $\nu$, that is, they occur for all but a finite set of values of $\nu$. Moreover, 
if $\rho\big(D_N(\lambda,\nu))|_d$ is diagonalizable, then the Jordan blocks specifically tie sectors $d$ and $d'$. 
\end{Theoreme}

\noindent(Even though the concept of a Jordan block {\em tying sectors $d$ and $d'$} is fairly intuitive, it was introduced formally in definition 4.1 of \cite{AMDSA}.) 
The proof of this theorem proceeds in two steps. The first identifies the Jordan blocks in the (representative) of the central element $F_N\in TL_N$. (The rest of the present section is devoted to extending this analysis to a central element in $\eptl$.) The second step shows that the Jordan structure of the matrix $\rho(D_N)$ coincides with that of $F_N$, except maybe for a finite subset of values of $u$. The proof of this step will not be repeated here as it applies without change to the periodic case. It relies on the following fact: let $A(u)$ be given by a finite sum $\sum f_ke^{iku}$ where the $f_k$ are commuting endomorphimsms on $V$, a vector space of finite dimension $\nu$. If one of the $f_k$'s has a non-trivial Jordan block, then so does $A(u)$ for all values of $u\in\mathbb R$ (or $\in\mathbb C$), except on a finite subset. 

This result may be counterintuitive as a small perturbation of a Jordan block is known to (usually) reduce the size of the Jordan block. The difference here is that the ``perturbation'' commutes with the Jordan block. A sketch of the proof is provided by the example of two commuting endomorphisms $F, G\in \textrm{End}(V)$ such that $F$ has a nontrivial Jordan block associated to its eigenvalue $f$. Let $W_f=\{v\in V\,|\, (F-f \cdot {id})^\nu v=0\}\subset V$ the associated generalized eigenspace of $F$. Since $F$ and $G$ commute, $(F-f\cdot {id})^\nu Gw=G(F-f\cdot {id})^\nu w=0$ for all $w\in W_f$ and thus $BW_f\subset W_f$. There must be an eigenvalue $g$ of $G$ such that $W_f\cap W_g\neq \{0\}$ where now $W_g=\{v\in V\,|\, (G-g \cdot {id})^\nu v=0\}\subset V$ is the generalized eigenspace of $G$. The nontrivial subspace $W_f\cap W_g$ is stable under both $F$ and $G$ and the restriction of these endomorphisms to it are of the form $f\cdot{id}+n_f$ and $g\cdot {id}+n_g$ where here $n_f$ and $n_g$ are {\em commuting} nilpotent matrices. Then a linear combination of the form $(e^{imu}F+e^{inu}G)|_{W_f\cap W_g}=(e^{imu}f+e^{inu}g)\cdot {id}+(e^{imu}n_f+e^{inu}n_g)$ with $m\neq n$ will have a nonzero nilpotent part for all values of $u$, except maybe on a finite subset. 

It is this property of commuting matrices that led to the proof of theorem \ref{thm:interstrip} in \cite{AMDSA}: It was shown that $F_N\in TL_N(\beta)$ is the top Fourier coefficient of $D_N(\lambda, u)$, which allowed us to extract the Jordan structure of $\rho\big(D_N(\lambda,\nu))$ from that of $\rho\big(F_N)$. The same strategy is used here to probe the Jordan structure of $\rho\big(T_N(\lambda,\nu))$.

As opposed to the case with open boundaries, the transfer matrix $T_N(\lambda,\nu)$ introduced in paragraph \ref{sec:transferMatrix} can have two types of Jordan blocks: within a sector with a given number of defects $d$ and between sectors with distinct $d$ and $d'$. Jordan blocks of the transfer matrix between sectors may be seen only if the action allows for an increase in the number of arches or, equivalently, a decrease in the number of defects. This is why the representation $\rho$, defined in \eqref{eq:reprRho}, will be used throughout this section.

Due the commutation property (i) of the transfer matrix $T_N$ (see paragraph \ref{sec:transferMatrix}), its coefficients in any expansion with respect to $\nu$ (Fourier, Taylor, ...) are mutually commuting. For a given $N$, there can be only a finite number of coefficients that are functionally independent since its Fourier expansion 
$$T_N(\lambda,z+\lambda/2)=\sum_{-N\leq k\leq N}f_k e^{-ikz}$$
is actually a Fourier polynomial. In this section, we use the top Fourier coefficient $f_N$ of $T_N(\lambda,\nu)$ to probe the Jordan structure between sectors of $\rho(T_N(\lambda,\nu))$. This will result in the following theorem:

\noindent \begin{Theoreme}\label{thm:inter} 
Let $\Lambda=\pi-\lambda=\pi a/b$ and $a,b$ coprime integers. Then Jordan blocks in $\rho\big(T_N(\lambda, \nu)\big)_{[d',d]}$ occur as follows: 
\smallskip 

\noindent\begin{tabular}{llc}
(a) $N$ even & (i) $a$ odd: high rank Jordan blocks if & ${\textstyle{\frac12}}(d-d')\equiv 0\hspace{-0.10cm}\mod\hspace{-0.05 cm} 2b \quad \textrm{{\bf or}} \quad {\textstyle{\frac12}}(d+d')\equiv 0\hspace{-0.10cm}\mod\hspace{-0.05 cm} 2b,$ \vspace{0.125cm}\\
& (ii) $a$ even: rank 2 Jordan blocks if &$|d-d'|<2b \quad \textrm{and} \quad {\textstyle{\frac12}}(d+d')\equiv 0\hspace{-0.10cm}\mod\hspace{-0.05 cm} 2b,$\vspace{0.125cm}\\
(b) $N$ odd & (i) $a$ odd: no Jordan cells, \vspace{0.125cm}\\
& (ii) $a$ even: rank 2 Jordan blocks if &$|d-d'|<2b \quad \textrm{and} \quad {\textstyle{\frac12}}(d+d')\equiv b\hspace{-0.10cm}\mod\hspace{-0.05 cm} 2b.$
\end{tabular}
\smallskip

\noindent The existence of Jordan blocks is independent of $\nu$, that is, they occur for all but a finite set of values of $\nu$. Moreover, if $\rho\big(T_N(\lambda,\nu))|_e$ is diagonalizable for all $e$ such that $d'<e\le d$ and either ${\textstyle{\frac12}}(d + e)\equiv 0\textrm{\rm\ mod\ }2b$ or ${\textstyle{\frac12}}(d - e)\equiv 0\textrm{\rm\ mod\ }2b$, then the Jordan blocks specifically tie sectors $d$ and $d'$. 
\end{Theoreme}

\noindent The assumption that $\rho(T_N(\lambda,\nu))|_{e}$ be diagonalizable is fairly strong. (Note that it can be weakened to the assumption that $\rho(T_N(\lambda,\nu))|_{e}$ have no non-trivial Jordan block in the eigenspace associated to the eigenvalue under consideration.) Our explorations using a symbolic manipulation software show that $\rho(T_N(\lambda,\nu))$ may have Jordan blocks between different sectors $d$ and $d'$ even though its restriction to $\tilde V_N^{d}$ is non-diagonalizable. We shall discuss the importance of this hypothesis in the context of logarithmic conformal field theory at the end of section \ref{sec:jordanFN}.

We will follow ideas developed in \cite{AMDSA} and use the following conventions. Let $c \in \eptl(\beta,\alpha)$ and $v,w \in \tilde B_N$. In a link pattern $v$, a {\em $1$-bubble} is an arch that does not surround any other bubble and an {\em $n$-bubble} one that contains at least one $(n-1)$-bubble. The link state with $N$ defects is noted $w^N$. A state with $m$ bubbles is labeled by integers $k_1, \dots, k_m$, with $1 \le k_i \le N$, that stand for midpoints of the bubbles that have to be closed (starting with $k_1$, then $k_2$, ...) to create the link state $v$. (Note that the $k_i$s actually label the edges between the tiles of $T_N$.) Such a state is denoted $w^N_{k_1, \dots, k_m}$. If all the $m$ bubbles of a state $w$ are concentric at position $x$, the shorthand notation $w^N_{x^m}$ will also be used. The matrix element $\rho(c)_{w,v}$ will often be denoted by $\langle w | c v \rangle$. Graphically, the exiting state $w$ in $\rho(c)_{w,v}$ is depicted by dashed curves. Finally we shall be using the subspace $\upto_d = \oplus_{e\le d} \tilde V_N^e$ (direct sum of vector spaces). Under the action of the representation $\rho$, the subspaces $\tilde V_N^d$ are not stable for $d>1$, but the subspaces $\upto_d$ are.

\subsection[The top Fourier coefficient $F_N$]{The top Fourier coefficient $\boldsymbol{F_N}$}

The element  $F_N(\Lambda) \in \eptl(\beta,\alpha)$ is 
\vspace{-0.3cm}

\begin{equation}\label{def:leFn} F_N(\Lambda) = \quad
\psset{linewidth=1pt}
\psset{unit=0.8}
\overbrace{
\begin{pspicture}(-0,0.375)(5,1.2)
\psdots(0.5,0)(1.5,0)(4.5,0)
\psdots(0.5,1)(1.5,1)(4.5,1)
\lw
\psline{-}(-0.15,0.5)(0.0,0.5)
\psline{-}(5.15,0.5)(5.0,0.5)
\unlw
\psline{-}(0,0)(1,0)(1,1)(0,1)(0,0)
\psline{-}(1,0)(2,0)(2,1)(1,1)(1,0)
\psline{-}(4,0)(5,0)(5,1)(4,1)(4,0)
\psline{-}(2,0)(2.5,0)\psline[linestyle=dashed,dash=2pt 2pt]{-}(2.5,0)(3.5,0)\psline{-}(3.5,0)(4,0)
\psline{-}(2,1)(2.5,1)\psline[linestyle=dashed,dash=2pt 2pt]{-}(2.5,1)(3.5,1)\psline{-}(3.5,1)(4,1)
\psset{linecolor=myc}
\psset{linecolor=myc}
\psline{-}(0,0.5)(1,0.5)
\psline{-}(0.5,0.)(0.5,0.35)
\psline{-}(0.5,0.65)(0.5,1)
\psline{-}(1,0.5)(2,0.5)
\psline{-}(1.5,0.)(1.5,0.35)
\psline{-}(1.5,0.65)(1.5,1)
\psline{-}(4,0.5)(5,0.5)
\psline{-}(4.5,0.)(4.5,0.35)
\psline{-}(4.5,0.65)(4.5,1)
\end{pspicture}}^N
\end{equation}
where \vspace{-0.1cm}

\begin{equation*}
\psset{unit=0.8}
\psset{linewidth=1pt}
\begin{pspicture}(-0.5,-0.1)(0.5,0.5)
\psline{-}(-0.5,-0.5)(0.5,-0.5)(0.5,0.5)(-0.5,0.5)(-0.5,-0.5)
\psset{linecolor=myc} 
\psline{-}(-0.5,0)(0.5,0)
\psline{-}(0,-0.5)(0,-0.15)
\psline{-}(0,0.15)(0,0.5)
\end{pspicture}\  = \ -i e^{i\lambda/2}	\displaystyle\lim_{u\to +i \infty} \frac1{\sin(\lambda-u)} \ \
\begin{pspicture}(-0.5,-0.1)(0.5,0.5)
\psline{-}(-0.5,-0.5)(0.5,-0.5)(0.5,0.5)(-0.5,0.5)(-0.5,-0.5)
\psarc[linewidth=0.5pt]{-}(-0.5,-0.5){0.25}{0}{90}
\rput(0,0){$u$}
\end{pspicture}\ \ =  \   e^{-i\Lambda/2}\ \ 
\begin{pspicture}(-0.5,-0.1)(0.5,0.5)
\psline{-}(-0.5,-0.5)(0.5,-0.5)(0.5,0.5)(-0.5,0.5)(-0.5,-0.5)
\psset{linecolor=myc} 
\psarc{-}(0.5,-0.5){0.5}{90}{180}
\psarc{-}(-0.5,0.5){0.5}{270}{360}
\end{pspicture}
\ \ +\  e^{i\Lambda/2}\ \ 
\begin{pspicture}(-0.5,-0.1)(0.5,0.5)
\psline{-}(-0.5,-0.5)(0.5,-0.5)(0.5,0.5)(-0.5,0.5)(-0.5,-0.5)
\psset{linecolor=myc}\unlw
\psarc{-}(-0.5,-0.5){0.5}{0}{90}
\psarc{-}(0.5,0.5){0.5}{180}{270}
\end{pspicture}\ \ 
\end{equation*}\\
and $\Lambda = \pi - \lambda$.  
As for the transfer matrix $T_N$, the outer edges of $F_N$ are identified.
$F_N$ possesses two crucial properties \cite{AMDSA}. First $e_i F_N = F_N e_i $ for all $1 \le i \le N$.  Since $ [F_N, \Omega^{\pm 1}] = 0$, $F_N$ is in the center of $\eptl(\beta,\alpha)$. Second, 
if the Fourier modes of $T_N$ are denoted by $f_k\in\eptl(\beta,\alpha)$ (with 
$T_N(\lambda,z+\lambda/2) = \sum_{k=-N}^N f_k e^{-i k z }$), then $F_N  = 2^{N}f_{N}$, as a simple computation shows.

The rest of the section is devoted to the study of Jordan cells of $F_N$ in the $\rho$ representation. In fact lemma 4.1 and proposition 4.10 of \cite{AMDSA} show that, because of the commuting property (i) of the transfer matrix $T_N$ (see paragraph \ref{sec:transferMatrix}), any Jordan cells in one of its Fourier modes will be present in $T_N(\lambda,\nu)$ for all but a finite set of values of $\nu$. 

\begin{Proposition} In the representation $\rho$, the central element $F_N$ acts as an upper-triangular matrix (in a basis ordered with increasing defect number) and its spectrum can be read from its restrictions to the subspaces $\tilde V_N^d$: $$\rho(F_N)|_d = 2 C_{d/2}\, \rho(id)|_d, \quad \textrm{for\ }d>0,\qquad\textrm{and}\qquad \rho(F_N)|_{d=0} = \alpha \, \rho(id)|_{d=0}.$$
\label{sec:eigenF}\end{Proposition}
\noindent{\scshape Proof\ \ } The proof rests upon the deceivingly simple identity
\begin{equation}
\label{eq:semicircle}
\psset{unit=0.6}
\psset{linewidth=1pt}
\begin{pspicture}(-0.5,0)(2.5,1.0)
\psline{-}(0,-0.5)(2,-0.5)(2,0.5)(0,0.5)(0,-0.5)
\psline{-}(1,-0.5)(1,0.5)
\psdots(0.5,0.5)(1.5,0.5)(0.5,-0.5)(1.5,-0.5)
\psset{linecolor=myc}\unlw
\psline{-}(0,0)(2,0)
\psline{-}(0.5,-0.5)(0.5,-0.15)\psline{-}(0.5,0.5)(0.5,0.15)
\psline{-}(1.5,-0.5)(1.5,-0.15)\psline{-}(1.5,0.5)(1.5,0.15)
\psset{linecolor=myc2}
\psarc{-}(1,0.5){0.5}{0}{180}
\end{pspicture} 
=
\begin{pspicture}(-0.5,0)(2.5,1.0)
\psline{-}(0,-0.5)(2,-0.5)(2,0.5)(0,0.5)(0,-0.5)
\psline{-}(1,-0.5)(1,0.5)
\psdots(0.5,0.5)(1.5,0.5)(0.5,-0.5)(1.5,-0.5)
\psset{linecolor=myc}\unlw
\psarc{-}(1,-0.5){0.5}{0}{180}
\psarc{-}(0,0.5){0.5}{270}{360}
\psarc{-}(2,0.5){0.5}{180}{270}
\psset{linecolor=myc2}
\psarc{-}(1,0.5){0.5}{0}{180}
\end{pspicture}
\end{equation} \vspace{-0.30cm}

\noindent that can be shown by expanding the two tiles of the left-hand side.
This identity implies that any pattern of nested bubbles will percolate through the tiles of $F_N(\Lambda)$ unchanged. An example is sufficient to convince oneself of the validity of this statement:
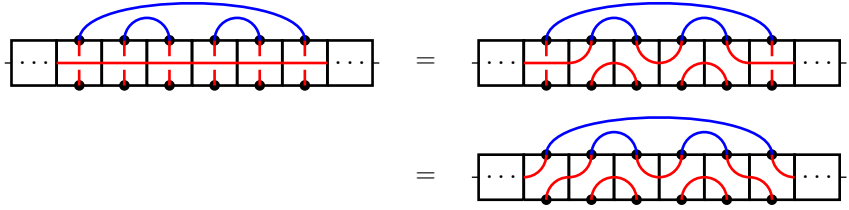
\begin{align*}
\psset{linewidth=1pt}
\psset{unit=0.6}
\begin{pspicture}[shift=-0.4](-1.2,0.0)(7.2,1.8)
\psdots(0.5,0)(1.5,0)(2.5,0)(3.5,0)(4.5,0)(5.5,0)
\psdots(0.5,1)(1.5,1)(2.5,1)(3.5,1)(4.5,1)(5.5,1)
\lw
\psline{-}(-1.15,0.5)(-1.0,0.5)
\psline{-}(7.15,0.5)(7.0,0.5)
\unlw
\rput(-0.5,0.5){$\ldots$}
\rput(6.5,0.5){$\ldots$}
\psline{-}(0,0)(-1,0)(-1,1)(0,1)(0,0)
\psline{-}(6,0)(7,0)(7,1)(6,1)(6,0)
\psline{-}(0,0)(1,0)(1,1)(0,1)(0,0)
\psline{-}(1,0)(2,0)(2,1)(1,1)(1,0)
\psline{-}(2,0)(3,0)(3,1)(2,1)(2,0)
\psline{-}(3,0)(4,0)(4,1)(3,1)(3,0)
\psline{-}(4,0)(5,0)(5,1)(4,1)(4,0)
\psline{-}(5,0)(6,0)(6,1)(5,1)(5,0)
\psset{linecolor=myc}
\psline{-}(0,0.5)(1,0.5)
\psline{-}(0.5,0.)(0.5,0.35)
\psline{-}(0.5,0.65)(0.5,1)
\psline{-}(1,0.5)(2,0.5)
\psline{-}(1.5,0.)(1.5,0.35)
\psline{-}(1.5,0.65)(1.5,1)
\psline{-}(2,0.5)(3,0.5)
\psline{-}(2.5,0.)(2.5,0.35)
\psline{-}(2.5,0.65)(2.5,1)
\psline{-}(3,0.5)(4,0.5)
\psline{-}(3.5,0.)(3.5,0.35)
\psline{-}(3.5,0.65)(3.5,1)
\psline{-}(4,0.5)(5,0.5)
\psline{-}(4.5,0.)(4.5,0.35)
\psline{-}(4.5,0.65)(4.5,1)
\psline{-}(5,0.5)(6,0.5)
\psline{-}(5.5,0.)(5.5,0.35)
\psline{-}(5.5,0.65)(5.5,1)
\psset{linecolor=myc2}
\psarc{-}(2,1.0){0.5}{0}{180}
\psarc{-}(4,1.0){0.5}{0}{180}
\psbezier{-}(0.5,1)(0.5,2.1)(5.5,2.1)(5.5,1)
\end{pspicture}
& \quad = \quad 
\psset{linewidth=1pt}
\psset{unit=0.6}
\begin{pspicture}[shift=-0.4](-1.2,0.0)(7.2,1.8)
\psdots(0.5,0)(1.5,0)(2.5,0)(3.5,0)(4.5,0)(5.5,0)
\psdots(0.5,1)(1.5,1)(2.5,1)(3.5,1)(4.5,1)(5.5,1)
\lw
\psline{-}(-1.15,0.5)(-1.0,0.5)
\psline{-}(7.15,0.5)(7.0,0.5)
\unlw
\rput(-0.5,0.5){$\ldots$}
\rput(6.5,0.5){$\ldots$}
\psline{-}(0,0)(-1,0)(-1,1)(0,1)(0,0)
\psline{-}(6,0)(7,0)(7,1)(6,1)(6,0)
\psline{-}(0,0)(1,0)(1,1)(0,1)(0,0)
\psline{-}(1,0)(2,0)(2,1)(1,1)(1,0)
\psline{-}(2,0)(3,0)(3,1)(2,1)(2,0)
\psline{-}(3,0)(4,0)(4,1)(3,1)(3,0)
\psline{-}(4,0)(5,0)(5,1)(4,1)(4,0)
\psline{-}(5,0)(6,0)(6,1)(5,1)(5,0)
\psset{linecolor=myc}
\psset{linecolor=myc}
\psline{-}(0,0.5)(1,0.5)
\psline{-}(0.5,0.)(0.5,0.35)
\psline{-}(0.5,0.65)(0.5,1)
\psline{-}(5,0.5)(6,0.5)
\psline{-}(5.5,0.)(5.5,0.35)
\psline{-}(5.5,0.65)(5.5,1)
\psarc{-}(1,1.0){0.5}{270}{360}
\psarc{-}(3,1.0){0.5}{180}{360}
\psarc{-}(5,1.0){0.5}{180}{270}
\psarc{-}(2,0.0){0.5}{0}{180}
\psarc{-}(4,0.0){0.5}{0}{180}
\psset{linecolor=myc2}
\psarc{-}(2,1.0){0.5}{0}{180}
\psarc{-}(4,1.0){0.5}{0}{180}
\psbezier{-}(0.5,1)(0.5,2.1)(5.5,2.1)(5.5,1)
\end{pspicture}
 \\[0.3cm] 
& \quad= \quad
\psset{linewidth=1pt}\psset{unit=0.6}
\begin{pspicture}[shift=-0.4](-1.2,0.0)(7.2,1.8)
\psdots(0.5,0)(1.5,0)(2.5,0)(3.5,0)(4.5,0)(5.5,0)
\psdots(0.5,1)(1.5,1)(2.5,1)(3.5,1)(4.5,1)(5.5,1)
\lw
\psline{-}(-1.15,0.5)(-1.0,0.5)
\psline{-}(7.15,0.5)(7.0,0.5)
\unlw
\rput(-0.5,0.5){$\ldots$}
\rput(6.5,0.5){$\ldots$}
\psline{-}(0,0)(-1,0)(-1,1)(0,1)(0,0)
\psline{-}(6,0)(7,0)(7,1)(6,1)(6,0)\psline{-}(0,0)(1,0)(1,1)(0,1)(0,0)
\psline{-}(1,0)(2,0)(2,1)(1,1)(1,0)
\psline{-}(2,0)(3,0)(3,1)(2,1)(2,0)
\psline{-}(3,0)(4,0)(4,1)(3,1)(3,0)
\psline{-}(4,0)(5,0)(5,1)(4,1)(4,0)
\psline{-}(5,0)(6,0)(6,1)(5,1)(5,0)
\psset{linecolor=myc}
\psarc{-}(6,1.0){0.5}{180}{270}
\psarc{-}(0,1.0){0.5}{270}{360}
\psarc{-}(1,0.0){0.5}{90}{180}
\psarc{-}(5,0.0){0.5}{0}{90}
\psarc{-}(1,1.0){0.5}{270}{360}
\psarc{-}(3,1.0){0.5}{180}{360}
\psarc{-}(5,1.0){0.5}{180}{270}
\psarc{-}(2,0.0){0.5}{0}{180}
\psarc{-}(4,0.0){0.5}{0}{180}
\psset{linecolor=myc2}
\psarc{-}(2,1.0){0.5}{0}{180}
\psarc{-}(4,1.0){0.5}{0}{180}
\psbezier{-}(0.5,1)(0.5,2.1)(5.5,2.1)(5.5,1)
\end{pspicture}
\end{align*}
In the above diagrammatic equation, six consecutive tiles of $F_N$ act on a pattern of nested bubbles. Equation \eqref{eq:semicircle} is first used on the interior bubbles, that is, on tiles $2$ and $3$, and on tiles $4$ and $5$. To obtain the last diagram, note that the two remaining tiles to be summed over are connected exactly as in \eqref{eq:semicircle}, namely the link emerging from the right edge of the left one enters the left edge of the right one. These two tiles can then be replaced by the two corresponding tiles of the right-hand side of \eqref{eq:semicircle}. 
Thus the $(N-d)/2$ bubbles of any $w\in\tilde B_N^d$ pass through $F_N$ unchanged and the component of $F_Nw$ 
 in $\tilde V_N^d$ is along $w$ itself. (We shall omit from now on the explicit reference to $\rho$ when the context is clear. Hence we wrote $F_Nw$ instead of $\rho(F_N)w$.) 
The matrix element $\langle w | F_N w\rangle$ is then equal to $\langle w^d | F_d w^{d}\rangle$. This constant is computed diagrammatically as follows. 
The leftmost tile is first expanded.
\begin{equation}\label{eq:1000} \langle w^d | F_d w^{d}\rangle = \ \
\psset{linewidth=1pt}
\psset{unit=0.6}
\overbrace{
\begin{pspicture}[shift=-0.4](-0,0.0)(4,1.5)
\psdots(0.5,0)(1.5,0)
(3.5,0)
\psdots(0.5,1)(1.5,1)
(3.5,1)
\lw
\psline{-}(-0.15,0.5)(0.0,0.5)
\psline{-}(4.15,0.5)(4.0,0.5)
\unlw
\psline{-}(0,0)(1,0)(1,1)(0,1)(0,0)
\psline{-}(1,0)(2,0)(2,1)(1,1)(1,0)
\psline{-}(2,0)(2.1,0)\psline[linestyle=dashed,dash=2pt 2pt](2.1,0)(2.9,0)
\psline{-}(3,0)(2.9,0)\psline[linestyle=dashed,dash=2pt 2pt](2.1,1)(2.9,1)
\psline{-}(3,1)(2.9,1)
\psline{-}(2,1)(2.1,1)
\psline{-}(3,0)(4,0)(4,1)(3,1)(3,0)
\psset{linecolor=myc2}
\psline{-}(0.5,1)(0.5,1.5)
\psline{-}(1.5,1)(1.5,1.5)
\psline{-}(3.5,1)(3.5,1.5)
\psset{linecolor=myc3}
\psline[linestyle=dashed,dash=2pt 2pt](0.5,0)(0.5,-0.5)
\psline[linestyle=dashed,dash=2pt 2pt](1.5,0)(1.5,-0.5)
\psline[linestyle=dashed,dash=2pt 2pt](3.5,0)(3.5,-0.5)
\psset{linecolor=myc}
\psline{-}(0,0.5)(1,0.5)
\psline{-}(0.5,0.)(0.5,0.35)
\psline{-}(0.5,0.65)(0.5,1)
\psline{-}(1,0.5)(2,0.5)
\psline{-}(1.5,0.)(1.5,0.35)
\psline{-}(1.5,0.65)(1.5,1)
\psline{-}(3,0.5)(4,0.5)
\psline{-}(3.5,0.)(3.5,0.35)
\psline{-}(3.5,0.65)(3.5,1)
\end{pspicture}}^d 
\ \ = e^{i\Lambda/2}\ \ 
\begin{pspicture}[shift=-0.4](-0,0.0)(4,1.5)
\psdots(0.5,0)(1.5,0)(3.5,0)
\psdots(0.5,1)(1.5,1)(3.5,1)
\lw
\psline{-}(-0.15,0.5)(0.0,0.5)
\psline{-}(4.15,0.5)(4.0,0.5)
\unlw
\psline{-}(0,0)(1,0)(1,1)(0,1)(0,0)
\psline{-}(1,0)(2,0)(2,1)(1,1)(1,0)
\psline{-}(2,0)(2.1,0)\psline[linestyle=dashed,dash=2pt 2pt](2.1,0)(2.9,0)
\psline{-}(3,0)(2.9,0)\psline[linestyle=dashed,dash=2pt 2pt](2.1,1)(2.9,1)
\psline{-}(3,1)(2.9,1)
\psline{-}(2,1)(2.1,1)
\psline{-}(3,0)(4,0)(4,1)(3,1)(3,0)
\psset{linecolor=myc2}
\psline{-}(0.5,1)(0.5,1.5)
\psline{-}(1.5,1)(1.5,1.5)
\psline{-}(3.5,1)(3.5,1.5)
\psset{linecolor=myc3}
\psline[linestyle=dashed,dash=2pt 2pt](0.5,0)(0.5,-0.5)
\psline[linestyle=dashed,dash=2pt 2pt](1.5,0)(1.5,-0.5)
\psline[linestyle=dashed,dash=2pt 2pt](3.5,0)(3.5,-0.5)
\psset{linecolor=myc}
\psarc{-}(0,0){0.5}{0}{90}
\psarc{-}(1,1){0.5}{180}{270}
\psline{-}(1,0.5)(2,0.5)
\psline{-}(1.5,0.)(1.5,0.35)
\psline{-}(1.5,0.65)(1.5,1)
\psline{-}(3,0.5)(4,0.5)
\psline{-}(3.5,0.)(3.5,0.35)
\psline{-}(3.5,0.65)(3.5,1)
\end{pspicture} \ \ + e^{-i\Lambda/2}\ \ 
\begin{pspicture}[shift=-0.4](-0,0.0)(4,1.5)
\psdots(0.5,0)(1.5,0)(3.5,0)
\psdots(0.5,1)(1.5,1)(3.5,1)
\lw
\psline{-}(-0.15,0.5)(0.0,0.5)
\psline{-}(4.15,0.5)(4.0,0.5)
\unlw
\psline{-}(0,0)(1,0)(1,1)(0,1)(0,0)
\psline{-}(1,0)(2,0)(2,1)(1,1)(1,0)
\psline{-}(2,0)(2.1,0)\psline[linestyle=dashed,dash=2pt 2pt](2.1,0)(2.9,0)
\psline{-}(3,0)(2.9,0)\psline[linestyle=dashed,dash=2pt 2pt](2.1,1)(2.9,1)
\psline{-}(3,1)(2.9,1)
\psline{-}(2,1)(2.1,1)
\psline{-}(3,0)(4,0)(4,1)(3,1)(3,0)
\psset{linecolor=myc2}
\psline{-}(0.5,1)(0.5,1.5)
\psline{-}(1.5,1)(1.5,1.5)
\psline{-}(3.5,1)(3.5,1.5)
\psset{linecolor=myc3}
\psline[linestyle=dashed,dash=2pt 2pt](0.5,0)(0.5,-0.5)
\psline[linestyle=dashed,dash=2pt 2pt](1.5,0)(1.5,-0.5)
\psline[linestyle=dashed,dash=2pt 2pt](3.5,0)(3.5,-0.5)
\psset{linecolor=myc}
\psarc{-}(1,0){0.5}{90}{180}
\psarc{-}(0,1){0.5}{270}{360}
\psline{-}(1,0.5)(2,0.5)
\psline{-}(1.5,0.)(1.5,0.35)
\psline{-}(1.5,0.65)(1.5,1)
\psline{-}(3,0.5)(4,0.5)
\psline{-}(3.5,0.)(3.5,0.35)
\psline{-}(3.5,0.65)(3.5,1)
\end{pspicture} \ \  .
\end{equation} \vspace{-0.2cm}

\noindent Recall that dashed patterns under the tiles of $F_d$ indicate what is the desired exiting pattern. Here the exiting state is $w^d$ and all defects entering on the upper edge of $F_d$ must connected to the lower one. Consider now the first term in the right-hand side above. Upon expansion of the second tile two terms are created:
$$
e^{i\Lambda/2}\ \ 
\psset{linewidth=1pt}
\psset{unit=0.6}\begin{pspicture}[shift=-0.4](-0,0.0)(4,1.5)
\psdots(0.5,0)(1.5,0)(3.5,0)
\psdots(0.5,1)(1.5,1)(3.5,1)
\lw
\psline{-}(-0.15,0.5)(0.0,0.5)
\psline{-}(4.15,0.5)(4.0,0.5)
\unlw
\psline{-}(0,0)(1,0)(1,1)(0,1)(0,0)
\psline{-}(1,0)(2,0)(2,1)(1,1)(1,0)
\psline{-}(2,0)(2.1,0)\psline[linestyle=dashed,dash=2pt 2pt](2.1,0)(2.9,0)
\psline{-}(3,0)(2.9,0)\psline[linestyle=dashed,dash=2pt 2pt](2.1,1)(2.9,1)
\psline{-}(3,1)(2.9,1)
\psline{-}(2,1)(2.1,1)
\psline{-}(3,0)(4,0)(4,1)(3,1)(3,0)
\psset{linecolor=myc2}
\psline{-}(0.5,1)(0.5,1.5)
\psline{-}(1.5,1)(1.5,1.5)
\psline{-}(3.5,1)(3.5,1.5)
\psset{linecolor=myc3}
\psline[linestyle=dashed,dash=2pt 2pt](0.5,0)(0.5,-0.5)
\psline[linestyle=dashed,dash=2pt 2pt](1.5,0)(1.5,-0.5)
\psline[linestyle=dashed,dash=2pt 2pt](3.5,0)(3.5,-0.5)
\psset{linecolor=myc}
\psarc{-}(0,0){0.5}{0}{90}
\psarc{-}(1,1){0.5}{180}{270}
\psline{-}(1,0.5)(2,0.5)
\psline{-}(1.5,0.)(1.5,0.35)
\psline{-}(1.5,0.65)(1.5,1)
\psline{-}(3,0.5)(4,0.5)
\psline{-}(3.5,0.)(3.5,0.35)
\psline{-}(3.5,0.65)(3.5,1)
\end{pspicture} 
\ \ = e^{2\cdot i\Lambda/2}\ \ 
\begin{pspicture}[shift=-0.4](-0,0.0)(4,1.5)
\psdots(0.5,0)(1.5,0)(3.5,0)
\psdots(0.5,1)(1.5,1)(3.5,1)
\lw
\psline{-}(-0.15,0.5)(0.0,0.5)
\psline{-}(4.15,0.5)(4.0,0.5)
\unlw
\psline{-}(0,0)(1,0)(1,1)(0,1)(0,0)
\psline{-}(1,0)(2,0)(2,1)(1,1)(1,0)
\psline{-}(2,0)(2.1,0)\psline[linestyle=dashed,dash=2pt 2pt](2.1,0)(2.9,0)
\psline{-}(3,0)(2.9,0)\psline[linestyle=dashed,dash=2pt 2pt](2.1,1)(2.9,1)
\psline{-}(3,1)(2.9,1)
\psline{-}(2,1)(2.1,1)
\psline{-}(3,0)(4,0)(4,1)(3,1)(3,0)
\psset{linecolor=myc2}
\psline{-}(0.5,1)(0.5,1.5)
\psline{-}(1.5,1)(1.5,1.5)
\psline{-}(3.5,1)(3.5,1.5)
\psset{linecolor=myc3}
\psline[linestyle=dashed,dash=2pt 2pt](0.5,0)(0.5,-0.5)
\psline[linestyle=dashed,dash=2pt 2pt](1.5,0)(1.5,-0.5)
\psline[linestyle=dashed,dash=2pt 2pt](3.5,0)(3.5,-0.5)
\psset{linecolor=myc}
\psarc{-}(0,0){0.5}{0}{90}
\psarc{-}(1,1){0.5}{180}{270}
\psarc{-}(1,0){0.5}{0}{90}
\psarc{-}(2,1){0.5}{180}{270}
\psline{-}(3,0.5)(4,0.5)
\psline{-}(3.5,0.)(3.5,0.35)
\psline{-}(3.5,0.65)(3.5,1)
\end{pspicture} \ \ + 1 \ \ 
\begin{pspicture}[shift=-0.4](-0,0.0)(4,1.5)
\psdots(0.5,0)(1.5,0)(3.5,0)
\psdots(0.5,1)(1.5,1)(3.5,1)
\lw
\psline{-}(-0.15,0.5)(0.0,0.5)
\psline{-}(4.15,0.5)(4.0,0.5)
\unlw
\psline{-}(0,0)(1,0)(1,1)(0,1)(0,0)
\psline{-}(1,0)(2,0)(2,1)(1,1)(1,0)
\psline{-}(2,0)(2.1,0)\psline[linestyle=dashed,dash=2pt 2pt](2.1,0)(2.9,0)
\psline{-}(3,0)(2.9,0)\psline[linestyle=dashed,dash=2pt 2pt](2.1,1)(2.9,1)
\psline{-}(3,1)(2.9,1)
\psline{-}(2,1)(2.1,1)
\psline{-}(3,0)(4,0)(4,1)(3,1)(3,0)
\psset{linecolor=myc2}
\psline{-}(0.5,1)(0.5,1.5)
\psline{-}(1.5,1)(1.5,1.5)
\psline{-}(3.5,1)(3.5,1.5)
\psset{linecolor=myc3}
\psline[linestyle=dashed,dash=2pt 2pt](0.5,0)(0.5,-0.5)
\psline[linestyle=dashed,dash=2pt 2pt](1.5,0)(1.5,-0.5)
\psline[linestyle=dashed,dash=2pt 2pt](3.5,0)(3.5,-0.5)
\psset{linecolor=myc}
\psarc{-}(1,0){0.5}{9}{180}
\psarc{-}(0,1){0.5}{270}{360}
\psarc{-}(2,1){0.5}{180}{270}
\psline{-}(3,0.5)(4,0.5)
\psline{-}(3.5,0.)(3.5,0.35)
\psline{-}(3.5,0.65)(3.5,1)
\end{pspicture}\ \  . \\ 
$$
The second diagram is a contribution to other matrix elements, namely to elements that have a bubble between positions $1$ and $2$. It drops out of the current computation. Similarly the expansion of the second term in the right-hand side of \eqref{eq:1000} produces a link between positions $1$ and $2$ on the upper edge and, consequently, two positions of the lower one will eventually be connected when the other tiles are expanded. Again this term must be ignored. By repeating this argument on the next tiles, we are left with only two terms:
\begin{align*} \langle w^d | F_d w^{d}\rangle & = \ \
\psset{linewidth=1pt}
\psset{unit=0.6}
\overbrace{
\begin{pspicture}(-0,0.395)(4,1.65)
\psdots(0.5,0)(1.5,0)
(3.5,0)
\psdots(0.5,1)(1.5,1)
(3.5,1)
\lw
\psline{-}(-0.15,0.5)(0.0,0.5)
\psline{-}(4.15,0.5)(4.0,0.5)
\unlw
\psline{-}(0,0)(1,0)(1,1)(0,1)(0,0)
\psline{-}(1,0)(2,0)(2,1)(1,1)(1,0)
\psline{-}(2,0)(2.1,0)\psline[linestyle=dashed,dash=2pt 2pt](2.1,0)(2.9,0)
\psline{-}(3,0)(2.9,0)\psline[linestyle=dashed,dash=2pt 2pt](2.1,1)(2.9,1)
\psline{-}(3,1)(2.9,1)
\psline{-}(2,1)(2.1,1)
\psline{-}(3,0)(4,0)(4,1)(3,1)(3,0)
\psset{linecolor=myc2}
\psline{-}(0.5,1)(0.5,1.5)
\psline{-}(1.5,1)(1.5,1.5)
\psline{-}(3.5,1)(3.5,1.5)
\psset{linecolor=myc3}
\psline[linestyle=dashed,dash=2pt 2pt](0.5,0)(0.5,-0.5)
\psline[linestyle=dashed,dash=2pt 2pt](1.5,0)(1.5,-0.5)
\psline[linestyle=dashed,dash=2pt 2pt](3.5,0)(3.5,-0.5)
\psset{linecolor=myc}
\psline{-}(0,0.5)(1,0.5)
\psline{-}(0.5,0.)(0.5,0.35)
\psline{-}(0.5,0.65)(0.5,1)
\psline{-}(1,0.5)(2,0.5)
\psline{-}(1.5,0.)(1.5,0.35)
\psline{-}(1.5,0.65)(1.5,1)
\psline{-}(3,0.5)(4,0.5)
\psline{-}(3.5,0.)(3.5,0.35)
\psline{-}(3.5,0.65)(3.5,1)
\end{pspicture}}^d 
\ = e^{id\Lambda/2}\ \ 
\begin{pspicture}(-0,0.395)(4,1.5)
\psdots(0.5,0)(1.5,0)(3.5,0)
\psdots(0.5,1)(1.5,1)(3.5,1)
\lw
\psline{-}(-0.15,0.5)(0.0,0.5)
\psline{-}(4.15,0.5)(4.0,0.5)
\unlw
\psline{-}(0,0)(1,0)(1,1)(0,1)(0,0)
\psline{-}(1,0)(2,0)(2,1)(1,1)(1,0)
\psline{-}(2,0)(2.1,0)\psline[linestyle=dashed,dash=2pt 2pt](2.1,0)(2.9,0)
\psline{-}(3,0)(2.9,0)\psline[linestyle=dashed,dash=2pt 2pt](2.1,1)(2.9,1)
\psline{-}(3,1)(2.9,1)
\psline{-}(2,1)(2.1,1)
\psline{-}(3,0)(4,0)(4,1)(3,1)(3,0)
\psset{linecolor=myc2}
\psline{-}(0.5,1)(0.5,1.5)
\psline{-}(1.5,1)(1.5,1.5)
\psline{-}(3.5,1)(3.5,1.5)
\psset{linecolor=myc3}
\psline[linestyle=dashed,dash=2pt 2pt](0.5,0)(0.5,-0.5)
\psline[linestyle=dashed,dash=2pt 2pt](1.5,0)(1.5,-0.5)
\psline[linestyle=dashed,dash=2pt 2pt](3.5,0)(3.5,-0.5)
\psset{linecolor=myc}
\psarc{-}(0,0){0.5}{0}{90}
\psarc{-}(1,0){0.5}{0}{90}
\psarc{-}(3,0){0.5}{0}{90}
\psarc{-}(1,1){0.5}{180}{270}
\psarc{-}(2,1){0.5}{180}{270}
\psarc{-}(4,1){0.5}{180}{270}
\end{pspicture} \ \ + e^{-id\Lambda/2}\ \ 
\begin{pspicture}(-0,0.395)(4,1.5)
\psdots(0.5,0)(1.5,0)(3.5,0)
\psdots(0.5,1)(1.5,1)(3.5,1)
\lw
\psline{-}(-0.15,0.5)(0.0,0.5)
\psline{-}(4.15,0.5)(4.0,0.5)
\unlw
\psline{-}(0,0)(1,0)(1,1)(0,1)(0,0)
\psline{-}(1,0)(2,0)(2,1)(1,1)(1,0)
\psline{-}(2,0)(2.1,0)\psline[linestyle=dashed,dash=2pt 2pt](2.1,0)(2.9,0)
\psline{-}(3,0)(2.9,0)\psline[linestyle=dashed,dash=2pt 2pt](2.1,1)(2.9,1)
\psline{-}(3,1)(2.9,1)
\psline{-}(2,1)(2.1,1)
\psline{-}(3,0)(4,0)(4,1)(3,1)(3,0)
\psset{linecolor=myc2}
\psline{-}(0.5,1)(0.5,1.5)
\psline{-}(1.5,1)(1.5,1.5)
\psline{-}(3.5,1)(3.5,1.5)
\psset{linecolor=myc3}
\psline[linestyle=dashed,dash=2pt 2pt](0.5,0)(0.5,-0.5)
\psline[linestyle=dashed,dash=2pt 2pt](1.5,0)(1.5,-0.5)
\psline[linestyle=dashed,dash=2pt 2pt](3.5,0)(3.5,-0.5)
\psset{linecolor=myc}
\psarc{-}(1,0){0.5}{90}{180}
\psarc{-}(2,0){0.5}{90}{180}
\psarc{-}(4,0){0.5}{90}{180}
\psarc{-}(0,1){0.5}{270}{360}
\psarc{-}(1,1){0.5}{270}{360}
\psarc{-}(3,1){0.5}{270}{360}
\end{pspicture} 
\ \ . \\ 
\end{align*}
These two remaining diagrams are precisely $\Omega^{-1}$ and $\Omega$ and both act as the identity on the link state with only defects. Therefore $\langle w^d | F_d w^{d}\rangle=2\cos (d\Lambda/2)$. The case $d = 0$ is particular: the $N/2$ bubbles all go through $F_N$  and a non-contractible loop is created, giving rise to a factor of $\alpha$. 
\hfill$\square$ 

\medskip

Because of the identity \eqref{eq:semicircle}, all matrix elements of $\rho(F_N)$ can be computed. Indeed, once the bubbles of a link state $v\in \tilde B_N^d$ have been pushed through $F_N$, it remains only to compute $F_dw^d$. It is thus sufficient to calculate the column of $\rho(F_s)$ corresponding to the components of the vector $F_sw^s$, for $0\le s\le N$ and $N \equiv s \textrm{\ mod\ } 2$, that is, the elements $\langle w | F_s w^s \rangle$ for $w$ a link state. That is the content of the next lemma.

\begin{Lemme}
Let $w^s_{k_1, \dots, k_m} \in B_s^d$ be a link state with only $1$-bubbles centered at $k_1, \dots, k_m$, $k_{i+1}>k_i+1$ and $k_i \in [1, s]$ for $1\le i \le m$. Then,
\begin{equation*}
\langle w^s_{k_1, \dots, k_m} | F_s w^s\rangle = \prod_{i=1}^m {S_{(k_{i+1}-k_{i}-1)/2}}/{S_{1/2}} 
\end{equation*}
where $k_{m+1} = k_1+s$. If $w$ contains one $n$-bubble with $n\geq 2$, then $\langle w | F_s w^s \rangle=0$.
\end{Lemme}

\noindent{\scshape Proof\ \ } Let $w^s_{k_1, \dots, k_m} \in B_s^d$ be a link state with $m\geq 1$. We zoom in first on one bubble joining nearest neighbour positions. (We remind the reader that, in the representation $\rho$, the number of defects may decrease by the action of $\eptl$.) The following diagram draws the incoming defects (top of drawing), the two corresponding tiles of $F_s$ and, in dashed curves, the bubble of the outgoing state:
$$
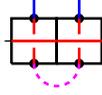
\psset{unit=0.6}
\begin{pspicture}(0,-0.25)(2,1.25)
\psdots(0.5,0)(1.5,0)
\psdots(0.5,1)(1.5,1)
\lw
\psline{-}(-0.15,0.5)(0.0,0.5)
\psline{-}(2.15,0.5)(2.0,0.5)
\unlw
\psline{-}(0,0)(1,0)(1,1)(0,1)(0,0)
\psline{-}(1,0)(2,0)(2,1)(1,1)(1,0)
\psset{linecolor=myc2}
\psline{-}(0.5,1)(0.5,1.5)
\psline{-}(1.5,1)(1.5,1.5)
\psset{linecolor=myc}
\psline{-}(0,0.5)(1,0.5)
\psline{-}(0.5,0.)(0.5,0.35)
\psline{-}(0.5,0.65)(0.5,1)
\psline{-}(1,0.5)(2,0.5)
\psline{-}(1.5,0.)(1.5,0.35)
\psline{-}(1.5,0.65)(1.5,1)
\psset{linecolor=myc3}
\psarc[linestyle=dashed,dash=2pt 2pt](1,0){0.5}{180}{360}
\end{pspicture}
$$
This dashed bubble means that the tiles of $F_s$ must be chosen so that these two positions on the lower edge are linked. Since $F_s$ has only a single row of tiles, there is a single choice possible for these two tiles, namely:
$$
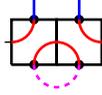
\psset{unit=0.6}
\begin{pspicture}(0,-0.25)(2,1.25)
\psdots(0.5,0)(1.5,0)
\psdots(0.5,1)(1.5,1)
\lw
\psline{-}(-0.15,0.5)(0.0,0.5)
\psline{-}(2.15,0.5)(2.0,0.5)
\unlw
\psline{-}(0,0)(1,0)(1,1)(0,1)(0,0)
\psline{-}(1,0)(2,0)(2,1)(1,1)(1,0)
\psset{linecolor=myc2}
\psline{-}(0.5,1)(0.5,1.5)
\psline{-}(1.5,1)(1.5,1.5)
\psset{linecolor=myc}
\psarc(1,0){0.5}{0}{180}
\psarc(0,1){0.5}{270}{360}
\psarc(2,1){0.5}{180}{270}
\psset{linecolor=myc3}
\psarc[linestyle=dashed,dash=2pt 2pt](1,0){0.5}{180}{360}
\end{pspicture}
$$
Suppose now that the link state $w^s_{k_1, \dots, k_m}$ contains at least one $n$-bubble with $n\geq 2$. Then there is no possible choice of the tiles that can draw this $n$-bubble and the matrix element $\langle w^s_{k_1, \dots, k_m} | F_s w^s\rangle$ must be zero. (The following example might help understand this statement. Here is the simplest $2$-bubble: 
$$
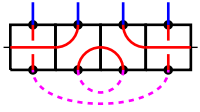
\psset{unit=0.6}
\begin{pspicture}(-1,-0.5)(3,1.25)
\psdots(0.5,0)(1.5,0)(-0.5,0)(2.5,0)
\psdots(0.5,1)(1.5,1)(-0.5,1)(2.5,1)
\lw
\psline{-}(-1.15,0.5)(-1.0,0.5)
\psline{-}(3.15,0.5)(3.0,0.5)
\unlw
\psline{-}(0,0)(1,0)(1,1)(0,1)(0,0)
\psline{-}(-1,0)(0,0)(0,1)(-1,1)(-1,0)
\psline{-}(1,0)(2,0)(2,1)(1,1)(1,0)
\psline{-}(2,0)(3,0)(3,1)(2,1)(2,0)
\psset{linecolor=myc}
\psline{-}(0,0.5)(-1,0.5)
\psline{-}(-0.5,0.)(-0.5,0.35)
\psline{-}(-0.5,0.65)(-0.5,1)
\psline{-}(2,0.5)(3,0.5)
\psline{-}(2.5,0.)(2.5,0.35)
\psline{-}(2.5,0.65)(2.5,1)
\psset{linecolor=myc2}
\psline{-}(0.5,1)(0.5,1.5)
\psline{-}(1.5,1)(1.5,1.5)
\psline{-}(-0.5,1)(-0.5,1.5)
\psline{-}(2.5,1)(2.5,1.5)
\psset{linecolor=myc}
\psarc(1,0){0.5}{0}{180}
\psarc(0,1){0.5}{270}{360}
\psarc(2,1){0.5}{180}{270}
\psset{linecolor=myc3}
\psarc[linestyle=dashed,dash=2pt 2pt](1,0){0.5}{180}{360}
\psbezier[linestyle=dashed,dash=2pt 2pt](-0.5,0)(-0.5,-1)(2.5,-1)(2.5,0.0)
\end{pspicture}
$$
It is easy to see here that no choice of the outer tiles (or of the tiles in $F_s$ that have not been drawn) can possibly connect the positions $1$ and $4$ on the lower edge since $F_s$ has a single row of tiles.)

It thus remains to study matrix elements $\langle w^s_{k_1, \dots, k_m} | F_s w^s\rangle$ where all $k_i$s represent $1$-bubble. With the above argument repeated for each of these $m$ bubbles, $2m$ tiles of $F_s$ are now fixed. For example
\begin{align*} \langle w^s_{k_1, \dots, k_m} | F_s w^s\rangle  &= \
\psset{linewidth=1pt}
\overbrace{
\psset{unit=0.6}
\begin{pspicture}(-0,0.375)(16,1.75)
\psdots(0.5,0)(1.5,0)(2.5,0)(3.5,0)(4.5,0)
(5.5,0)(6.5,0)(7.5,0)(8.5,0)(9.5,0)
(10.5,0)
(13.5,0)(14.5,0)(15.5,0)
\psdots(0.5,1)(1.5,1)(2.5,1)(3.5,1)(4.5,1)
(5.5,1)(6.5,1)(7.5,1)(8.5,1)(9.5,1)
(10.5,1)
(13.5,1)(14.5,1)(15.5,1)
\lw
\psline{-}(-0.15,0.5)(0.0,0.5)
\psline{-}(16.15,0.5)(16.0,0.5)
\unlw
\psline{-}(0,0)(1,0)(1,1)(0,1)(0,0)
\psline{-}(1,0)(2,0)(2,1)(1,1)(1,0)
\psline{-}(2,0)(3,0)(3,1)(2,1)(2,0)
\psline{-}(3,0)(4,0)(4,1)(3,1)(3,0)
\psline{-}(4,0)(5,0)(5,1)(4,1)(4,0)
\psline{-}(5,0)(6,0)(6,1)(5,1)(5,0)
\psline{-}(6,0)(7,0)(7,1)(6,1)(6,0)
\psline{-}(7,0)(8,0)(8,1)(7,1)(7,0)
\psline{-}(8,0)(9,0)(9,1)(8,1)(8,0)
\psline{-}(9,0)(10,0)(10,1)(9,1)(9,0)
\psline{-}(10,0)(11,0)(11,1)(10,1)(10,0)
\psline{-}(11,0)(11.5,0)
\psline{-}(11,1)(11.5,1)
\psline[linestyle=dashed,dash=2pt 2pt]{-}(11.5,0)(12.5,0)
\psline[linestyle=dashed,dash=2pt 2pt]{-}(11.5,1)(12.5,1)
\psline{-}(13,0)(12.5,0)
\psline{-}(13,1)(12.5,1)
\psline{-}(13,0)(14,0)(14,1)(13,1)(13,0)
\psline{-}(14,0)(15,0)(15,1)(14,1)(14,0)
\psline{-}(15,0)(16,0)(16,1)(15,1)(15,0)
\psset{linecolor=myc2}
\psline{-}(0.5,1)(0.5,1.5)
\psline{-}(1.5,1)(1.5,1.5)
\psline{-}(2.5,1)(2.5,1.5)
\psline{-}(3.5,1)(3.5,1.5)
\psline{-}(4.5,1)(4.5,1.5)
\psline{-}(5.5,1)(5.5,1.5)
\psline{-}(6.5,1)(6.5,1.5)
\psline{-}(7.5,1)(7.5,1.5)
\psline{-}(8.5,1)(8.5,1.5)
\psline{-}(9.5,1)(9.5,1.5)
\psline{-}(10.5,1)(10.5,1.5)
\psline{-}(13.5,1)(13.5,1.5)
\psline{-}(14.5,1)(14.5,1.5)
\psline{-}(15.5,1)(15.5,1.5)
\psset{linecolor=myc3}
\psline[linestyle=dashed,dash=2pt 2pt](0.5,0)(0.5,-0.5)
\psarc[linestyle=dashed,dash=2pt 2pt](2,0){0.5}{180}{360}
\psline[linestyle=dashed,dash=2pt 2pt](3.5,0)(3.5,-0.5)
\psline[linestyle=dashed,dash=2pt 2pt](4.5,0)(4.5,-0.5)
\psline[linestyle=dashed,dash=2pt 2pt](5.5,0)(5.5,-0.5)
\psarc[linestyle=dashed,dash=2pt 2pt](7,0){0.5}{180}{360}
\psarc[linestyle=dashed,dash=2pt 2pt](9,0){0.5}{180}{360}
\psline[linestyle=dashed,dash=2pt 2pt](10.5,0)(10.5,-0.5)
\psarc[linestyle=dashed,dash=2pt 2pt](14,0){0.5}{180}{360}
\psline[linestyle=dashed,dash=2pt 2pt](15.5,0)(15.5,-0.5)
\psset{linecolor=myc}
\psline{-}(0,0.5)(1,0.5)
\psline{-}(0.5,0.)(0.5,0.35)
\psline{-}(0.5,0.65)(0.5,1)
\psline{-}(1,0.5)(2,0.5)
\psline{-}(1.5,0.)(1.5,0.35)
\psline{-}(1.5,0.65)(1.5,1)
\psline{-}(2,0.5)(3,0.5)
\psline{-}(2.5,0.)(2.5,0.35)
\psline{-}(2.5,0.65)(2.5,1)
\psline{-}(3,0.5)(4,0.5)
\psline{-}(3.5,0.)(3.5,0.35)
\psline{-}(3.5,0.65)(3.5,1)
\psline{-}(4,0.5)(5,0.5)
\psline{-}(4.5,0.)(4.5,0.35)
\psline{-}(4.5,0.65)(4.5,1)
\psline{-}(5,0.5)(6,0.5)
\psline{-}(5.5,0.)(5.5,0.35)
\psline{-}(5.5,0.65)(5.5,1)
\psline{-}(6,0.5)(7,0.5)
\psline{-}(6.5,0.)(6.5,0.35)
\psline{-}(6.5,0.65)(6.5,1)
\psline{-}(7,0.5)(8,0.5)
\psline{-}(7.5,0.)(7.5,0.35)
\psline{-}(7.5,0.65)(7.5,1)
\psline{-}(8,0.5)(9,0.5)
\psline{-}(8.5,0.)(8.5,0.35)
\psline{-}(8.5,0.65)(8.5,1)
\psline{-}(9,0.5)(10,0.5)
\psline{-}(9.5,0.)(9.5,0.35)
\psline{-}(9.5,0.65)(9.5,1)
\psline{-}(10,0.5)(11,0.5)
\psline{-}(10.5,0.)(10.5,0.35)
\psline{-}(10.5,0.65)(10.5,1)
\psline{-}(13,0.5)(14,0.5)
\psline{-}(13.5,0.)(13.5,0.35)
\psline{-}(13.5,0.65)(13.5,1)
\psline{-}(14,0.5)(15,0.5)
\psline{-}(14.5,0.)(14.5,0.35)
\psline{-}(14.5,0.65)(14.5,1)
\psline{-}(15,0.5)(16,0.5)
\psline{-}(15.5,0.)(15.5,0.35)
\psline{-}(15.5,0.65)(15.5,1)
\end{pspicture}}^s \\[0.3cm] 
& = \ 
\psset{unit=0.6}
\begin{pspicture}(-0,0.375)(16,1.95)
\psdots(0.5,0)(1.5,0)(2.5,0)(3.5,0)(4.5,0)
(5.5,0)(6.5,0)(7.5,0)(8.5,0)(9.5,0)
(10.5,0)
(13.5,0)(14.5,0)(15.5,0)
\psdots(0.5,1)(1.5,1)(2.5,1)(3.5,1)(4.5,1)
(5.5,1)(6.5,1)(7.5,1)(8.5,1)(9.5,1)
(10.5,1)
(13.5,1)(14.5,1)(15.5,1)
\lw
\psline{-}(-0.15,0.5)(0.0,0.5)
\psline{-}(16.15,0.5)(16.0,0.5)
\unlw
\psline{-}(0,0)(1,0)(1,1)(0,1)(0,0)
\psline{-}(1,0)(2,0)(2,1)(1,1)(1,0)
\psline{-}(2,0)(3,0)(3,1)(2,1)(2,0)
\psline{-}(3,0)(4,0)(4,1)(3,1)(3,0)
\psline{-}(4,0)(5,0)(5,1)(4,1)(4,0)
\psline{-}(5,0)(6,0)(6,1)(5,1)(5,0)
\psline{-}(6,0)(7,0)(7,1)(6,1)(6,0)
\psline{-}(7,0)(8,0)(8,1)(7,1)(7,0)
\psline{-}(8,0)(9,0)(9,1)(8,1)(8,0)
\psline{-}(9,0)(10,0)(10,1)(9,1)(9,0)
\psline{-}(10,0)(11,0)(11,1)(10,1)(10,0)
\psline{-}(11,0)(11.5,0)
\psline{-}(11,1)(11.5,1)
\psline[linestyle=dashed,dash=2pt 2pt]{-}(11.5,0)(12.5,0)
\psline[linestyle=dashed,dash=2pt 2pt]{-}(11.5,1)(12.5,1)
\psline{-}(13,0)(12.5,0)
\psline{-}(13,1)(12.5,1)
\psline{-}(13,0)(14,0)(14,1)(13,1)(13,0)
\psline{-}(14,0)(15,0)(15,1)(14,1)(14,0)
\psline{-}(15,0)(16,0)(16,1)(15,1)(15,0)
\psset{linecolor=myc2}
\psline{-}(0.5,1)(0.5,1.5)
\psline{-}(1.5,1)(1.5,1.5)
\psline{-}(2.5,1)(2.5,1.5)
\psline{-}(3.5,1)(3.5,1.5)
\psline{-}(4.5,1)(4.5,1.5)
\psline{-}(5.5,1)(5.5,1.5)
\psline{-}(6.5,1)(6.5,1.5)
\psline{-}(7.5,1)(7.5,1.5)
\psline{-}(8.5,1)(8.5,1.5)
\psline{-}(9.5,1)(9.5,1.5)
\psline{-}(10.5,1)(10.5,1.5)
\psline{-}(13.5,1)(13.5,1.5)
\psline{-}(14.5,1)(14.5,1.5)
\psline{-}(15.5,1)(15.5,1.5)
\psset{linecolor=myc3}
\psline[linestyle=dashed,dash=2pt 2pt](0.5,0)(0.5,-0.5)
\psarc[linestyle=dashed,dash=2pt 2pt](2,0){0.5}{180}{360}
\psline[linestyle=dashed,dash=2pt 2pt](3.5,0)(3.5,-0.5)
\psline[linestyle=dashed,dash=2pt 2pt](4.5,0)(4.5,-0.5)
\psline[linestyle=dashed,dash=2pt 2pt](5.5,0)(5.5,-0.5)
\psarc[linestyle=dashed,dash=2pt 2pt](7,0){0.5}{180}{360}
\psarc[linestyle=dashed,dash=2pt 2pt](9,0){0.5}{180}{360}
\psline[linestyle=dashed,dash=2pt 2pt](10.5,0)(10.5,-0.5)
\psarc[linestyle=dashed,dash=2pt 2pt](14,0){0.5}{180}{360}
\psline[linestyle=dashed,dash=2pt 2pt](15.5,0)(15.5,-0.5)
\psset{linecolor=myc}
\psline{-}(0,0.5)(1,0.5)
\psline{-}(0.5,0.)(0.5,0.35)
\psline{-}(0.5,0.65)(0.5,1)
\psarc{-}(2,0){0.5}{0}{180}
\psarc{-}(1,1){0.5}{270}{360}
\psarc{-}(3,1){0.5}{180}{270}
\psline{-}(3,0.5)(4,0.5)
\psline{-}(3.5,0.)(3.5,0.35)
\psline{-}(3.5,0.65)(3.5,1)
\psline{-}(4,0.5)(5,0.5)
\psline{-}(4.5,0.)(4.5,0.35)
\psline{-}(4.5,0.65)(4.5,1)
\psline{-}(5,0.5)(6,0.5)
\psline{-}(5.5,0.)(5.5,0.35)
\psline{-}(5.5,0.65)(5.5,1)
\psarc{-}(7,0){0.5}{0}{180}
\psarc{-}(6,1){0.5}{270}{360}
\psarc{-}(8,1){0.5}{180}{270}
\psarc{-}(9,0){0.5}{0}{180}
\psarc{-}(8,1){0.5}{270}{360}
\psarc{-}(10,1){0.5}{180}{270}
\psline{-}(10,0.5)(11,0.5)
\psline{-}(10.5,0.)(10.5,0.35)
\psline{-}(10.5,0.65)(10.5,1)
\psarc{-}(14,0){0.5}{0}{180}
\psarc{-}(13,1){0.5}{270}{360}
\psarc{-}(15,1){0.5}{180}{270}
\psline{-}(15,0.5)(16,0.5)
\psline{-}(15.5,0.)(15.5,0.35)
\psline{-}(15.5,0.65)(15.5,1)
\end{pspicture} \ \ .
\end{align*}
\vskip0.35cm

\noindent where the first $k_i$s have been chosen as $k_1=2, k_2=7$ and $k_3=9$. The remaining tiles must be chosen so that the remaining positions on the lower edge are connected to defects of the incoming state $w^s$. The choice of tiles between two consecutive $1$-bubbles will have to be done together to respect this requirement. However the choice for tiles between a given pair of consecutive bubbles and the choice for those between another pair are independent. The matrix element thus factorizes as
$$\langle w^s_{k_1, \dots, k_m} | F_s w^s\rangle =
A(s+k_1-k_m)\prod_{i=1}^{m-1} A(k_{i+1}-k_{i}) $$
where the notation 
$$ A(n)  = \,
\psset{linewidth=1pt}
\overbrace{
\psset{unit=0.6}
\begin{pspicture}[shift=-0.8](-0.5,-0.4)(7.5,1.95)
\psdots(0.5,0)(1.5,0)(2.5,0)(3.5,0)(4.5,0)(5.5,0)(6.5,0)
\psdots(0.5,1)(1.5,1)(2.5,1)(3.5,1)(4.5,1)(5.5,1)(6.5,1)(-0.5,1)(7.5,1)
\psline{-}(0,0)(1,0)(1,1)(0,1)(0,0)
\psline{-}(1,0)(2,0)(2,1)(1,1)(1,0)
\psline{-}(2,0)(3,0)(3,1)(2,1)(2,0)
\psline{-}(3,0)(4,0)(4,1)(3,1)(3,0)
\psline{-}(4,0)(5,0)(5,1)(4,1)(4,0)
\psline{-}(5,0)(6,0)(6,1)(5,1)(5,0)
\psline{-}(6,0)(7,0)(7,1)(6,1)(6,0)
\psset{linecolor=myc2} 
\psline{-}(-0.5,1)(-0.5,1.5)
\psline{-}(0.5,1)(0.5,1.5)
\psline{-}(1.5,1)(1.5,1.5)
\psline{-}(2.5,1)(2.5,1.5)
\psline{-}(3.5,1)(3.5,1.5)
\psline{-}(4.5,1)(4.5,1.5)
\psline{-}(5.5,1)(5.5,1.5)
\psline{-}(6.5,1)(6.5,1.5)
\psline{-}(7.5,1)(7.5,1.5)
\psset{linecolor=myc3} 
\psline[linestyle=dashed,dash=2pt 2pt](0.5,0)(0.5,-0.5)
\psline[linestyle=dashed,dash=2pt 2pt](1.5,0)(1.5,-0.5)
\psline[linestyle=dashed,dash=2pt 2pt](2.5,0)(2.5,-0.5)
\psline[linestyle=dashed,dash=2pt 2pt](3.5,0)(3.5,-0.5)
\psline[linestyle=dashed,dash=2pt 2pt](4.5,0)(4.5,-0.5)
\psline[linestyle=dashed,dash=2pt 2pt](5.5,0)(5.5,-0.5)
\psline[linestyle=dashed,dash=2pt 2pt](6.5,0)(6.5,-0.5)
\psset{linecolor=myc} 
\psline{-}(0,0.5)(1,0.5)
\psline{-}(0.5,0.)(0.5,0.35)
\psline{-}(0.5,0.65)(0.5,1)
\psline{-}(1,0.5)(2,0.5)
\psline{-}(1.5,0.)(1.5,0.35)
\psline{-}(1.5,0.65)(1.5,1)
\psline{-}(2,0.5)(3,0.5)
\psline{-}(2.5,0.)(2.5,0.35)
\psline{-}(2.5,0.65)(2.5,1)
\psline{-}(3,0.5)(4,0.5)
\psline{-}(3.5,0.)(3.5,0.35)
\psline{-}(3.5,0.65)(3.5,1)
\psline{-}(4,0.5)(5,0.5)
\psline{-}(4.5,0.)(4.5,0.35)
\psline{-}(4.5,0.65)(4.5,1)
\psline{-}(5,0.5)(6,0.5)
\psline{-}(5.5,0.)(5.5,0.35)
\psline{-}(5.5,0.65)(5.5,1)
\psline{-}(6,0.5)(7,0.5)
\psline{-}(6.5,0.)(6.5,0.35)
\psline{-}(6.5,0.65)(6.5,1)
\psarc{-}(0,1){0.5}{180}{270}
\psarc{-}(7,1){0.5}{270}{0}
\end{pspicture}}^n 
$$

\noindent has been introduced and where it is understood that, in the underlying summation over all possible tiles, only the contributions where $(n-2)$ defects reach the lower edge are kept. With this last constraint, $A(n)$ is then a number. The computation is now straightforward and mimics that of the previous lemma.
\begin{align*} A(n) 
& = e^{i (n-2)\Lambda / 2}\, \,
\psset{unit=0.6}
\begin{pspicture}(-0.5,0.375)(7.5,1.75)
\psdots(0.5,0)(1.5,0)(2.5,0)(3.5,0)(4.5,0)(5.5,0)(6.5,0)
\psdots(0.5,1)(1.5,1)(2.5,1)(3.5,1)(4.5,1)(5.5,1)(6.5,1)(-0.5,1)(7.5,1)
\psline{-}(0,0)(1,0)(1,1)(0,1)(0,0)
\psline{-}(1,0)(2,0)(2,1)(1,1)(1,0)
\psline{-}(2,0)(3,0)(3,1)(2,1)(2,0)
\psline{-}(3,0)(4,0)(4,1)(3,1)(3,0)
\psline{-}(4,0)(5,0)(5,1)(4,1)(4,0)
\psline{-}(5,0)(6,0)(6,1)(5,1)(5,0)
\psline{-}(6,0)(7,0)(7,1)(6,1)(6,0)
\psset{linecolor=myc2} \unlw
\psline{-}(-0.5,1)(-0.5,1.5)
\psline{-}(0.5,1)(0.5,1.5)
\psline{-}(1.5,1)(1.5,1.5)
\psline{-}(2.5,1)(2.5,1.5)
\psline{-}(3.5,1)(3.5,1.5)
\psline{-}(4.5,1)(4.5,1.5)
\psline{-}(5.5,1)(5.5,1.5)
\psline{-}(6.5,1)(6.5,1.5)
\psline{-}(7.5,1)(7.5,1.5)
\psset{linecolor=myc3} 
\psline[linestyle=dashed,dash=2pt 2pt](0.5,0)(0.5,-0.5)
\psline[linestyle=dashed,dash=2pt 2pt](1.5,0)(1.5,-0.5)
\psline[linestyle=dashed,dash=2pt 2pt](2.5,0)(2.5,-0.5)
\psline[linestyle=dashed,dash=2pt 2pt](3.5,0)(3.5,-0.5)
\psline[linestyle=dashed,dash=2pt 2pt](4.5,0)(4.5,-0.5)
\psline[linestyle=dashed,dash=2pt 2pt](5.5,0)(5.5,-0.5)
\psline[linestyle=dashed,dash=2pt 2pt](6.5,0)(6.5,-0.5)
\psset{linecolor=myc} 
\psarc{-}(0,0){0.5}{0}{90}
\psarc{-}(1,1){0.5}{180}{270}
\psarc{-}(1,0){0.5}{0}{90}
\psarc{-}(2,1){0.5}{180}{270}
\psarc{-}(2,0){0.5}{0}{90}
\psarc{-}(3,1){0.5}{180}{270}
\psarc{-}(3,0){0.5}{0}{90}
\psarc{-}(4,1){0.5}{180}{270}
\psarc{-}(4,0){0.5}{0}{90}
\psarc{-}(5,1){0.5}{180}{270}
\psarc{-}(5,0){0.5}{0}{90}
\psarc{-}(6,1){0.5}{180}{270}
\psarc{-}(6,0){0.5}{0}{90}
\psarc{-}(7,1){0.5}{180}{270}
\psarc{-}(0,1){0.5}{180}{270}
\psarc{-}(7,1){0.5}{270}{0}
\end{pspicture} + 
e^{i (n-4)\Lambda / 2}\, \,
\begin{pspicture}(-0.5,0.375)(7.5,1.75)
\psdots(0.5,0)(1.5,0)(2.5,0)(3.5,0)(4.5,0)(5.5,0)(6.5,0)
\psdots(0.5,1)(1.5,1)(2.5,1)(3.5,1)(4.5,1)(5.5,1)(6.5,1)(-0.5,1)(7.5,1)
\psline{-}(0,0)(1,0)(1,1)(0,1)(0,0)
\psline{-}(1,0)(2,0)(2,1)(1,1)(1,0)
\psline{-}(2,0)(3,0)(3,1)(2,1)(2,0)
\psline{-}(3,0)(4,0)(4,1)(3,1)(3,0)
\psline{-}(4,0)(5,0)(5,1)(4,1)(4,0)
\psline{-}(5,0)(6,0)(6,1)(5,1)(5,0)
\psline{-}(6,0)(7,0)(7,1)(6,1)(6,0)
\psset{linecolor=myc2} \unlw
\psline{-}(-0.5,1)(-0.5,1.5)
\psline{-}(0.5,1)(0.5,1.5)
\psline{-}(1.5,1)(1.5,1.5)
\psline{-}(2.5,1)(2.5,1.5)
\psline{-}(3.5,1)(3.5,1.5)
\psline{-}(4.5,1)(4.5,1.5)
\psline{-}(5.5,1)(5.5,1.5)
\psline{-}(6.5,1)(6.5,1.5)
\psline{-}(7.5,1)(7.5,1.5)
\psset{linecolor=myc3} 
\psline[linestyle=dashed,dash=2pt 2pt](0.5,0)(0.5,-0.5)
\psline[linestyle=dashed,dash=2pt 2pt](1.5,0)(1.5,-0.5)
\psline[linestyle=dashed,dash=2pt 2pt](2.5,0)(2.5,-0.5)
\psline[linestyle=dashed,dash=2pt 2pt](3.5,0)(3.5,-0.5)
\psline[linestyle=dashed,dash=2pt 2pt](4.5,0)(4.5,-0.5)
\psline[linestyle=dashed,dash=2pt 2pt](5.5,0)(5.5,-0.5)
\psline[linestyle=dashed,dash=2pt 2pt](6.5,0)(6.5,-0.5)
\psset{linecolor=myc} 
\psarc{-}(0,0){0.5}{0}{90}
\psarc{-}(1,1){0.5}{180}{270}
\psarc{-}(1,0){0.5}{0}{90}
\psarc{-}(2,1){0.5}{180}{270}
\psarc{-}(2,0){0.5}{0}{90}
\psarc{-}(3,1){0.5}{180}{270}
\psarc{-}(3,0){0.5}{0}{90}
\psarc{-}(4,1){0.5}{180}{270}
\psarc{-}(4,0){0.5}{0}{90}
\psarc{-}(5,1){0.5}{180}{270}
\psarc{-}(5,0){0.5}{0}{90}
\psarc{-}(6,1){0.5}{180}{270}
\psarc{-}(6,1){0.5}{270}{0}
\psarc{-}(7,0){0.5}{90}{180}
\psarc{-}(0,1){0.5}{180}{270}
\psarc{-}(7,1){0.5}{270}{0}
\end{pspicture} \\[0.3cm]  & \,
\psset{unit=0.6}
+ 
e^{i (n-6)\Lambda / 2}\, \,
\begin{pspicture}(-0.5,0.375)(7.5,1.95)
\psdots(0.5,0)(1.5,0)(2.5,0)(3.5,0)(4.5,0)(5.5,0)(6.5,0)
\psdots(0.5,1)(1.5,1)(2.5,1)(3.5,1)(4.5,1)(5.5,1)(6.5,1)(-0.5,1)(7.5,1)
\psline{-}(0,0)(1,0)(1,1)(0,1)(0,0)
\psline{-}(1,0)(2,0)(2,1)(1,1)(1,0)
\psline{-}(2,0)(3,0)(3,1)(2,1)(2,0)
\psline{-}(3,0)(4,0)(4,1)(3,1)(3,0)
\psline{-}(4,0)(5,0)(5,1)(4,1)(4,0)
\psline{-}(5,0)(6,0)(6,1)(5,1)(5,0)
\psline{-}(6,0)(7,0)(7,1)(6,1)(6,0)
\psset{linecolor=myc2} \unlw
\psline{-}(-0.5,1)(-0.5,1.5)
\psline{-}(0.5,1)(0.5,1.5)
\psline{-}(1.5,1)(1.5,1.5)
\psline{-}(2.5,1)(2.5,1.5)
\psline{-}(3.5,1)(3.5,1.5)
\psline{-}(4.5,1)(4.5,1.5)
\psline{-}(5.5,1)(5.5,1.5)
\psline{-}(6.5,1)(6.5,1.5)
\psline{-}(7.5,1)(7.5,1.5)
\psset{linecolor=myc3} 
\psline[linestyle=dashed,dash=2pt 2pt](0.5,0)(0.5,-0.5)
\psline[linestyle=dashed,dash=2pt 2pt](1.5,0)(1.5,-0.5)
\psline[linestyle=dashed,dash=2pt 2pt](2.5,0)(2.5,-0.5)
\psline[linestyle=dashed,dash=2pt 2pt](3.5,0)(3.5,-0.5)
\psline[linestyle=dashed,dash=2pt 2pt](4.5,0)(4.5,-0.5)
\psline[linestyle=dashed,dash=2pt 2pt](5.5,0)(5.5,-0.5)
\psline[linestyle=dashed,dash=2pt 2pt](6.5,0)(6.5,-0.5)
\psset{linecolor=myc} 
\psarc{-}(0,0){0.5}{0}{90}
\psarc{-}(1,1){0.5}{180}{270}
\psarc{-}(1,0){0.5}{0}{90}
\psarc{-}(2,1){0.5}{180}{270}
\psarc{-}(2,0){0.5}{0}{90}
\psarc{-}(3,1){0.5}{180}{270}
\psarc{-}(3,0){0.5}{0}{90}
\psarc{-}(4,1){0.5}{180}{270}
\psarc{-}(4,0){0.5}{0}{90}
\psarc{-}(5,1){0.5}{180}{270}
\psarc{-}(5,1){0.5}{270}{0}
\psarc{-}(6,0){0.5}{90}{180}
\psarc{-}(6,1){0.5}{270}{0}
\psarc{-}(7,0){0.5}{90}{180}
\psarc{-}(0,1){0.5}{180}{270}
\psarc{-}(7,1){0.5}{270}{0}
\end{pspicture} + \dots +  
e^{-i (n-2)\Lambda / 2}\, \,
\begin{pspicture}(-0.5,0.375)(7.5,1.95)
\psdots(0.5,0)(1.5,0)(2.5,0)(3.5,0)(4.5,0)(5.5,0)(6.5,0)
\psdots(0.5,1)(1.5,1)(2.5,1)(3.5,1)(4.5,1)(5.5,1)(6.5,1)(-0.5,1)(7.5,1)
\psline{-}(0,0)(1,0)(1,1)(0,1)(0,0)
\psline{-}(1,0)(2,0)(2,1)(1,1)(1,0)
\psline{-}(2,0)(3,0)(3,1)(2,1)(2,0)
\psline{-}(3,0)(4,0)(4,1)(3,1)(3,0)
\psline{-}(4,0)(5,0)(5,1)(4,1)(4,0)
\psline{-}(5,0)(6,0)(6,1)(5,1)(5,0)
\psline{-}(6,0)(7,0)(7,1)(6,1)(6,0)
\psset{linecolor=myc2} \unlw
\psline{-}(-0.5,1)(-0.5,1.5)
\psline{-}(0.5,1)(0.5,1.5)
\psline{-}(1.5,1)(1.5,1.5)
\psline{-}(2.5,1)(2.5,1.5)
\psline{-}(3.5,1)(3.5,1.5)
\psline{-}(4.5,1)(4.5,1.5)
\psline{-}(5.5,1)(5.5,1.5)
\psline{-}(6.5,1)(6.5,1.5)
\psline{-}(7.5,1)(7.5,1.5)
\psset{linecolor=myc3} 
\psline[linestyle=dashed,dash=2pt 2pt](0.5,0)(0.5,-0.75)
\psline[linestyle=dashed,dash=2pt 2pt](1.5,0)(1.5,-0.75)
\psline[linestyle=dashed,dash=2pt 2pt](2.5,0)(2.5,-0.75)
\psline[linestyle=dashed,dash=2pt 2pt](3.5,0)(3.5,-0.75)
\psline[linestyle=dashed,dash=2pt 2pt](4.5,0)(4.5,-0.75)
\psline[linestyle=dashed,dash=2pt 2pt](5.5,0)(5.5,-0.75)
\psline[linestyle=dashed,dash=2pt 2pt](6.5,0)(6.5,-0.75)
\psset{linecolor=myc} 
\psarc{-}(0,1){0.5}{270}{0}
\psarc{-}(1,0){0.5}{90}{180}
\psarc{-}(1,1){0.5}{270}{0}
\psarc{-}(2,0){0.5}{90}{180}
\psarc{-}(2,1){0.5}{270}{0}
\psarc{-}(3,0){0.5}{90}{180}
\psarc{-}(3,1){0.5}{270}{0}
\psarc{-}(4,0){0.5}{90}{180}
\psarc{-}(4,1){0.5}{270}{0}
\psarc{-}(5,0){0.5}{90}{180}
\psarc{-}(5,1){0.5}{270}{0}
\psarc{-}(6,0){0.5}{90}{180}
\psarc{-}(6,1){0.5}{270}{0}
\psarc{-}(7,0){0.5}{90}{180}
\psarc{-}(0,1){0.5}{180}{270}
\psarc{-}(7,1){0.5}{270}{0}
\end{pspicture}
\\ \\ &= e^{-i (n-2)\Lambda / 2} \sum_{j=0}^{n-2} e^{i j \Lambda} = \frac{S_{(n-1)/2}}{S_{1/2}}.
\end{align*} \vspace{-1.2cm}

\hfill$\square$ 

\bigskip

We now construct a basis of eigenstates of $\rho(F_N)$. 
Let $w$ be a link state with $s$ defects. The identity $\eqref{eq:semicircle}$ shows that $\rho(F_N)w$ is a linear combination of link states that share all the bubbles in $w$. In fact, contributions to $\rho(F_N)w$ are link states that may have others bubbles located where $w$ had defects. Let us denote by $\mathcal W^w$ the subspace of $\tilde V_N$ generated by link states that share all the bubbles of $w$, and by $\mathcal W^w_d$ the restriction of $\mathcal W^w$ to states with $d$ defects. From \eqref{eq:semicircle}, $\mathcal W^w$ is invariant under the repeated action of $F_N$, so eigenvectors of $\rho(F_N)|_{\mathcal W^w}$ are eigenstates of $\rho(F_N)$. The subspace $\mathcal W^w_s$ is one-dimensional and satisfies $$\rho(F_N - 2C_{s/2} id) \mathcal W^w_s \subset \cup_{e<s} \mathcal W^w_e.$$ More generally, $$\dim \mathcal W^w_d = \left(\begin{smallmatrix} s \\  \tfrac{s-d}2\end{smallmatrix}\right), \qquad \rho(F_N - 2C_{d/2} id) \mathcal W^w_{d} \subset \cup_{e<d} \mathcal W^w_e.$$ The matrix $\rho(F_N)|_{\mathcal W^w}$ is upper-triangular, in a link basis with increasing number of defects, and can be seen to be identical to the matrix $\rho(F_s)$, in the basis obtained from the first by removing the bubbles in $w$ (those that are shared by all elements in $\mathcal W^w$). 
Only one eigenstate of $\rho(F_N)|_{\mathcal W^w}$ has a component along $w$. To get the full set of eigenvectors of $\rho(F_N)$, it is sufficient to compute the unique eigenstate with eigenvalue $2C_{s/2}$ of $\rho(F_s)$ with a nonzero component along the unique state with $s$ defects, for all $s$ satisfying $s \le N$ and $s = N \!\!\mod 2$. 
(For two distinct link states $w_1$ and $w_2$ with $s$ defects, the spaces $\mathcal W^{w_1}$ and $\mathcal W^{w_2}$ are different, but the matrices $\rho(F_N)|_{\mathcal W^{w_1}}$ and $\rho(F_N)|_{\mathcal W^{w_2}}$ are both identical to  $\rho(F_s)$ in the appropriate bases.)
We will denote this eigenstate $\psi^s$.  As just said, $\psi^s$ has eigenvalue $2 C_{s/2}$ and, by convention, we choose $\langle w^s | \psi^s\rangle =1$. In the next proposition, we calculate other components of $\psi^s$. An example of the procedure described above will be given after lemma \ref{lem:asdf}.

It will be shown that Jordan blocks between sectors occur only if $\Lambda = \pi a/b$ with $a, b \in \mathbb Z^\times$. For the purpose of the next propositions, a generic $\Lambda$ is $\pi$ times any complex number that is not a (real) rational number. We shall note soon that this definition is too restrictive.

\begin{Lemme}\label{lem:007} Let $w^s_{m^m}$ be the state $\in \tilde B_s^{s-2m}$ with $m$ concentric bubbles against the left imaginary boundary and let $X^s_m = \langle w^s_{m^m} | \psi^s\rangle$. Then, for generic $\Lambda$, the components $X_m^s$ are
\begin{align}
X^s_m & = (-1)^m \prod_{k=1}^m \frac{S_{(s+1-2k)/2}}{4 S_{1/2}S_{k/2}S_{(s-k)/2}}, \quad \textrm{for $m = 0$ up to} \, \, \left\{ \begin{array}{l l} (s-1)/2 & \quad  \textrm{for $s$ odd,}\\ (s-2)/2  & \quad  \textrm{for $s$ even,} \end{array} \right. \label{eq:xms}
\\
X^s_{s/2} & = \frac{(-1)^{s/2}}{\alpha - 2C_{s/2}}\prod_{k=1}^{s/2-1} \frac{S_{(s+1-2k)/2}}{4 S_{1/2}S_{k/2}S_{(s-k)/2}}.\label{eq:xms2}
\end{align}\label{sec:components}
\end{Lemme}
\noindent{\scshape Proof\ \ } Because $F_s$ commutes with the translation operator $\Omega$, the eigenstate $\psi^s$ is invariant under translation. The component along $w^s_{m^m}$ of the equation $F_s\psi^s=2C_{s/2}\psi^s$ for the eigenvector $\psi^s$ is then
\begin{align*}
2 C_{s/2} X^s_m &= \langle w^s_{m^m} | F_s \psi^s\rangle = \langle w^s_{m^m} | F_s w^s_{m^m} \rangle \langle w^s_{m^m} | \psi^s \rangle +  \langle w^s_{m^{m}} | F_s w^s_{m^{m-1}} \rangle \langle w^s_{m^{m-1}} | \psi^s \rangle \\
& = \langle w^{s-2m}| F_{s-2m}w^{s-2m}\rangle \langle w^s_{m^m} | \psi^s \rangle +   \langle w^{s-2m+2}_1| F_{s-2m+2}w^{s-2m+2}\rangle  \langle w^s_{{m}^{m-1}} | \psi^s \rangle \\
& = 2C_{s/2-m} \langle w^s_{m^m} | \psi^s \rangle +   A(s-2m+2) \langle w^s_{{(m-1)}^{m-1}} | \psi^s \rangle\\ 
&= 2C_{s/2-m} X^s_m + \frac{S_{(s-2m+1)/2}}{S_{1/2}}X^s_{m-1}
\end{align*}
where $s>1$ was assumed. To obtain the second equality, we have used the fact that the only two link states $w$ with non-zero contribution to $\langle w^s_{m^m} | F_s w\rangle $ are $w^s_{m^m}$ and $w^s_{m^{m-1}}$. 
(The only way $F_s$ can change $m-1$ nested bubbles into a pattern with a larger number of nested bubbles is to create a new simple bubble at the centre of the nested pattern and increase the size of the $m-1$ existing ones by pushing their legs by one position:
$$\psset{linewidth=1pt}
\psset{unit=0.6}
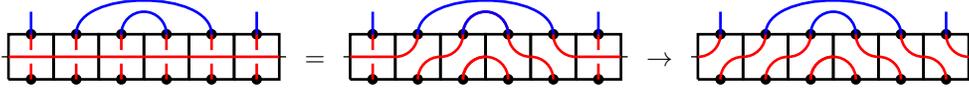
\begin{pspicture}[shift=-0.3](-0.2,0.0)(6.2,1.8)
\psdots(0.5,0)(1.5,0)(2.5,0)(3.5,0)(4.5,0)(5.5,0)
\psdots(0.5,1)(1.5,1)(2.5,1)(3.5,1)(4.5,1)(5.5,1)
\lw
\psline{-}(-0.15,0.5)(0.0,0.5)
\psline{-}(6.15,0.5)(6.0,0.5)
\unlw
\psline{-}(0,0)(1,0)(1,1)(0,1)(0,0)
\psline{-}(1,0)(2,0)(2,1)(1,1)(1,0)
\psline{-}(2,0)(3,0)(3,1)(2,1)(2,0)
\psline{-}(3,0)(4,0)(4,1)(3,1)(3,0)
\psline{-}(4,0)(5,0)(5,1)(4,1)(4,0)
\psline{-}(5,0)(6,0)(6,1)(5,1)(5,0)
\psset{linecolor=myc}
\psline{-}(0,0.5)(1,0.5)
\psline{-}(0.5,0.)(0.5,0.35)
\psline{-}(0.5,0.65)(0.5,1)
\psline{-}(1,0.5)(2,0.5)
\psline{-}(1.5,0.)(1.5,0.35)
\psline{-}(1.5,0.65)(1.5,1)
\psline{-}(2,0.5)(3,0.5)
\psline{-}(2.5,0.)(2.5,0.35)
\psline{-}(2.5,0.65)(2.5,1)
\psline{-}(3,0.5)(4,0.5)
\psline{-}(3.5,0.)(3.5,0.35)
\psline{-}(3.5,0.65)(3.5,1)
\psline{-}(4,0.5)(5,0.5)
\psline{-}(4.5,0.)(4.5,0.35)
\psline{-}(4.5,0.65)(4.5,1)
\psline{-}(5,0.5)(6,0.5)
\psline{-}(5.5,0.)(5.5,0.35)
\psline{-}(5.5,0.65)(5.5,1)
\psset{linecolor=myc2}
\psarc{-}(3,1.0){0.5}{0}{180}
\psbezier{-}(1.5,1)(1.5,2.0)(4.5,2.0)(4.5,1)
\psline{-}(0.5,1)(0.5,1.5)
\psline{-}(5.5,1)(5.5,1.5)
\end{pspicture}
\ = \ 
\begin{pspicture}[shift=-0.3](-0.2,0.0)(6.2,1.8)
\psdots(0.5,0)(1.5,0)(2.5,0)(3.5,0)(4.5,0)(5.5,0)
\psdots(0.5,1)(1.5,1)(2.5,1)(3.5,1)(4.5,1)(5.5,1)
\lw
\psline{-}(-0.15,0.5)(0.0,0.5)
\psline{-}(6.15,0.5)(6.0,0.5)
\unlw
\psline{-}(0,0)(1,0)(1,1)(0,1)(0,0)
\psline{-}(1,0)(2,0)(2,1)(1,1)(1,0)
\psline{-}(2,0)(3,0)(3,1)(2,1)(2,0)
\psline{-}(3,0)(4,0)(4,1)(3,1)(3,0)
\psline{-}(4,0)(5,0)(5,1)(4,1)(4,0)
\psline{-}(5,0)(6,0)(6,1)(5,1)(5,0)
\psset{linecolor=myc}
\psset{linecolor=myc}
\psline{-}(0,0.5)(1,0.5)
\psline{-}(0.5,0.)(0.5,0.35)
\psline{-}(0.5,0.65)(0.5,1)
\psline{-}(5,0.5)(6,0.5)
\psline{-}(5.5,0.)(5.5,0.35)
\psline{-}(5.5,0.65)(5.5,1)
\psarc{-}(3,0.0){0.5}{0}{180}
\psarc{-}(2,0){0.5}{90}{180}
\psarc{-}(3,0){0.5}{90}{180}
\psarc{-}(1,1){0.5}{270}{360}
\psarc{-}(2,1){0.5}{270}{360}
\psarc{-}(3,1.0){0.5}{0}{180}
\psarc{-}(4,1){0.5}{180}{270}
\psarc{-}(5,1){0.5}{180}{270}
\psarc{-}(4,0){0.5}{0}{90}
\psset{linecolor=myc2}
\psarc{-}(3,1.0){0.5}{0}{180}
\psbezier{-}(1.5,1)(1.5,2.0)(4.5,2.0)(4.5,1)
\psline{-}(0.5,1)(0.5,1.5)
\psline{-}(5.5,1)(5.5,1.5)
\end{pspicture}
\ \rightarrow \ 
\begin{pspicture}[shift=-0.3](-0.2,0.0)(6.2,1.8)
\psdots(0.5,0)(1.5,0)(2.5,0)(3.5,0)(4.5,0)(5.5,0)
\psdots(0.5,1)(1.5,1)(2.5,1)(3.5,1)(4.5,1)(5.5,1)
\lw
\psline{-}(-0.15,0.5)(0.0,0.5)
\psline{-}(6.15,0.5)(6.0,0.5)
\unlw
\psline{-}(0,0)(1,0)(1,1)(0,1)(0,0)
\psline{-}(1,0)(2,0)(2,1)(1,1)(1,0)
\psline{-}(2,0)(3,0)(3,1)(2,1)(2,0)
\psline{-}(3,0)(4,0)(4,1)(3,1)(3,0)
\psline{-}(4,0)(5,0)(5,1)(4,1)(4,0)
\psline{-}(5,0)(6,0)(6,1)(5,1)(5,0)
\psset{linecolor=myc}
\psarc{-}(3,0.0){0.5}{0}{180}
\psarc{-}(1,0){0.5}{90}{180}
\psarc{-}(2,0){0.5}{90}{180}
\psarc{-}(3,0){0.5}{90}{180}
\psarc{-}(0,1){0.5}{270}{360}
\psarc{-}(1,1){0.5}{270}{360}
\psarc{-}(2,1){0.5}{270}{360}
\psarc{-}(4,1){0.5}{180}{270}
\psarc{-}(5,1){0.5}{180}{270}
\psarc{-}(6,1){0.5}{180}{270}
\psarc{-}(4,0){0.5}{0}{90}
\psarc{-}(5,0){0.5}{0}{90}
\psset{linecolor=myc2}
\psarc{-}(3,1.0){0.5}{0}{180}
\psbezier{-}(1.5,1)(1.5,2.0)(4.5,2.0)(4.5,1)
\psline{-}(0.5,1)(0.5,1.5)
\psline{-}(5.5,1)(5.5,1.5)
\end{pspicture}
$$Again, because $F_s$ has a single row of tiles, it can increase the number of nested bubbles only by one.)
For the third equality, the identity \eqref{eq:semicircle} allows for the removal of all bubbles entering $F_s$. Finally, for the fourth, we used the invariance under translation of $\psi^s$ to write $ \langle w^s_{{m}^{m-1}} | \psi^s \rangle = \langle w^s_{(m-1)^{m-1}} | \psi^s \rangle$. The result is a recurrence relation for $X_m^s$:
\begin{equation*}
X_m^s = - \frac{X_{m-1}^s S_{(s-2m+1)/2}}{4 S_{1/2} S_{m/2} S_{(s-m)/2}}
\end{equation*}
which can be used, along with the initial condition $X_0^s = \langle w^s | \psi^s\rangle=1$, to complete the proof of the first part. If $s$ is even, $ \langle w^s_{(s/2)^{s/2}} | F_s w^s_{(s/2)^{s/2}} \rangle = \alpha$ and the recurrence relation reads instead
\begin{equation*} 2C_{s/2} X^s_{s/2} = \alpha X^s_{s/2} + X^s_{s/2-1}
\end{equation*}
and the rest of the argument is identical.
\hfill$\square$ 

\medskip

\noindent It is clear that the expressions for the coefficients $X_m^s$ in \eqref{eq:xms} and \eqref{eq:xms2} are valid as long as they are finite for the value of $\Lambda$ of interest. Calculating other components is also possible. However, the $X^s_m$s will be sufficient to probe the Jordan structure of $F_N$ and of $T_N$ and will be related to singularities in the denominators of equations \eqref{eq:xms} and \eqref{eq:xms2}. Because 
$$X_m^s = X_{m-1}^s \cdot \frac{A(s-2m+2)}{\gamma_s -\gamma_{s-2m}}$$
where $\gamma_d = 2C_{d/2}$ is the eigenvalue of $\rho(F_N)$ in the sector $d$, the definition of genericity for $\Lambda$ should be partially relaxed to allow rational numbers $a/b$ for which the eigenvalues $\gamma_{s-2m}, \gamma_{s-2m+2}, \dots, \gamma_{s-2}$ are all distinct from $\gamma_s$.

\subsection[The Jordan structures of $\rho(F_N)$ and $\rho(T_N(\lambda,\nu))$]{The Jordan structures of $\boldsymbol{\rho(F_N)}$ and $\boldsymbol{\rho(T_N(\lambda,\nu))}$}\label{sec:jordanFN}

In the previous section, we were successful in computing the eigenvalues of $\rho(F_N)$ and some components $X_m^s$ of the corresponding eigenvectors. In this section we show how the $X_m^s$s allow us to unravel the Jordan structure of $\rho(F_N)$.
\begin{Lemme}\label{lem:asdf} For generic $\Lambda$, the eigenvector $\psi^s$ satisfies $e_i \psi^s = 0$ for all $i = 1, \dots, s$.
\end{Lemme}
\noindent{\scshape Proof\ \ } Because $F_s( e_i \psi^s ) =  e_i F_s \psi^s = 2 C_{s/2} e_i \psi^s$,  the vector $e_i \psi^s$ is an eigenstate of $F_s$ with eigenvalue $2C_{s/2}$. 
The eigenvector $\psi^s$ is a linear combination of states with $s$ entries, with the coefficient of the unique link state with $s$ defects, $\langle w^s| \psi^s \rangle$, set to $1$. Therefore $\psi^s\in\upto_s$ but $e_i\psi^s\in\upto_{s-2}$ since the generator $e_i$ has closed two defects in the only link state with $s$ defects in this linear combination giving $\psi^s$. But $e_i\psi^s\in \upto_{s-2}$ cannot be an eigenstate of $F_N$ with eigenvalue $2C_{s/2}$, as the eigenvalues on the subspace $\upto_{s-2}$ are $2C_{k/2}, 0\le k\le s-2$, which are all different from $2C_{s/2}$ for a generic $\Lambda$. The vector $e_i\psi^s$ must then be zero.
\hfill$\square$ 

\medskip

A change of bases will now show that, for generic $\Lambda$, there cannot be Jordan blocks between sectors. 
For this goal a linear transformation $\Psi^d : \tilde V_N^d \rightarrow \upto_d$ is defined as follows. If $v \in \tilde B_N^d$, the vector $\Psi^dv$ is obtained as follows. First remove all bubbles of $v$. Second replace the $d$ defects of $v$ by the linear combination $\psi^d$. Finally reinsert the bubbles in their original locations. 

Here is an example of the procedure. For $N=2$, in the basis $\{ \,
\begin{pspicture}(0.1,0.0)(0.4,0.3)
\psdots[dotsize=0.08](0.1,0)(0.3,0.0)
\psset{linecolor=myc2}
\psarc{-}(0.2,0){0.1}{0}{180}
\end{pspicture},\,
\begin{pspicture}(0.1,0.0)(0.4,0.3)
\psdots[dotsize=0.08](0.1,0)(0.3,0.0)
\psset{linecolor=myc2}
\psarc{-}(0.0,0){0.1}{0}{90}
\psarc{-}(0.4,0){0.1}{90}{180}
\end{pspicture},\,
\begin{pspicture}(0.1,0.0)(0.4,0.3)
\psdots[dotsize=0.08](0.1,0)(0.3,0.0)
\psset{linecolor=myc2}
\psline{-}(0.3,0)(0.3,0.3)
\psline{-}(0.1,0)(0.1,0.3)\end{pspicture}
\}$,
the matrix $\rho(F_2)$ and the state $\psi^2$ are
\begin{equation*}
\rho(F_2) = \begin{pmatrix} 
\alpha & 0 & 1 \\ 0 & \alpha & 1 \\ 0 & 0 & 2C_1
\end{pmatrix}, 
\qquad \psi^{s=2} = \psset{unit=0.6}
\begin{pspicture}(0,0.0)(2,0.8)
\psdots(0.5,0)(1.5,0)
\psset{linecolor=myc2}
\psline{-}(0.5,0)(0.5,0.5)
\psline{-}(1.5,0)(1.5,0.5)
\end{pspicture}
- (\alpha - 2C_1)^{-1} \left( 
\begin{pspicture}(0,0.0)(2,0.8)
\psdots(0.5,0)(1.5,0)
\psset{linecolor=myc2}
\psarc{-}(1,0.0){0.5}{0}{180}
\end{pspicture}
+\
\begin{pspicture}(0,0.0)(2,0.8)
\psdots(0.5,0)(1.5,0)
\psset{linecolor=myc2}
\psarc{-}(0,0.0){0.5}{0}{90}
\psarc{-}(2,0.0){0.5}{90}{180}
\end{pspicture}
\ \right).
\end{equation*}
Then for $N=6$ and $v = \begin{pspicture}(0.1,0.0)(1.2,0.3)
\psdots[dotsize=0.08](0.1,0)(0.3,0.0)(0.5,0.0)(0.7,0.0)(0.9,0.0)(1.1,0.0)
\psset{linecolor=myc2}
\psarc{-}(0.6,0){0.1}{0}{180}
\psbezier{-}(0.3,0.0)(0.3,0.25)(0.9,0.25)(0.9,0.0)
\psline{-}(0.1,0)(0.1,0.20)
\psline{-}(1.1,0)(1.1,0.20)
\end{pspicture}$, the previous procedure produces the state $\Psi^{d=2} v$:
\begin{align*}\psset{unit=0.6}
\begin{pspicture}(0,0.0)(6.2,0.8)
\psdots(0.5,0)(1.5,0)(2.5,0)(3.5,0)(4.5,0)(5.5,0)
\psset{linecolor=myc2}
\psarc{-}(3,0.0){0.5}{0}{180}
\psbezier{-}(1.5,0)(1.5,1.0)(4.5,1.0)(4.5,0)
\psline{-}(0.5,0)(0.5,0.5)
\psline{-}(5.5,0)(5.5,0.5)
\end{pspicture}
\ & \psset{unit=0.6} \rightarrow \ 
\begin{pspicture}(0,0.0)(6.2,0.8)
\psdots(0.5,0)(1.5,0)(2.5,0)(3.5,0)(4.5,0)(5.5,0)
\psset{linecolor=myc2}
\psline{-}(0.5,0)(0.5,0.5)
\psline{-}(5.5,0)(5.5,0.5)
\end{pspicture} 
\\ & \hspace{-3cm}\psset{unit=0.6}
\rightarrow \ 1 \cdot \!\!
\begin{pspicture}(0,0.0)(6.2,0.8)
\psdots(0.5,0)(1.5,0)(2.5,0)(3.5,0)(4.5,0)(5.5,0)
\psset{linecolor=myc2}
\psline{-}(0.5,0)(0.5,0.5)
\psline{-}(5.5,0)(5.5,0.5)
\end{pspicture}\!\! - (\alpha - 2C_1)^{-1} \Big(
\begin{pspicture}(0,0.0)(6.2,0.8)
\psdots(0.5,0)(1.5,0)(2.5,0)(3.5,0)(4.5,0)(5.5,0)
\psset{linecolor=myc2}
\psbezier{-}(0.5,0)(0.5,1.2)(5.5,1.2)(5.5,0)
\end{pspicture}
+
\begin{pspicture}(0,0.0)(6.2,0.8)
\psdots(0.5,0)(1.5,0)(2.5,0)(3.5,0)(4.5,0)(5.5,0)
\psset{linecolor=myc2}
\psarc{-}(0,0){0.5}{0}{90}
\psarc{-}(6,0){0.5}{90}{180}
\end{pspicture}
\Big)\\&\hspace{-3cm}\psset{unit=0.6}
\rightarrow \ 1 \cdot \!\!
\begin{pspicture}(0,0.0)(6.2,0.8)
\psdots(0.5,0)(1.5,0)(2.5,0)(3.5,0)(4.5,0)(5.5,0)
\psset{linecolor=myc2}
\psline{-}(0.5,0)(0.5,0.5)
\psline{-}(5.5,0)(5.5,0.5)
\psarc{-}(3,0.0){0.5}{0}{180}
\psbezier{-}(1.5,0)(1.5,1.0)(4.5,1.0)(4.5,0)
\end{pspicture}\!\! - (\alpha - 2C_1)^{-1} \Big(
\begin{pspicture}(0,0.0)(6.2,0.8)
\psdots(0.5,0)(1.5,0)(2.5,0)(3.5,0)(4.5,0)(5.5,0)
\psset{linecolor=myc2}
\psarc{-}(3,0.0){0.5}{0}{180}
\psbezier{-}(1.5,0)(1.5,1.0)(4.5,1.0)(4.5,0)
\psbezier{-}(0.5,0)(0.5,1.2)(5.5,1.2)(5.5,0)
\end{pspicture}
+
\begin{pspicture}(0,0.0)(6.2,0.8)
\psdots(0.5,0)(1.5,0)(2.5,0)(3.5,0)(4.5,0)(5.5,0)
\psset{linecolor=myc2}
\psarc{-}(3,0.0){0.5}{0}{180}
\psbezier{-}(1.5,0)(1.5,1.0)(4.5,1.0)(4.5,0)
\psarc{-}(0,0){0.5}{0}{90}
\psarc{-}(6,0){0.5}{90}{180}
\end{pspicture}
\Big)
\end{align*}
where each step is shown: the removal of the bubbles, the replacement of the $2$-defect state by $\psi^{s=2}$ and, finally, the reinsertion of the original bubbles at their positions. 
The resulting $\Psi^dv$ is a linear combination of link states in $\upto_d$ 
and its component in the subspace $\tilde V_N^d$ in the original basis is $v$ itself. 
The set of link states obtained by acting with $\Psi^d$ on the states of $\tilde B_N^d$ will be noted $\Psi \tilde B_N^d$. 
Clearly $\cup_d\Psi \tilde B_N^d$ is a basis of $\tilde V_N$
constituted of eigenvectors 
of $F_N$ for generic $\Lambda$. 

\begin{Proposition}
For $\Lambda$ generic, the subspace spanned by $\Psi \tilde B_N^d$ is stable under $\eptl(\beta, \alpha)$. \end{Proposition}
\noindent{\scshape Proof\ \ } 
 The proof consists in showing that the generators $e_i$s and the translations $\Omega^{\pm}$ permute the elements of the basis $\Psi \tilde B_N^d$ or send them to zero. The easiest is the translations $\Omega^{\pm}$. The procedure defining $\Psi^d$ simply commutes with the translations and $\Omega^\pm(\Psi^d w)=\Psi^d(\Omega^\pm w)$ for all link states $w$ in $\tilde B_N^d$.

The study of the action of $e_i$ on an element of $\Psi^d(w)\in\Psi \tilde B_N^d$ splits into three subcases depending on whether the positions $i$ and $i+1$ upon which $e_i$ acts are occupied by zero, one or two defects of the original $w$. If  two defects of $w$ occupy positions $i$ and $i+1$, then $e_i$ acts in $\Psi^d(w)$ on two positions of $\psi^d$ and the result is zero by the previous lemma. If $e_i$ acts on a defect and a bubble of $w$, the resulting link state is simply $\Psi^d(e_iw)$ and one of the defects has moved. The last case is when $e_i$ does not act on any of the positions where $w$ had defects. By definition of $\Psi^d$ all terms of the linear combinations giving $\Psi^d(w)$ in the link basis share the original bubbles of $w$. So the action of $e_i$ on any of the terms of this linear combination will shuffle some bubbles of the original $w$ or add a factor of $\beta$, and this occurs in exactly the same way for all the terms. Therefore, in this case, $e_i(\Psi^d(w))=\Psi^d(e_iw)$ is still an element of $\Psi \tilde B_N^d$, maybe up to a factor $\beta$.\hfill$\square$

\medskip

From the first case of the last proposition, any $c \in \eptl(\beta,\alpha)$ has 
$\rho'(c)_{e,d} = 0$ for $e < d$, where $\rho'(c)$ is the matrix representing $c$ in the basis $\Psi \tilde B_N$ 
and $\rho'(c)_{e,d}$ is the block, above the diagonal, for which the domain is 
$\textrm{span\,}\Psi \tilde B_N^d$ and the image is $\textrm{span\,}\Psi \tilde B_N^e$.
For $\Lambda$ generic, there are thus no Jordan blocks between sectors. 

The next proposition gives a criteria for the existence of Jordan blocks for some non-generic values $\Lambda$. For this proof, we will denote by $\gamma_d$ the (unique) eigenvalue of $\rho(F_N)|_d$ (computed in proposition \ref{sec:eigenF}).

\begin{Proposition} Let $\Lambda$ (non-generic) and  $d_1, \dots, d_n$ be integers such that $0\le d_1< \dots < d_n \le N$, $N-{d_i}\equiv 0\hspace{-0.10cm}\mod\hspace{-0.05 cm} \,2$ for $i = 1, \dots, n$, $\gamma \equiv \gamma_{d_1} = \gamma_{d_2} = \dots = \gamma_{d_n}$ and $A(d_1+2), A(d_1+4), \dots, A(d_n) \neq 0$. Then there is at least one Jordan cell of size $n$ in $\rho(F_N)$ connecting sectors  $d_1, \dots, d_n$. \label{sec:intersectorial}
\end{Proposition}
\noindent{\scshape Proof\ \ } 
We are interested in the Jordan structure between sectors $d_1, d_2, \dots, d_n$, so we restrict our study to $\rho(F_N)|_{[d_1,d_n]}$. 
This restriction contains also all sectors between $d_1$ and $d_n$ and therefore sectors with $d$'s distinct from the $d_i$'s, $1\leq i\leq n$. Restricted to the sectors that are between the various $d_i$'s, the central element acts as the identity times some eigenvalues different from $\gamma$.
We call them $\gamma_i, i = 1, \dots, m$, and each appears in $n_i$ different sectors. Because Jordan blocks can only occur between sectors, $\rho(F_N)|_{[d_1,d_n]}$ satisfies the polynomial equation
\begin{equation*} (\rho(F_N - \gamma\cdot id)|_{[d_1,d_n]})^{n} \times \prod_{i=1}^m \big(\rho(F_N - \gamma_i\cdot id)|_{[d_1,d_n]}\big)^{n_i} = 0.
\end{equation*}
To prove that a Jordan block of size $n$ exists, we show that
\begin{equation*}\rho(G)|_{[d_1,d_n]} \neq 0 \quad \textrm{for} \quad G \equiv ( F_N - \gamma\cdot id)^{n-1} \times \prod_{i=1}^m (F_N - \gamma_i\cdot id)^{n_i},
\end{equation*} 
by computing one non-zero matrix element. Taking for simplicity $N/2<x<N$, we will show indeed that $ \langle w^N_{x^{(N-d_1)/2}} | G \, w^N_{x^{(N-d_n)/2}} \rangle$ is non-zero. From its definition, $G$ is a polynomial in $F_N$ with degree $(d_n-d_1)/2$ and leading coefficient $1$. Both the initial and final states have been chosen to have only concentric bubbles, so each application of $F_N$ adds one concentric bubbles and the only term contributing is $F_N^{(d_n-d_1)/2}$:
\begin{align*} 
\langle w^N_{x^{(N-d_1)/2}} | G w^N_{x^{(N-d_n)/2}} \rangle &= \langle w^N_{x^{(N-d_1)/2}} | F_N^{(d_n-d_1)/2} w^N_{x^{(N-d_n)/2}} \rangle \\
& = \langle w^{d_n}_{y^{(d_n-d_1)/2}} | F_{d_n}^{(d_n-d_1)/2} w^{d_n} \rangle \qquad \textrm{with} \quad y = x-(N-d_n)/2 \\
& = \prod_{i=0}^{(d_n-d_1-2)/2} \langle w^{d_n}_{y^{i+1}}| F_{d_n} w^{d_n}_{y^{i}} \rangle = \prod_{i=0}^{(d_n-d_1-2)/2} \langle w^{d_n-2i}_{y-i}| F_{d_n-2i} w^{d_n-2i} \rangle \\
& =\prod_{i=0}^{(d_n-d_1-2)/2} A(d_n-2i) = \prod_{i=0}^{{(d_n-d_1-2)/2}} \frac{S_{(d_n-2i-1)/2}}{S_{1/2}} \neq 0.
\end{align*}
\hfill$\square$ \\
\indent In this proof, we used a starting state with only concentric bubbles, but in fact for any initial state $w_1$ in $\tilde V_N^{d_n}$, one can find a state in $w_2$ in $\tilde V_N^{d_1}$ such that $\langle w_2 | G w_1\rangle \neq 0$. (Note that our numerical explorations show that, for $d_n\neq N$, there is more than one Jordan blocks connecting the sectors $d_n$ and $d_1$.)

In the derivation of \eqref{eq:xms}, we found that
$$ X_{(d_n-d_1)/2}^{d_n}  =  \frac{\displaystyle{\prod_{i=0}^{(d_n-d_1-2)/2} A(d_n-2i)}}{\displaystyle{ \prod_{d=d_1, d_1 + 2, \dots, d_n-2}(\gamma_{d_n} - \gamma_{d})}} .$$
The condition that $A(d_1+2), A(d_1+4), \dots, A(d_n)$ be non-zero is therefore equivalent to requiring that $X_{(d_n-d_1)/2}^{d_n}$ has a pole $(\gamma_{d_n} - \gamma_{d_i})^{-1}$ for each $d_i = d_1, d_2, \dots, d_{n-1}$.

The constraint $\prod_{i=0}^{(d_n-d_1-2)/2} A(d_n-2i) \neq 0$ appears to be the simplest formulation of the condition for Jordan blocks to appear, but it can be translated in terms of constraints for $N, d$ and $\Lambda$. The following corollary is obtained by first identifying, for a given parity of $N$, the numbers $n$ for which $A(n)$ vanishes (for $n\equiv N\,\textrm{mod\,}2$) and then selecting a maximal consecutive range of $n$s such that the  $A(n)$s do not vanish. The second step consists in choosing sectors $d, d', d'', \dots$ in this range that share the same eigenvalue $\gamma_d$. Note that $\gamma_d=\gamma_{d'}$ if and only if $(d'+d)\Lambda/4$ or $(d'-d)\Lambda/4$ is $\pi$ times an integer.

\begin{Corollaire}\label{coro:leCoroFN}
Let $\Lambda = \pi a / b$ with $a$ and $b$ coprime integers. Then
\begin{itemize}
\item[(a)] $N$ even
	\begin{itemize}
	\item[(i)] $a$ odd: $A(n)$ with $n$ an even integer is never zero. A given sector $d$, with $0\le d \le 2b$ and $d$ even, is coupled to all sectors $4bj-d$, $4bj+d$, for $j\ge 1$, with at least one high-rank Jordan block in $F_N$. 
The sector $d = 0$ also couples to these sectors if $\alpha = 2 C_{d/2}$.
	\item[(ii)] $a$ even (and $b$ odd) : $A(n)$ with $n$ even is zero if and only if $n=(2s+1)b+1$ for some integer $s\ge0$. Sectors in $\rho(F_N)$ 	can be tied in a Jordan cell if they are labeled by even numbers in the range $\big[(2s+1)b+1, (2s+3)b+1\big [$ for some fixed $s\ge0$. In the above range, this occurs only for pairs $(d,d')$ with $d'=4(s+1)b-d$ and the Jordan cell is of rank $2$. The sector $d = 0$ couples to the sector $d'$ with a rank $2$ Jordan cell if $\alpha = 2C_{d'/2}$ and $d'<b$. 
	\end{itemize}
\item[(b)] $N$ odd: $A(n)$ with $n$ odd is zero if and only if $n = 2sb+1$ for some integer $s\ge0$. Sectors $\rho(F_N)$ can be tied in a Jordan cell if they are labeled by odd integers in the range $\big[2sb+1, 2(s+1)b+1\big[$ for some fixed $s\ge0$.
	\begin{itemize}
	\item[(i)] $a$ odd: All eigenvalues for the sectors in this range are distinct and there are no Jordan blocks.
	\item[(ii)] $a$ even (and $b$ odd): In the above range, this occurs only for pairs $(d,d')$ with $d'=2(2s+1)b-d$ and the Jordan cell is of rank $2$.
	\end{itemize}
\end{itemize}
\end{Corollaire}

\noindent Of course a given sector does not couple to itself even if $d' = d$ satisfies the conditions of this corollary.

The case $N=3$ provides a simple example: 
in the basis $\{ \,
\begin{pspicture}(0.1,0.0)(0.6,0.3)
\psset{linewidth=1pt}
\psdots[dotsize=0.08](0.1,0)(0.3,0.0)(0.5,0.0)
\psset{linecolor=myc2}
\psline{-}(0.5,0)(0.5,0.3)
\psarc{-}(0.2,0){0.1}{0}{180}
\end{pspicture},\,
\begin{pspicture}(0.1,0.0)(0.6,0.3)
\psset{linewidth=1pt}
\psset{unit=1.0}
\psdots[dotsize=0.08](0.1,0)(0.3,0.0)(0.5,0.0)
\psset{linecolor=myc2}
\psline{-}(0.1,0)(0.1,0.3)
\psarc{-}(0.4,0){0.1}{0}{180}
\end{pspicture},\
\begin{pspicture}(0.1,0.0)(0.6,0.3)
\psset{linewidth=1pt}
\psset{unit=1.0}
\psdots[dotsize=0.08](0.1,0)(0.3,0.0)(0.5,0.0)
\psset{linecolor=myc2}
\psline{-}(0.3,0)(0.3,0.3)
\psarc{-}(0.0,0){0.1}{0}{90}
\psarc{-}(0.6,0){0.1}{90}{180}
\end{pspicture}\,,\,
\begin{pspicture}(0.1,0.0)(0.6,0.3)
\psset{linewidth=1pt}
\psset{unit=1.0}
\psdots[dotsize=0.08](0.1,0)(0.3,0.0)(0.5,0.0)
\psset{linecolor=myc2}
\psline{-}(0.5,0)(0.5,0.3)
\psline{-}(0.3,0)(0.3,0.3)
\psline{-}(0.1,0)(0.1,0.3)
\end{pspicture}\}$,
$$\rho(F_3) = 2\left(\begin{matrix} C_{1/2} & 0 & 0 & C_{1/2} \\ 0 & C_{1/2} & 0 & C_{1/2} \\ 0 & 0 & C_{1/2} & C_{1/2} \\ 0 & 0 & 0 & C_{3/2} \end{matrix} \right) \quad \textrm{and} \quad \textrm{Jordan form of } \rho(F_3)  \Big|_{\substack{\Lambda = 0}}= \left(\begin{matrix} 2 & 0 & 0 & 0 \\ 0 & 2 & 0 & 0 \\ 0 & 0 & 2 & 1 \\ 0 & 0 & 0 & 2 \end{matrix} \right).$$
The eigenvalues $2C_{1/2}$ and $2C_{3/2}$ are equal for $\Lambda = 0$ $(\beta = -2)$ and $\Lambda = \pi$ $(\beta = 2)$. The state 
$v = \big( -\frac{1}{4S^2_{1/2}}, -\frac{1}{4S^2_{1/2}}, -\frac{1}{4S^2_{1/2}}, 1\big)$ is an eigenvector of $\rho(F_3)$ for generic $\Lambda$ and it diverges for $\Lambda = 0$ only. The matrix $\rho(F_3)|_{\Lambda = \pi}$ is the zero matrix and is diagonal. One can also check that  $\rho(G)|_{\Lambda = 0} = \rho(F_3 -2\, id)|_{\Lambda = 0} \neq 0$.

Using lemma 4.1 and proposition 4.10 of \cite{AMDSA}, the above result is extended to the transfer matrix $T_N$. Omitting the particularities for $d = 0$ gives theorem \ref{thm:inter}.

Even though corollary \ref{coro:leCoroFN} provides the condition for the existence of cells in $F_N$ between sectors $d$ and $d'$ (with $d\neq d'$), theorem \ref{thm:inter} adds the further hypothesis that $\left.\rho(T_N(\lambda,\nu))\right|_{e}$
be diagonalizable for $e$'s such that $d'<e\le d$ and either of the conditions ${\textstyle{\frac12}}(d\pm e)\equiv 0\hspace{-0.10cm}\mod\hspace{-0.05 cm} 2b$ is satisfied. Without this hypothesis, the cell in $\rho(T_N(\lambda,\nu))$ could be in fact between two vectors with non-zero components in $\tilde V_N^{d}$.
This changes somewhat the conditions characterizing the Virasoro representations appearing in the continuum limit. Indeed one of the key features of the periodic loop model is that the transfer matrix commutes with itself for all values of the anisotropy parameter. (See section \ref{sec:transferMatrix}.) All its Fourier coefficients are thus conserved quantities and this property should be preserved through the limiting process. So certain generators like $L_0$ might have Jordan cells within the subspace corresponding to the sector $d$ while other conserved quantities, like the one arising from the limit of $F_N$, will have cells only between sectors $d$ and $d'\neq d$.

%
\section{The quantum algebra $\boldsymbol{\uq}$ and the homomorphism $\boldsymbol{\tilde i_N^d}$} \label{sec:tildeiNd}
%

The problem of finding Jordan cells of the loop Hamiltonian $\mathcal H$ within a sector with a fixed number $d$ of defects, that is, for $\mathcal H$ in the representation $\omega_d$, requires further tools. This section introduces them. The first paragraph presents a brief review of the quantum algebra $\uq$, tailored for the present needs. Only lemmas \ref{lem:41} and \ref{sec:firstcomm2} are new there; they extend for all $v\in\mathbb C^\times$ results that appeared first in \cite{Deguchi}. (The familiar reader will want to go over this paragraph quickly.) The second paragraph recalls the properties of an intertwiner $\tilde i_N^d$ between the representations of the loop and XXZ models. In the previous section, it is the singularities of the central element $F_N$ that revealed the Jordan cells between sectors. In the next one, it will be those of the intertwiner $\tilde i_N^d$ that will lead us to the Jordan cells within a sector.

\subsection[The quantum algebra $\uq$]{The quantum algebra $\boldsymbol{\uq}$}\label{sec:uqsl2}

The quantum algebra $\uq$ (see \cite{ChariPressley, KS1997}) is defined by generators $id, S^\pm, q^{\pm S_z}$ satisfying the relations
\begin{equation} q^{S^z} S^\pm q^{-S^z} = q^{\pm1} S^\pm, \qquad \qquad [S^+,S^-] = \frac{q^{2S^z}-q^{-2S^z}}{q-q^{-1}}, \qquad\qquad q^{S^z}q^{-S^z}=q^{-S^z}q^{S^z}=id. \label{eq:defUqsl2}\end{equation}
This algebra is infinite-dimensional. 

The representation $\uq\rightarrow \textrm{End}\cn$ that will be used throughout is given by \begin{align*} 
q^{S^z} &= q^{\sigma^z /2} \otimes q^{\sigma^z /2} \otimes \dots \otimes q^{\sigma^z /2} = \prod_{j=1} ^Nq^{\sigma_j^z/2}, \\ 
S^\pm &= \sum_{k=1}^N  \underbrace{ v^{\mp 1}  q^{-\sigma^z /2} \otimes \dots  \otimes v^{\mp 1} q^{-\sigma^z /2}}_{k-1} \otimes ~\sigma^\pm \otimes \underbrace{v^{\pm 1}q^{\sigma^z /2}\otimes \dots \otimes v^{\pm 1} q^{\sigma^z /2}}_{N-k} \\ 
& = v^{\pm(N+1)} \sum_{k=1}^{N} v^{\mp 2k} \Bigg(\prod_{i=1}^{k-1} q^{-\sigma_i^z/2}\Bigg) \sigma^\pm_k \Bigg(\prod_{j=k+1}^{N} q^{\sigma_j^z/2}\Bigg) \equiv  \sum_{k=1}^N  S^\pm_k.
\end{align*}
It can be proved that this is indeed a representation by checking equation (\ref{eq:defUqsl2}) for $N=1$ and by using the coproduct 
\begin{equation*} 
\Delta(q^{S^z}) = q^{S^z} \otimes q^{S^z}, \qquad \qquad \Delta(S^{\pm}) = v^{\mp 1} q^{-S^z} \otimes S^{\pm} + v^{\pm 1} S^{\pm} \otimes q^{S^z}
\end{equation*}
to build representations for $N>1$. Note that this representation also depends on the twist parameter $v$. Since most relations involving elements of $\uq$ will be acting on $\cn$, no distinction is made between abstract generators and their representation as endomorphims of $\cn$. This representation is useful because it satisfies $[S^\pm, \bar e_i] = 0$ for $i = 1, \dots, N-1$. However, $[S^\pm, \bar e_N] \neq 0$ in general and the commutation of $U_q(sl_2)$ with the XXZ Hamiltonian that holds for the {\em open} spin chain is lost for the {\em periodic} one. 

Another representation of the algebra $U_q(sl_2)$ is obtained by replacing $q$ by $q^{-1}$ in the definition of the $S^\pm$ generators:
\begin{align*} 
T^\pm &= \sum_{k=1}^N  \underbrace{ v^{\mp 1} q^{\sigma^z /2} \otimes \dots  \otimes v^{\mp 1} q^{\sigma^z /2}}_{k-1} \otimes ~\sigma^\pm \otimes \underbrace{v^{\pm 1}q^{-\sigma^z /2}\otimes \dots \otimes v^{\pm 1} q^{-\sigma^z /2}}_{N-k} \\ 
& = v^{\pm(N+1)} \sum_{k=1}^{N} v^{\mp 2k} \Bigg(\prod_{i=1}^{k-1} q^{\sigma_i^z/2}\Bigg) \sigma^\pm_k \Bigg(\prod_{j=k+1}^{N} q^{-\sigma_j^z/2}\Bigg) \equiv \sum_{k=1}^N  T^\pm_k.
 \end{align*}
In this last representation, $T^\pm$ no longer commutes with $\bar e_i$, $i = 1, \dots , N-1$, but instead, $[T^\pm, \bar e_i^*] = 0$, where $\bar e_i^* \equiv \left.\bar e_i\right|_{q \rightarrow q^{-1}}$. Taking powers of the generators $S^\pm$ and $T^\pm$ gives
\begin{align}
(S^\pm)^x &= [x]! v^{\pm x(N+1)} \sum_{1\le j_1 < j_2 < \dots < j_x \le N} v^{\mp 2 \sum_{k=1}^x j_k} \Bigg\{\prod_{i=1}^{x+1} \Bigg( \prod_{l= j_{i-1}+1}^{j_i-1} q^{-(x/2 + 1 - i) \sigma^z_l}\Bigg) \Bigg\}   \sigma^\pm_{j_1} \sigma^\pm_{j_2} \dots  \sigma^\pm_{j_x}, \label{eq:Spx}\\
(T^\pm)^x &= [x]! v^{\pm x(N+1)} \sum_{1\le j_1 < j_2 < \dots  < j_x \le N} v^{\mp 2 \sum_{k=1}^x j_k} \Bigg\{\prod_{i=1}^{x+1} \Bigg( \prod_{l= j_{i-1}+1}^{j_i-1} q^{(x/2 + 1 - i) \sigma^z_l}\Bigg) \Bigg\}   \sigma^\pm_{j_1} \sigma^\pm_{j_2} \dots  \sigma^\pm_{j_x},\nonumber
\end{align}
where $j_0 \equiv 0$ and $j_{x+1} \equiv N+1$ and the $q$-numbers and $q$-factorials are, as usual, $[n] = (q^n - q^{-n}) / (q-q^{-1})$ and $[n]! = \prod_{k=1}^n [k]$.
Similarly $q$-binomials will also be used: $\left[ \begin{smallmatrix} m \\ j \end{smallmatrix} \right] = \frac{[m]!}{[j]![m-j]!}$. 
Because $[x]!$ can be zero if $q$ is a root of unity, it is usual to introduce the renormalized generators (also known as divided powers)
\begin{equation*} S^{\pm (x)} = (S^\pm)^x / [x]! \qquad\textrm{and} \qquad T^{\pm (x)} = (T^\pm)^x / [x]!
\end{equation*}
which are non-zero endomorphisms for every value of $q\in \mathbb C^\times$ and for all $x\le N$. Note that the values of $q$ for which $(S^\pm)^x$ and $(T^\pm)^x$ vanish are independent of $v$. The renormalized generators satisfy the commutation relations
\begin{equation}
[S^{+(m)}, S^{-(n)}] = \sum_{j=1}^{\textrm{min}(m,n)}\frac1{[j]!} S^{-(n-j)}S^{+(m-j)} \prod_{k=0}^{j-1}[2S^z + m -n - k].
\label{eq:bigcomm}\end{equation}
This equation can be derived directly from the defining relations of $U_q(sl_2)$ and is therefore independent of the twist parameter $v$. Because $q$-numbers and $q$-binomials are invariant under $q \rightarrow q^{-1}$, equation (\ref{eq:bigcomm}) also gives $[T^{+(m)}, T^{+(n)}]$ if we replace every $S$ by a $T$. We finally quote the following commutation relations without proofs (see \cite{Deguchi} and references therein): 
\begin{equation}\label{eq:SandTcommute}
[S^{+(m)},T^{+(n)}]=[S^{-(m)},T^{-(n)}]=0.
\end{equation}

We end this paragraph by extending the domain of validity of some relations that were proved in \cite{Deguchi} for the special value $v=1$. We use the shorthand notation $S^z \equiv n \textrm{\,mod\,} P$ to indicate that the identities hold if the action is restricted to the subspace where $S^z$ acts as the identity times an integer congruent to $n$ modulo $P$. In the following lemmas, $H$ is the XXZ Hamiltonian introduced in paragraph \ref{sec:transferMatrix}.

\begin{Lemme}\label{lem:41} If $q^{2P} = 1$, $S^z \equiv n \hspace{-0.15cm}\mod\hspace{-0.05 cm}  P$ and $q^{2n} v^{\pm 2N}  = 1$, then
\begin{align}
\bar \Omega S^{\pm(P)} \bar \Omega^{-1} = q^P S^{\pm(P)}, \qquad \bar \Omega T^{\mp(P)} \bar \Omega^{-1} = q^P T^{\mp(P)}, \qquad [S^{\pm(P)}, H] = [T^{\mp(P)}, H] = 0. 
\end{align}
\label{sec:firstcomm}
\end{Lemme}
\noindent{\scshape Proof\ \ } The proof of the commutations with $H$ will be a direct consequence of the commutation or anticommutation relation with $\bar \Omega$ (as $q^P \in \{+1, -1\}$). Indeed, 
\begin{equation*}
[S^{\pm(P)}, H] = [S^{\pm(P)}, \bar e_N] = S^{\pm(P)} (\bar \Omega^{-1} \bar e_1 \bar \Omega) - (\bar \Omega^{-1} \bar e_1 \bar \Omega) S^{\pm(P)} = 0
\end{equation*}
because $S^{\pm(P)}$ commutes with $\bar e_1$. To extend the argument to $T^{\pm(P)}$, we note that
\begin{equation}\bar e_j^* = \bar e_j + \frac{(q-q^{-1})}2(\sigma_j^z- \sigma_{j+1}^z) \label{eq:barej} \end{equation}
so that $H = \sum_{j=1}^{N} \bar e_j = \sum_{j=1}^{N} \bar e_j^*$, and because $[T^{\pm(P)}, \bar e_1^*] = 0$, the same argument carries through. It only remains to prove the commutation relations with $\bar \Omega$. From the definitions,
\begin{align*}
\bar \Omega S^{\pm} \bar \Omega^{-1} &= q^{-\sigma_N^z} \left( (S^{\pm}-S^{\pm}_N) + q^{2S^z}  v^{\pm 2N}  S^{\pm}_N \right) \\
\bar \Omega T^{\pm} \bar \Omega^{-1} &= q^{\sigma_N^z} \left( (T^{\pm}-T^{\pm}_N) + q^{-2S^z}  v^{\pm 2N}  T^{\pm}_N \right)
\end{align*}
where the second is obtained from the first by changing $q$ for $q^{-1}$. A slightly tedious computation yields
\begin{align}
(\bar \Omega S^{\pm} \bar \Omega^{-1})^x = q^{-x \sigma_N^z} \left((S^{\pm})^x + q^{\mp(x-1)}[x] (S^{\pm})^{x-1}S^{\pm}_N (q^{2(S^z \pm x)}  v^{\pm 2N} - 1) \right) \label{eq:OSxO},\\
(\bar \Omega T^{\pm} \bar \Omega^{-1})^x = q^{x \sigma_N^z} \left((T^{\pm})^x + q^{\pm(x-1)}[x] (T^{\pm})^{x-1}T^{\pm}_N (q^{-2(S^z \pm x)}  v^{\pm 2N} - 1) \right) \label{eq:OTxO}.
\end{align}
Note that these two identities hold for all $q, v$ and for all $S^z$-eigenspaces. Dividing on both sides by $[x]!$, setting $x=P$ and choosing $v$ such that $q^{2S^z}  v^{\pm 2N} = 1$ in the first equation but $q^{2S^z}  v^{\mp 2N} =1$ in the second, we obtain the correct result, as $q^{\pm x \sigma_N^z} \rightarrow q^P$ and the dependence on the last spin disappears.
\hfill$\square$ \\

\begin{Lemme}\label{sec:firstcomm2} If $q^{2P} = 1$, $S^z \equiv n \hspace{-0.15cm}\mod\hspace{-0.05 cm}  P$, $k \in \mathbb N$
\begin{itemize}
\item and $q^{2(n\pm k)} v^{\pm 2N}  = 1$, then 
\begin{align}
\bar \Omega \left( T^{\mp(k)}S^{\pm(k+P)} \right) \bar \Omega^{-1} = q^P  \left( T^{\mp(k)}S^{\pm(k+P)} \right), \qquad [ T^{\mp(k)}S^{\pm(k+P)}, H] = 0;\label{eq:TSSO}
\end{align}
\item and $q^{-2(n\pm k)} v^{\pm 2N}  = 1$, then 
\begin{align}
\bar \Omega \left( S^{\mp(k)}T^{\pm(k+P)} \right) \bar \Omega^{-1} &= q^P   \left( S^{\mp(k)}T^{\pm(k+P)} \right), \qquad  [ S^{\mp(k)}T^{\pm(k+P)}, H] = 0.\label{eq:STTO}
\end{align}
\end{itemize}
\end{Lemme}
\noindent{\scshape Proof\ \ } To prove the first, we start with equations (\ref{eq:OSxO}) and (\ref{eq:OTxO}) and compute
\begin{align*}
\bar \Omega \left( T^{\mp(k)}S^{\pm (k+x)} \right) \bar \Omega^{-1} = &\left(T^{\mp(k)} + q^{\mp(k-1)} T^{\mp(k-1)}T^{\mp}_N (q^{-2S^z}  v^{\mp 2N} - 1) \right) q^{-x \sigma_N^z} \\ & \qquad \times \left(S^{\pm(k+x)} + q^{\mp(x+k-1)}S^{\pm(k+x-1)}S^{\pm}_N (q^{2(S^z \pm (k+x))}  v^{\pm 2N} - 1) \right).
\end{align*}
If $x=P$, $q^{2P}=1$, $S^z \equiv n \hspace{-0.15cm}\mod\hspace{-0.05 cm}  P$, the second term of the second parenthesis vanishes. Because 
$$(q^{-2S^z}  v^{\mp 2N} - 1) S^{\pm(k+x)} = S^{\pm(k+x)} (q^{-2(S^z \pm (k+x)) }  v^{\mp 2N} - 1) \rightarrow 0,$$
the second term of the first parenthesis is also zero and equation (\ref{eq:TSSO}) follows. Proving the commutation with $H$ is not as straightforward as before. For $j = 1, \dots, N-1$,
\begin{equation*}
[T^{\mp(k)}S^{\pm (k+P)}, \bar e_j] = [T^{\mp(k)}, \bar e_j] S^{\pm (k+P)} = \frac{(q-q^{-1})}2 [T^{\mp(k)},  \sigma_{j+1}^z- \sigma_{j}^z] S^{\pm (k+P)},
\end{equation*}
where we have used (\ref{eq:barej}) and  $[T^{\pm(P)}, \bar e_1^*] = 0$ at the second equality. Then,
\begin{equation}
[T^{\mp(k)}S^{\pm (k+P)}, \sum_{j=1}^{N-1}\bar e_j] = \frac{(q-q^{-1})}2 [T^{\mp(k)},  \sigma_{N}^z- \sigma_{1}^z] S^{\pm (k+P)}. 
\label{eq:partialcomm}\end{equation}
The term $j=N$ is problematic and has to be treated differently,
\begin{align*}
[T^{\mp(k)}S^{\pm (k+P)}, \bar e_N] &= q^P \bar \Omega [T^{\mp(k)}S^{\pm (k+P)}, \bar e_1] \bar \Omega^{-1} = q^P  \frac{(q-q^{-1})}2 \bar \Omega [T^{\mp(k)},  \sigma_{2}^z- \sigma_{1}^z] S^{\pm (k+P)} \bar \Omega^{-1} \\
&= q^P  \frac{(q-q^{-1})}2 [\bar \Omega T^{\mp(k)}\bar \Omega^{-1} ,  \sigma_{1}^z- \sigma_{N}^z] \bar \Omega S^{\pm (k+P)} \bar \Omega^{-1}.
\end{align*}
With the conditions given in the propositions, equations \eqref{eq:OSxO} and \eqref{eq:OTxO} lead to the following substitutions:
\begin{equation*}
\bar \Omega S^{\pm (k+P)} \bar \Omega^{-1} \rightarrow q^P q^{-k \sigma_N^z} S^{\pm(k+P)}, \qquad \bar \Omega T^{\mp(k)} \bar \Omega^{-1} \rightarrow q^{k \sigma_N^z}  T^{\mp(k)}.
\end{equation*}
Then the $q^{k \sigma_N^z}$s cancel out and 
\begin{equation*}
[T^{\mp(k)}S^{\pm (k+P)}, \bar e_N] =  \frac{(q-q^{-1})}2 [ T^{\mp(k)},  \sigma_{1}^z- \sigma_{N}^z]S^{\pm (k+P)} 
\end{equation*}
and, from (\ref{eq:partialcomm}), $[T^{\mp(k)}S^{\pm (k+P)}, H] = 0$. The equations \eqref{eq:STTO}
are obtained by replacing $q$ by $q^{-1}$ everywhere, as $H$ and $\bar \Omega$ are invariant under this transformation.
\hfill$\square$ 

\subsection[The intertwiner $\tilde i_N^d$]{The intertwiner $\boldsymbol{\tilde i_N^d}$}\label{sec:linkAndSpin}

Recall that the representation $\omega_d$ of $\eptl$ acts on $\tilde V_N^d$ whose dimension is $\left(\begin{smallmatrix}N\\ (N-d)/2\end{smallmatrix}\right)$. Note also that the representation $\tau$ that acts on $\cn$ commutes with $S^z$. Therefore the subspace $W_n\subset \cn$ where $S^z$ acts as $n \cdot id$ is stable under $\tau$. This subspace is of dimension $\left(\begin{smallmatrix}N\\ (N-2n)/2\end{smallmatrix}\right)$. The coincidence between their dimension is not fortuitous: $\omega_d$ and $\left.\tau\right|_{W_{d/2}}$ are intimately related and the relation is captured by an intertwiner $\tilde i_N^d$ that we now define. The intertwiner is similar to a map between some quotient of the affine Temperley-Lieb algebra $PTL_{2N}$, that is, the algebra generated by the $e_i$s only (without the $\Omega^\pm$), and the eigenspace $W_0$ introduced a long time ago by Martin and Saleur \cite{MartinSaleur} (see their equations (33--34)). In \cite{AY:alpha}, we proved the intertwining properties of $\tilde i_N^d$. The proof that equation \eqref{eq:elementE} and the last one of \eqref{eq:TLPN} are preserved by the intertwiner is the most delicate in our work.

Let $w \in \tilde B_N^d$ be a link state containing $n = (N-d)/2$ arches and let $\psi(w) = \{ (i_1, j_1), (i_2, j_2), \dots, (i_n, j_n)\}$, where the $i_m$s are positions where arches start and the $j_m$s the positions where they end. The $i_m$s are chosen in the interval $1, \dots, N$ while the $j_m$s satisfy $i_m+1 \le j_m \le N+i_m-1$. The linear transformation $\tilde{i}^d_N: \tilde{V}_N^d \rightarrow  W_{d/2}\subset\cn$ is defined by its action on the basis elements of $\tilde B_N^d$ as:
\begin{equation}
\tilde{i}_N^d(w) = \prod_{(i,j) \in \psi(w)} \hspace{-0.2cm}\tilde{T}_{i,j} \hspace{0.2cm}| 0 \rangle \qquad \textrm{where} \qquad \tilde{T}_{i,j} = v^{j-i}  (-q)^{\frac12} \sigma^-_j + v^{-(j-i)}(-q)^{-\frac12} \sigma^-_i, 
\label{eq:homomorphisme}\end{equation}
and $ | 0 \rangle =  | + + \dots + \, \rangle$ is the state with all spins up. Some states have boundary arches, that is, arches that cross the imaginary boundaries depicted by dotted lines in paragraph \ref{sec:EPTL}. This happens if a point $i$ is connected to a point $j\ge N+1$. As usual we then use the convention that  $\sigma^\pm_j = \sigma^\pm_{j\,\textrm{mod} \, N}$.

In \cite{AY:alpha} we proved the following properties of $\tilde i_N^d$.

\begin{Proposition}\label{thm:leszerosdei} The linear map $\tilde{i}^d_N: \tilde{V}_N^d \rightarrow  W_{d/2}\subset\cn$ is an intertwiner for the representations $\omega_d$ and $\left.\tau\right|_{W_{d/2}}$ of $\eptl(\beta,\alpha)$, namely:
$$\left.\tau\right|_{W_{d/2}}\circ \tilde i_N^d=\tilde i_N^d\circ\omega_d$$
for $\beta=-q-q^{-1}$ and $\alpha=v^N+v^{-N}$. In the link state basis for $\tilde V_N^d$ and the usual spin basis for $W_{d/2}$, the determinant of the matrix $I_N^d$ representing the intertwiner $\tilde i_N^d$ is, up to a sign,
\begin{equation}\det I_N^d(q,v)=\prod_{k=1}^{(N-d)/2}\langle k+d/2\rangle^{\left(\begin{smallmatrix}N\\ (N-d)/2-k\end{smallmatrix}\right)}\label{eq:detINd}
\end{equation}
where $\langle x\rangle=q^xv^N-q^{-x}v^{-N}$. 
\end{Proposition}

\noindent Let us recall that the letter $v$ was chosen to describe both the free parameter in the representation $\tau$ and the twist parameter in the representation $\omega_{d,v}$. The previous proposition relates these two representations evaluated at the same value of $v$.

The pair $(N,d)$ is said to be {\em critical} if it belongs to 
$$\left\{ (N,d)\,|\,  q^{2k+d}v^{2N}=1  \textrm{\ for some }k, 1\le k\le (N-d)/2\right\}.$$
Clearly the criticality of $(N,d)$ depends on $q$ and $v$. 
If $(N,d)$ is a critical pair, the intertwiner $\tilde i_N^d$ has therefore a non-trivial kernel. Otherwise it is an isormorphism.

%
\section[Jordan blocks within sectors]{Jordan blocks within sectors of loop models}\label{sec:BJintra}
%

{\it Throughout this section, the action of the periodic Temperley-Lieb algebra on the space $\tilde V_N^d$ is that of representation $\omega_d$. The present section is independent of the previous one. From now on, we will often write $\mathcal H$ for $\omega_d(\mathcal H)$ and use greek letters for elements of $\cn$ and latin ones for those in $\tilde V_N^d$.
}

\medskip

Finding Jordan cells in
$\mathcal H = \omega_d(\mathcal H)=\omega_d(\sum_{1\le i\le N}e_i)$
 requires new techniques. The method used in section \ref{sec:BJentre} and based on our previous work \cite{AMDSA} fails here, simply because the central element $F_N$ acts diagonally in the representation $\omega_d$. The new method relies on the behaviour of the intertwiner $\tilde i_N^d$ at critical pairs $(q_c,v_c)$. Specifically, our construction of Jordan cells will be valid for the values of $N, d, q$ and $v$ constrained by the following hypotheses:
\begin{Hypotheses}\label{hyp:hypHourra}\begin{itemize}
\item[(i)] $q$ is a root of unity with $P\in [2,N/2]$ the smallest integer such that $q^{2P}=1$;
\item[(ii)] the twist parameter $v$ satisfies $(qv^2)^N=q^{2k}$ for an integer $k\in [0,P-1]$;
\item[(iii)] the number of defects $d$ is $N-2P$;
\item[(iv)] the number of sites $N$ satisfies the inequality $N \ge 2P+k$.
\end{itemize}
\end{Hypotheses}
(The integer $P$ plays the role of $p'$ introduced in \cite{PRZ} for logarithmic minimal models $\mathcal{LM}(p,p')$.)

The steps of the construction are as follows. Paragraph \ref{sec:observations} will show that, if $\mathcal H$ has a Jordan cell, the eigenvector of this cell and all the first Jordan partners, except maybe the very ``last'' one, must belong to the kernel of $\tilde i_N^d(q_c,v_c)$. 
The next step will be to study the relationship between two matrices, $I_0$ and $M_0$, appearing in the expansion of $I_N^d$ and its inverse in a neighbourhood of the critical $q_c$. Lemmas \ref{sec:firstcomm} and \ref{sec:firstcomm2} showed that some elements of $\uq$ commute with $H$. These will be used to explore the (image by $\tilde i_N^d$ of the) generalized eigenspaces of $\mathcal H$. The last paragraph constructs explicitly Jordan cells of $\mathcal H$ for the infinite family of critical pairs $(N,d)$ and values $(q,v)$ satisfying hypotheses \ref{hyp:hypHourra}. The final result is that under these hypotheses, the loop Hamiltonian $\mathcal H$ has a $2 \times 2$ Jordan cell with eigenvalue $\lambda = 0$.

%
\subsection{Basic observations and identities} \label{sec:observations}
%

Both $\omega_d(\mathcal H)$ and the XXZ Hamiltonian $H$ are representatives of the abstract element $\sum_{1\le i\le N}e_i\in \eptl$. The spin Hamiltonian $H$ is hermitian if $q$ and $v$ are both on the unit circle and is then diagonalizable. If $\tilde i_N^d:\tilde V_N^d\rightarrow W_{d/2}$ is an isomorphism and therefore invertible, any eigenvector $|\nu\rangle$ of $H$ is mapped by $(\tilde i_N^d)^{-1}$ onto an eigenvector of $\mathcal H$ with the same eigenvalue. In this case $\mathcal H$ is thus also diagonalizable.
 Suppose the pair $(N,d)$ is critical and that $\tilde i_N^d(q_c,v_c)$ is singular (see proposition \ref{thm:leszerosdei}). For fixed $v$, the map $\tilde i_N^d$ is singular at a finite set of points. Fix $v=v_c$ and choose $D_0\subset \mathbb C$ to be a neighbourhood of $q_c$ that contains only $q_c$ as singular point. By the previous argument the matrices $\mathcal H(q)$ and $H(q)$, $q\in D_0$ and $q\neq q_c$, have the same eigenvalues $\{\lambda_s(q),1\le s\le p\}$ for some $p$, with the same (algebraic) multiplicities $m_s$. The set $D_0$ may be chosen such that the number $p$ of distinct eigenvalues is constant on $D_0\setminus \{q_c\}$ \cite{Kato}. By continuity the coincidence of eigenvalues and multiplicities will also hold for the pair $\mathcal H(q_c)$ and $H(q_c)$, even though the number of distinct eigenvalues in $\mathcal H(q_c)$ may be less than $p$.
(The symbol ``$\lambda$'' is used to denote a generic eigenvalue and to parametrize $q=e^{i\lambda}$. These two quantities are distinct and the context should clearly indicate which is meant.)

Suppose now that $\mathcal H$ has a non-trivial $n\times n$ Jordan cell ($n>1$) associated with the eigenvalue $\lambda$ and that vectors $v_i\in\tilde V_N^d$, $1\le i\le n$, have been chosen so they satisfy the canonical relations:
\begin{equation*}
\mathcal Hv_1=\lambda v_1\qquad\textrm{and}\qquad
\mathcal Hv_i=\lambda v_i+v_{i-1}, \quad 2\le i\le n.
\end{equation*}
Let $|\nu_i\rangle=\tilde i_N^d(v_i)$ be their images in $\cn $. The subspaces $U_i=\textrm{sp\,}\{|\nu_1\rangle, |\nu_2\rangle, \dots,|\nu_i\rangle\}$, $1\le i\le n$, are all stable under the action of $H$. Since $H$ is diagonalizable on any subspace stable under its action, the restrictions of $H$ to the $U_i$s must be diagonalizable. For example, it must be diagonalizable on $U_2$ where its action is 
$$(H-\lambda\cdot id)|\nu_1\rangle = 0,\qquad 
(H-\lambda\cdot id)|\nu_2\rangle = |\nu_1\rangle.$$
Since all Jordan cells in $H$ are $1\times 1$, $|\nu_1\rangle$ must vanish. This argument may be repeated for the following pairs $|\nu_{i-1}\rangle$ and $|\nu_i\rangle$. Only $|\nu_n\rangle=\tilde i_N^d(v_n)$ might thus be distinct from zero. (Even though it is not used in the following, we note that, in all the cases we considered, our computer explorations show that this image 
$\tilde i_N^d(v_n)$ is indeed non-zero.)

The map $\tilde i_N^d$ is singular along curves in the complex plane $(q,v)$ or, more precisely, in $(\mathbb C^{\times})^2$. Suppose the structure of $\mathcal H$ at a given singular point $(q_c, v_c)$ is to be studied. Fix $v_c$ once for all and consider the matrix $I_N^d$ representing $\tilde i_N^d$ in the link and spin bases. It is now 
a polynomial in $q$ and $q^{-1}$ if $(N-d)/2$ is even, and in $q^{\frac12}$ and $q^{-\frac12}$ if $(N-d)/2$ is odd (see equation \eqref{eq:homomorphisme}). For simplicity we discuss the case where it is a polynomial in $q$ and $q^{-1}$. An expansion is then possible:
$$I_N^d(q)=I_N^d(q, v_c)=I_0+(q-q_c)I_1+\dots$$
in a neighborhood of $q_c$. 
(Note that multiplication by a power of $q$ transforms $I_N^d$ into a polynomial in $q$ and this factor does not change the singular behavior of $I_N^d$ at $(q_c,v_c)$. Some expressions might be simpler with this additional factor.) The zeros of the determinant of $I_N^d(q)$ were identified in proposition \ref{thm:leszerosdei}. They occur if $q^{2k+d}v_c^{2N}=1$ for some $k$, $1\le k\le (N-d)/2$. For $q\neq q_c$ in a neighborhood of $q_c$, the inverse $I_N^d(q)$ can be written as $A/\det I_N^d$ where $A$ is the matrix of cofactors which are polynomials in $q$. Because the polynomial $\det I_N^d$ has a zero at $q_c$ of a certain (positive) degree, there must be an integer $\iota$ such that the function $(I_N^d)^{-1}(q)$ can be cast into the following Laurent series
\begin{equation}(I_N^d)^{-1}(q)=\frac1{(q-q_c)^\iota}M_0+\frac1{(q-q_c)^{\iota-1}}M_1+\dots \label{eq:defM0}\end{equation}
where $M_0, M_1,\, \dots$ are constant matrices and $M_0$ is non-zero. It is important to remember between which spaces these act:
$$I_0,I_1:\tilde V_N^d\rightarrow W_{d/2}\qquad\textrm{and}\qquad
M_0,M_1:W_{d/2}\rightarrow \tilde V_N^d.$$
Since, in the neighborhood of $q_c$, they satisfy $I_N^d(q)(I_N^d)^{-1}(q)=id=(I_N^d)^{-1}(q)I_N^d(q)$ for $q\neq q_c$, these matrices must satisfy 
\begin{equation}\label{eq:imM0}I_0M_0=0=M_0I_0.\end{equation}
We conclude that $\im I_0\subset \ker M_0$ and  $\im M_0\subset \ker I_0$. For the family of examples presented in paragraph \ref{sec:jordan}, these inclusions will actually be equalities. This family includes the case $I_4^0$ around $(q_c,v_c)=(i,1)$ that will be given as an example below.

Finally, since both the loop and the XXZ Hamiltonians are polynomials in $q$, they enjoy similar expansions around $q_c$:
$$\mathcal H(q)=\mathcal H_0+(q-q_c)\mathcal H_1+\dots\qquad\textrm{and}\qquad
H(q)=H_0+(q-q_c) H_1+\dots$$
The intertwining property of $\tilde i_N^d$ gives, for the leading orders in $q-q_c$, the relations:
\begin{equation}\label{eq:h0m0inter}\mathcal H_0M_0=M_0H_0\qquad\textrm{and}\qquad \mathcal H_0M_1+\mathcal H_1M_0=M_1H_0+M_0H_1.\end{equation}

\noindent Here is an example for $N=4$ and $d=0$, where
\begin{equation}\label{eq:exemplei42}
I_4^0(q,v)=
\left(
\begin{smallmatrix}
 -q v^2 & v^2 & v^{-2} & -q^{-1}v^{-2} & v^{-2} & v^2 \\
 0 & -q v^4 & 0 & 1 & 0 & -q^{-1}v^{-4} \\
 1 & 0 & -q v^4 & 0 & -q^{-1}v^{-4} & 0 \\
-q^{-1}v^{-2} & v^{-2} & v^2 & -q v^2 & v^2 & v^{-2} \\
 1 & 0 & -q^{-1}v^{-4} & 0 & -q v^4 & 0 \\
 0 & -q^{-1}v^{-4} & 0 & 1 & 0 & -q v^4
\end{smallmatrix}
\right),
\end{equation}
in the basis
$$\left\{\ 
| +-+- \rangle,\ \ 
| ++-- \rangle,\ \ 
| -++- \rangle,\ \ 
| -+-+ \rangle,\ \ 
| +--+ \rangle,\ \ 
| --++ \rangle
\ \right\}.$$
For $v = v_c = 1$, all non-zero elements of $I_4^0(q,v=1)$ are either $1$, $-q$ or $-q^{-1}$. Its determinant has a zero at $q_c=i$ of degree $1$ and its expansion is the following polynomial where $I_0$, $I_1$ and $I_2$ can be easily read off: 
\begin{equation*}
q\cdot I_4^0(q,v=1)=\left(
\begin{smallmatrix}
 1 & i & i & -1 & i & i \\
 0 & 1 & 0 & i & 0 & -1 \\
 i & 0 & 1 & 0 & -1 & 0 \\
 -1 & i & i & 1 & i & i \\
 i & 0 & -1 & 0 & 1 & 0 \\
 0 & -1 & 0 & i & 0 & 1
\end{smallmatrix}\right)
+(q-i)
\left(
\begin{smallmatrix}
 -2 i & 1 & 1 & 0 & 1 & 1 \\
 0 & -2 i & 0 & 1 & 0 & 0 \\
 1 & 0 & -2 i & 0 & 0 & 0 \\
 0 & 1 & 1 & -2 i & 1 & 1 \\
 1 & 0 & 0 & 0 & -2 i & 0 \\
 0 & 0 & 0 & 1 & 0 & -2 i
\end{smallmatrix}
\right)
+(q-i)^2
\left(
\begin{smallmatrix}
 -1 & 0 & 0 & 0 & 0 & 0 \\
 0 & -1 & 0 & 0 & 0 & 0 \\
 0 & 0 & -1 & 0 & 0 & 0 \\
 0 & 0 & 0 & -1 & 0 & 0 \\
 0 & 0 & 0 & 0 & -1 & 0 \\
 0 & 0 & 0 & 0 & 0 & -1
\end{smallmatrix}
\right).
\end{equation*}
The inverse can be similarly expanded to get the first $M_0$ and $M_1$: 
\begin{equation*}
\big(q\cdot I_4^0(q,v=1) \big)^{-1}=\frac{1}{q-i}\cdot{\textstyle{\frac1{8}}}\hskip-0.1cm
\left(
\begin{smallmatrix}
 0 & 0 & 0 & 0 & 0 & 0 \\
 -1 & i & -i & 1 & -i & i \\
 1 & -i & i & -1 & i & -i \\
 0 & 0 & 0 & 0 & 0 & 0 \\
 1 & -i & i & -1 & i & -i \\
 -1 & i & -i & 1 & -i & i
\end{smallmatrix}
\right)+{\textstyle{\frac1{16}}}\hskip-0.1cm
\left(
\begin{smallmatrix}
 4 & -4 i & -4 i & -4 & -4 i & -4 i \\
 -3 i & 3 & 1 & -i & 1 & -5 \\
 -i & 1 & 3 & -3 i & -5 & 1 \\
 -4 & -4 i & -4 i & 4 & -4 i & -4 i \\
 -i & 1 & -5 & -3 i & 3 & 1 \\
 -3 i & -5 & 1 & -i & 1 & 3
\end{smallmatrix}
\right)+\dots
\end{equation*}
from which one reads $\iota = 1$.

%
\subsection[The left nullspace of $I_N^d$]{The left nullspace of $\boldsymbol{I_N^d}$}\label{sec:kern}
%

This paragraph is a description of the left nullspace of the matrix $I_N^d$ and will take the form of two corollaries. 
We recall that the left kernel of an $n\times n$ matrix $A$ is $\{v\in\mathbb C^n\, |\, v^\dagger A=0\}$ and that the subspace $\textrm{left}\ker A$ is an orthogonal  complement of $\im A$ in $\mathbb C^n$ for the usual inner product. 

The purpose of this description is two-fold. First, constructing vectors in the left nullspace of $I_N^d$ will allow us to determine that under the hypotheses \ref{hyp:hypHourra}, $\iota = 1$ and $\im I_0= \ker M_0$. Second, proving the existence of Jordan cells will require a projection along this left nullspace, and corollary \ref{sec:coro2} will provide us with the appropriate state to project on.
To understand this better, write $W_{d/2}$ as the direct sum $\im I_N^d\oplus \textrm{left}\ker I_N^d$. Since $\im I_0 \subset \ker M_0$, if $M_0\nu$ (with $\nu\in W_{d/2}$) is non-zero, then $\nu$ has a non-zero component in the left kernel of $I_N^d$ for this direct sum splitting. Moreover, in the case $\im I_0 = \ker M_0$, $M_0\nu$ is non-zero if and only if $\nu$ has such a non-zero component.

Let $w_{N^y}^N$ be the link state with $y = (N-d)/2$ concentric arches straddling the imaginary boundary, that is, all centered at position $N$. Note that $\tilde V_N^d$, under the action of $\omega_d$, is cyclic with generator $w_{N^y}^N$. This means that any link state $w$ in $\tilde B_N^d$ can be written as $\left(\prod_i e_{k_i} \right) w_{N^y}^N$ for some $k_i$s. 
In fact, transforming $w_{N^y}^N$ into $w$ can be done first by moving the defects of $w_{N^y}^N$ to their desired positions in $w$ and then reassembling the bubbles to create the nesting of bubbles appearing in $w$. It is not too difficult to see that both steps can be performed without the use of $e_N$. Therefore the set allowed for the indices $k_i$ can even be restricted to $\{1, \dots, N-1\}$. Because $S^+$ commutes with all $e_i$s with $1\le i\le N-1$, any link state $w$ satisfies $S^{+(x)}\tilde i_N^d(w) = \left(\prod_i \bar e_{k_i} \right) S^{+(x)}\tilde i_N^d(w_{N^y}^N)$.

The linear map $\tilde i_N^d$ is singular by proposition \ref{thm:leszerosdei} for certain pairs $(N,d)$ associated to $(q_c,v_c)$. Suppose now that $q$ and $v$ are such that $S^{+(x)}\tilde i_N^d(w_{N^y}^N)$ is zero for some integer $x\leq y$. By the previous observations, $S^{+(x)}\tilde i_N^d(w)=0$ for any link state $w$ and therefore $S^{+(x)}\tilde i_N^d$ acts as zero on $\tilde V_N^d$. In other words, $\im \tilde i_N^d\subset \ker S^{+(x)}$ and the vectors $\langle 0 | \sigma^+_{j_1}\sigma^+_{j_2}\dots\sigma^+_{j_{y-x}} S^{+(x)}$, for all choices of $j_1,j_2,\dots,j_{y-x}$, are in the left nullspace of the matrix $I_N^{N-2y}$. Proposition \ref{sec:ouach} in appendix \ref{app:0} is thus devoted to the computation of $S^{+(x)}\tilde i_N^d(w_{N^y}^N)$ 
which has the immediate corollary:

\begin{Corollaire} Let $1\le x \le y\le \lfloor N/2 \rfloor$. The vector $S^{+(x)}\tilde i_N^d(w_{N^y}^N)$ is zero if and only if there is an integer $x_c$ in the range $1\le x_c \le x \le y$ such that $\langle d/2+x_c\rangle=0$.
Moreover, if $q, v$ and the integer $x_c$, $1\le x_c\le y$, are such that $\langle d/2 + x_c \rangle = 0$, 
the limit $S^{+(x)}\tilde i_N^d(w_{N^y}^N)/(q-q_c)$ as $q\rightarrow q_c$ exists and is non-zero, and
the states $\langle 0 | \sigma^+_{j_1}, \dots, \sigma^+_{j_{y-x}} S^{+(x)}$ are in the left nullspace of $\tilde i_N^{N-2y}$, for all $x\in \{ x_c, x_c+1, \dots, y \}$ and all choices $1\le j_1<j_2<\dots<j_{y-x}\le N$.
\label{sec:coro}\end{Corollaire}

\noindent Note that lemmas \ref{sec:firstcomm} and \ref{sec:firstcomm2} both require that $q$ be a root of unity, but corollary \ref{sec:coro} does not. 

We end this paragraph with a last corollary that provides us with yet another state in the left kernel of $I_N^d$ and comes as a direct consequence of Proposition \ref{sec:nu3}. 
\begin{Corollaire} Under the hypotheses \ref{hyp:hypHourra}, the state $\langle 0|T^{+(N-k)}S^{-(N-P-k)}$ is in the left nullspace of $I_N^d$.
\label{sec:coro2}\end{Corollaire}

%
\subsection[The images and kernels of $I_0$ and $M_0$]{The images and kernels of $\boldsymbol{I_0}$ and $\boldsymbol{M_0}$}\label{sec:imker}
%

From now on, the hypotheses \ref{hyp:hypHourra} will restrict the values of $(N,d)$ and $(q,v)$ considered. 
We start this paragraph by explaining the role of these hypotheses.

First, the Jordan cell to be constructed will belong to the generalized eigenspace of $\mathcal H$ of eigenvalue $\lambda=0$. The three following vectors
\begin{align*}
|\nu_1 \rangle &= S^{+(k)}T^{-(P+k)}|0 \rangle, \\
|\nu_2 \rangle &= S^{+(P+k)}T^{-(2P+k)}|0 \rangle, \\
|\nu_3 \rangle &= T^{+(N-P-k)}S^{-(N-k)}|0 \rangle,
\end{align*}
belong to the subspace $W_{N/2-P}$. Because of hypotheses (i) and (ii), lemmas \ref{sec:firstcomm} and \ref{sec:firstcomm2}, and the fact that $H|0 \rangle=0$, these three vectors, if non-zero, will also be eigenvectors of $H$ with eigenvalue $0$. Hypothesis (iv) is necessary for $| \nu_2 \rangle$ to be non-zero. These three vectors will play a central role in what follows.

Second, paragraph \ref{sec:observations} has argued that a non-trivial Jordan cell may occur only if $\tilde i_N^d$ is singular and proposition \ref{thm:leszerosdei} restricts the values of $(N,d)$ and of $(q,v)$ where $\tilde i_N^d$ is singular. Indeed, up to a sign, the determinant of $I_N^d$ is 
\begin{equation*}
\det I_{N}^{d} = \prod_{j=1}^{(N-d)/2} \langle j+d/2\rangle^{\left(\begin{smallmatrix} N \\ (N-d)/2-j\end{smallmatrix}\right)}=
\prod_{j=1}^{P} \langle j+N/2-P\rangle^{\left(\begin{smallmatrix} N \\ P-j\end{smallmatrix}\right)}
\end{equation*}
if the number of defects is $N-2P$. The factors $\langle j+N/2-P\rangle$ are zero if and only if $q^{2j+N-2P}v^{2N}=1$. With the hypotheses \ref{hyp:hypHourra} this happens only for the term $j=P-k$ in the product  and the zero at $q=q_c$ is of degree $\left(\begin{smallmatrix} N \\ k \end{smallmatrix}\right)$. All other terms are non-zero.  For the rest of the section, the twist parameter $v$ is fixed to one of the values $v_c$ satisfying the hypotheses \ref{hyp:hypHourra}, but not $q$, as the proof of the existence of Jordan blocks will require information from the neighboorhood of $q=q_c$.

Third, the hypotheses \ref{hyp:hypHourra} lead to a finer description of the fundamental subspaces of $I_0$ and $M_0$, namely the inclusion $\im I_0\subset \ker M_0$ can then be replaced with the equality $\im I_0=\ker M_0$. The rest of the present paragraph is devoted to proving this statement.

\begin{Lemme} Assume the hypotheses \ref{hyp:hypHourra}. The left kernel of $\tilde i_N^d(q_c)$ has dimension $\left(\begin{smallmatrix}N\\ k\end{smallmatrix}\right)$ and is spanned by 
$$\langle \mu|=\langle 0|S^{+(P)}\qquad \textrm{and} \qquad \langle \mu_{j_1, \dots, j_k}|=\langle 0|\sigma^+_{j_1}\sigma^+_{j_2} \dots  \sigma^+_{j_k}S^{+(P-k)}$$
for $1\leq j_1<j_2<\dots <j_k\leq N$.
\end{Lemme}

\noindent{\scshape Proof\ } Note that the maps $S^{+(P)}$ and $\sigma^+_{j_1}\sigma^+_{j_2} \dots  \sigma^+_{j_k}S^{+(P-k)}$ are non-zero, each being a sum of $\sigma_{i_1}^+\sigma_{i_2}^+\dots\sigma_{i_P}^+$ with weights that are polynomials in $q$ and do not vanish if $q=q_c$ is a $2P$-root of unity. Therefore the vectors $\langle \mu|$ and $\langle \mu_{j_1, \dots, j_k}|$ are non-zero for all $q$ in a neighborhood of $q_c$. 

Setting first $x = y = P$ and then $x = P-k, y = P$ in corollary \ref{sec:coro}, we find that, close to $q_c$, the function $\langle \mu|\tilde i_N^d(w_{N^P}^N)$ behaves as $(q-q_c)^1$ times a non-zero constant, and $\langle \mu_{j_1, \dots, j_k} |\tilde i_N^d(w_{N^P}^N)$ as $(q-q_c)^i$ times a non-zero constant for some $i \ge 1$. Because $\tilde V_N^d$ is cyclic with generator $w_{N^P}^N$ and under the action of only the $e_i$s with $1\le i\le N-1$ (as argued in paragraph \ref{sec:kern}), the previous statement also holds true for $\langle \mu| I_N^d$ and $\langle \mu_{j_1, \dots, j_k} | I_N^d$, where the constant terms are now vectors. The vectors $\langle \mu|$ and $\langle \mu_{j_1, \dots, j_k}|$ thus belong to the left kernel of $\tilde i_N^d(q_c)$ .

The states $\langle \mu |$ and $\langle \mu_{j_1, \dots, j_k} |$, $1\le j_1<j_2<\dots <j_k\le N$, are not all independent in the limit $q \rightarrow q_c$. For $k=0$, this is trivial. For $k>0$, the linear combination
$$\sum_{1\le j_1< \dots  < j_k \le N} (qv^{2})^{(N+1)k/2-\sum_{i=1}^k j_i} \langle \mu_{j_1, \dots , j_k}|= \langle 0 | S^{+(k)}S^{+(P-k)} = \left[ \begin{matrix} P \\ k \end{matrix}\right] \times \langle 0 | S^{+(P)}$$
vanishes at $q = q_c$. The next goal is to show that this is the only linear combination of elements in  $\{\langle \mu |, \langle \mu_{j_1, \dots, j_k} |$, $1\le j_1<j_2<\dots <j_k\le N\}$, up to a constant, that is zero. 

Let us suppose for now that this last statement is true and that $\langle \mu_{j_1, \dots, j_k} | I_N^d$ has a zero of degree $1$ at $q = q_c$. Suppose further that the first $\left( \begin{smallmatrix}N \\ k \end{smallmatrix} \right)$ vectors in the spin basis of $\left.\cn\right|_{S^z=N/2-P}$ are replaced by $\left( \begin{smallmatrix}N \\ k \end{smallmatrix} \right)$ independent states in the previous set in such a way that the new set $\mathcal B$ is still a basis 
at $q_c$ and in a neighborhood of it. Let $\bar I^d_N$ be the matrix representing $\tilde i_N^d$ in this new basis. Its determinant is likely to be different, but only by a polynomial that does not vanish at $q_c$. Since all the rows of $\bar I_N^d$ corresponding to the states $\langle \mu|$ and $\langle \mu_{j_1, \dots, j_k}|$ have an overall factor $(q-q_c)$, the zero at $q_c$ of degree $\left( \begin{smallmatrix}N \\ k \end{smallmatrix} \right)$ of $\det I_N^d$ is accounted for by the product of the $(q-q_c)$ in each of the $\left(\begin{smallmatrix}N \\ k \end{smallmatrix} \right)$ first columns. Could there be another vector $\langle \mu'|$, linearly independent from $\langle \mu|$ and all the $\langle \mu_{j_1, \dots, j_k}|$s, in the left nullspace of $\tilde i_N^d$? 
If such a vector existed, a further change of bases from $\mathcal B$ to $\mathcal B'$ could be performed where $\langle \mu'|$ would take the place in $\mathcal B$ of one vector of the original link basis.
Again, in this new basis $\mathcal B'$, the matrix would still have polynomial entries and its determinant would have a further zero at $q_c$. But the zero at $q_c$ is of degree $\left( \begin{smallmatrix}N \\ k \end{smallmatrix} \right)$, as seen above, and the existence of such a $\langle \mu'|$ must be ruled out. For the same reason, every vector $\langle \mu_{j_1, \dots, j_k} | I_N^d$ must 
vanish at $q=q_c$ with a zero of degree $1$. 
If the existence of $\left( \begin{smallmatrix}N \\ k \end{smallmatrix} \right)$ independent states in the aforementionned set holds, the left nullspace of $\tilde i_N^d(q_c)$ is therefore $\left( \begin{smallmatrix}N \\ k \end{smallmatrix} \right)$-dimensional and spanned by the states $\langle \mu|$ and $\langle \mu_{j_1, \dots, j_k}|$.

We now prove that the span of $\{\langle 0|S^{+(P)} \}\cup \{\langle 0|\sigma^+_{j_1}\sigma^+_{j_2} \dots \sigma^+_{j_k}  S^{+(P-k)}, 1\le j_1<\dots <j_k\le N\}$ is the left nullspace of $\tilde i_N^d$. The question is reduced to computing the dimension of this subspace or, more simply, of $\textrm{span}\,\{ S^{-(P)}|0\rangle, S^{-(P-k)}\sigma^-_{j_1}\sigma^-_{j_2} \dots \sigma^-_{j_k}|0\rangle, 1\le j_1<\dots <j_k\le N\}$. This can be answered using the representation theory of $U_q(sl_2)$.

The representation of $S^\pm$ and $q^{S^z}$ introduced in paragraph \ref{sec:uqsl2} is somewhat unusual. However this representation is equivalent to the usual one, as studied for example in \cite{Jimbo}. (It can be brought to the latter form by the change of bases discussed in paragraph \ref{sec:transferMatrix}: $S^\pm \rightarrow \mathcal OS^\pm \mathcal O^{-1}$ and similarly for the other generators, with $\mathcal O=v^{\sum_{1\le j\le N}j\sigma^z_j}$. Note that Jimbo works with the generators $T_i$ of the Hecke algebra. Then $e_i=T_i-q^{-1}$ satisfies the relations of the left column of \eqref{eq:TLPN}.) With this observation, one can use what is known about the (usual) action of $U_q(sl_2)$ on $\cn $, from the pioneering work \cite{PasquierSaleur} to recent ones \cite{Bushlanov, GainutVasseur}. The latter works use a slightly different algebra, where the generators $S^\pm$ and $q^{S^z}$ satisfy $(S^{+})^P=(S^{-})^P=0$ and $(q^{S^z})^{2P}=1$. It also includes generators that act as the renormalized generators $S^{+(P)}$ and $S^{-(P)}$ (called divided operators in \cite{GainutVasseur}). The irreducible and projective representations of this algebra $\mathcal LU_q(sl_2)$ have been described in \cite{Bushlanov} and the decomposition of $\cn $ in terms of these has been given in \cite{GainutVasseur}. In this decomposition, the vector $|0\rangle$ belongs to one of these representations. Any other irreducible or projective representation $M$ appearing in $\cn $ enjoys the following property: for $0\le k<P$, if $M$ intersects non-trivially the subspace where $S^z=N/2-k$, then the restriction of $(S^-)^{P-k}$ to this intersection is of maximal rank. 
The vector $(S^-)^k|0\rangle$ in the subspace where $S^z=N/2-k$ belongs to the indecomposable representation of $\mathcal LU_q(sl_2)$ to which $|0\rangle$ belongs. This vector is clearly in the kernel of $(S^-)^{P-k}$ since $(S^-)^{P-k}((S^-)^{k}|0\rangle)=(S^-)^{P}|0\rangle$ and $(S^-)^{P}$ is zero. A complement of $\textrm{span}\{(S^-)^{k}|0\rangle\}$ in the same eigenspace of $S^z$ can be chosen so that it belongs to a sum of representations satisfying the above property. Therefore, on this complement $(S^-)^{P-k}$ is of maximal rank and the kernel of $(S^-)^{P-k}$ on the eigenspace $S^z=N/2-k$ 
is precisely one-dimensional and spanned by $(S^-)^k|0\rangle$. Therefore $\dim \textrm{span}\, (\{S^{-(P)}|0\rangle\}\cup \{S^{-(P-k)}\sigma^-_{j_1}\sigma^-_{j_2} \dots \sigma^-_{j_k}|0\rangle, 0\le j_1<j_2<\dots <j_k\le N\})$ is $(\begin{smallmatrix}N\\ k\end{smallmatrix})$ as claimed.\hfill $\square$

The next result is the main result of the current paragraph.

\begin{Lemme} Under the hypothesis \ref{hyp:hypHourra}, $\ker M_0 = \mathrm{im}\, I_0$. 
\end{Lemme}

\noindent{\scshape Proof\ } Because $M_0 I_0 = 0$, we know that $\im I_0 \subset \ker M_0$ and 
$$\dim \ker M_0 \ge \dim \im I_0 = \left( \begin{smallmatrix} N \\ P \end{smallmatrix} \right) - \dim \ker I_0 =  \left( \begin{smallmatrix} N \\ P \end{smallmatrix} \right) - \left( \begin{smallmatrix} N \\ k \end{smallmatrix} \right).$$
It is therefore sufficient to prove that $\dim \ker M_0 \le  \left( \begin{smallmatrix} N \\ P \end{smallmatrix} \right) - \left( \begin{smallmatrix} N \\ k \end{smallmatrix} \right)$. Let $u_i$, $i = 1, \dots, \left( \begin{smallmatrix} N \\ P \end{smallmatrix} \right)$, be the row vectors of the matrix $\bar I_N^d$ 
in the basis $\mathcal B$ introduced in the proof of the previous lemma. We have completely characterized the left kernel of $I_0$ and this allows us to write this Taylor expansion for the $u_i$s:
$$u_i = \left\{\begin{array}{ll}\displaystyle{\sum_{n = 1}^\infty}  (q-q_c)^n c_i^{n}, \qquad i = 1, \dots,  \left( \begin{smallmatrix} N \\ k \end{smallmatrix} \right), \\
\displaystyle{\sum_{n = 0}^\infty}  (q-q_c)^n c_i^{n}, \qquad i =  \left( \begin{smallmatrix} N \\ k \end{smallmatrix} \right) +1,  \dots,  \left( \begin{smallmatrix} N \\ P \end{smallmatrix} \right)
 \end{array}\right.$$ 
where all leading terms are non-zero.
The basis $\mathcal B$ can be properly normalized to ensure that $\det \bar I_0 = \det I_0 = (q-q_c)^{\left( \begin{smallmatrix} N \\ k \end{smallmatrix} \right)} A(q)$, with $A(q_c) \neq 0$. A factor of $(q-q_c)^1$ can be factored from each of the first $\left( \begin{smallmatrix} N \\ k \end{smallmatrix} \right)$ vectors above and the matrix obtained by removing these factors is regular. The set $\{ c_i^1, i = 1, \dots, \left( \begin{smallmatrix} N \\ k \end{smallmatrix} \right)\} \cup \{ c_i^0, i = \left( \begin{smallmatrix} N \\ k \end{smallmatrix} \right) + 1, \dots, \left( \begin{smallmatrix} N \\ P \end{smallmatrix} \right)\}$ is then a basis of $W_{d/2}$.

Let us denote by $w_j$ the columns of the matrix $(\bar I_N^d)^{-1}$. They can be expanded in Laurent series as 
$$ w_j = \sum_{m = m_j}^{\infty} (q-q_c)^m d_j^m,$$
and because $I_N^d$ is singular at $q = q_c$, some of the $m_j$s must be negative. Then $\iota = -\min \{m_j\}$. Because $\bar I_N^d (\bar I_N^d)^{-1} = id$, 
$$ u_iw_j  = \delta_{i,j} = \left\{\begin{array}{ll} c_i^1d_j^{m_j}  (q-q_c)^{m_j + 1} + \mathcal O (q-q_c)^{m_j + 2}, & \quad i = 1, \dots,  \left( \begin{smallmatrix} N \\ k \end{smallmatrix} \right), \vspace{0.1cm}\\ 
c_i^0d_j^{m_j}  (q-q_c)^{m_j} + \mathcal O (q-q_c)^{m_j + 1}, & \quad  i =  \left( \begin{smallmatrix} N \\ k \end{smallmatrix} \right) +1,  \dots,  \left( \begin{smallmatrix} N \\ P \end{smallmatrix} \right).
\end{array} \right.$$
For a given $j$, suppose that $m_j < -1$. On the right-hand side, the coefficients of all the terms with negative powers of $q-q_c$ must vanish. But if $d_j^{m_j} \neq 0$, $ c_i^1 (d_j^{m_j})$ and $c_i^0 (d_j^{m_j})$ cannot all vanish because the $c_i^0$s and $c_i^1$s form together a basis of $W_{d/2}$. Hence $m_j$ must be $\ge -1$. Because at least one must be negative, $\iota = 1$.

Going back to the original basis, we now know that 
$$\det \left((I_N^d)^{-1}\right)  = \det \left(\frac{M_0}{q-q_c} + M_1 + \dots \right) = B(q) (q-q_c)^{-\left( \begin{smallmatrix} N \\ k \end{smallmatrix} \right)}$$
with $B(q_c) \neq 0$. 
Suppose that this determinant is computed in the basis where $M_0$ is in its Jordan form $\bar M_0$. Then each factor $(q-q_c)^{-1}$ can come only from a non-zero row of $\bar M_0$. There must be at least $\left(\begin{smallmatrix}N\\ k\end{smallmatrix}\right)$ of them and $\dim  \im \bar M_0=\dim \im M_0\ge \left(\begin{smallmatrix}N\\ k\end{smallmatrix}\right)$, and thus $\dim\ker M_0\le \left(\begin{smallmatrix}N\\ P\end{smallmatrix}\right)-\left(\begin{smallmatrix}N\\ k\end{smallmatrix}\right)$. \hfill $\square$

%
\subsection{The construction of a Jordan cell}\label{sec:jordan}
%

This last paragraph constructs explicitly a Jordan cell of rank $2$ in the generalized eigenspace of $\mathcal H$ associated to the zero eigenvalue. Again the hypotheses \ref{hyp:hypHourra} are assumed. Both the matrices appearing in the expansion of $I_N^d(q)$ and $(I_N^d)^{-1}(q)$ and the vector $| \chi \rangle =x_1 | \nu_1 \rangle - x_2 | \nu_2 \rangle$ play an important role in the proof of theorem \ref{thm:intra} below. The constant $x_1$ and $x_2$ are 
\begin{equation} x_1 = \langle 0| S^{+(P)}|\nu_2\rangle =  2q_c^{P^2 + Pk}\begin{pmatrix} N \\ 2P+k\end{pmatrix}, \qquad x_2 = \langle 0| S^{+(P)}|\nu_1\rangle = q_c^{Pk} \begin{pmatrix} N \\ P+k\end{pmatrix} \label{eq:valeurx1x2}\end{equation}
and the last equalities are valid only at $q = q_c$. These constants follow from the following expressions:
\begin{equation*}
S^{-(x)}|0 \rangle = (qv^{-2})^{(N+1)x/2} \sum_{1 \le j_1 < j_2 < \dots < j_x \le N} (qv^{-2})^{-\sum_{k=1}^x j_k} \sigma_{j_1}^-\dots \sigma_{j_x}^- | 0 \rangle,
\end{equation*}
\begin{equation}
T^{+(x)} \sigma_{j_1}^-\dots \sigma_{j_x}^- | 0 \rangle = (qv^{-2})^{-(N+1)x/2+\sum_{k=1}^x j_k} |0 \rangle. \label{eq:plustot}
\end{equation}
With these results, we find $\langle 0 |T^{+(x)}S^{-(x)} |0\rangle = \langle 0 |S^{+(x)}T^{-(x)} |0\rangle = \left( \begin{smallmatrix}N \\ x\end{smallmatrix} \right)$ 
and
\begin{equation}x_1 =
\langle 0| S^{+(P)}|\nu_2\rangle
= \left[ \begin{matrix} 2P+k \\ P \end{matrix}\right] \begin{pmatrix} N \\ 2P+k\end{pmatrix}, \qquad  \quad x_2 =
\langle 0| S^{+(P)}|\nu_1\rangle
= \left[ \begin{matrix} P+k \\ P \end{matrix}\right] \begin{pmatrix} N \\ P+k\end{pmatrix}.\label{eq:x1x2}\end{equation}
The $q$-binomials in equation \eqref{eq:x1x2} are computed using the relation
\begin{equation}
\lim_{q \rightarrow q_c}\left[\begin{matrix} Ps + a \\ P \end{matrix} \right] = sq_c^{Pa} \times \left\{
\begin{array}{ll} 1 & \quad \textrm{for} \, s \, \textrm{odd}, \\ q_c^{P^2} & \quad \textrm{for} \, s \, \textrm{even}.\end{array}\right.
\label{eq:xt2}\end{equation}

Under hypotheses \ref{hyp:hypHourra}, the state $| \chi \rangle$ has two nice properties. The first is that it is non-zero. This could be shown directly, 
but we will prefer a more roundabout way to be given later on. The second property is that $M_0 |\chi\rangle = 0$. Indeed, because $\im M_0^T = \textrm{left} \ker I_0$ and $\textrm{left} \ker I_0 = \textrm{span}\{\langle \mu|, \langle \mu_{j_1, \dots, j_k} |\}$, $M_0 | \chi \rangle = 0$ if and only if $\langle \mu | \chi \rangle = 0$ and $\langle \mu_{j_1, \dots, j_k} | \chi \rangle = 0$, and both equalities hold at $q = q_c$ from the definition of $|\chi \rangle$.

The vector $|\chi\rangle$ is crucial as it leads to a Jordan cell of size $2$ in $\mathcal H$ in the generalized eigenspace associated with the eigenvalue $0$. Let $\mathcal H_0=\mathcal H(q=q_c)$, $\textrm{spec }\mathcal H_0$ the spectrum of $\mathcal H_0$ and $\tilde V_N^d=\oplus_{\lambda\in\textrm{spec }\mathcal H_0}V_\lambda$ the decomposition into its generalized eigenspaces $V_\lambda$. Similarly, since $\textrm{spec }\mathcal H_0=\textrm{spec }H_0$, a similar decomposition $W_{d/2}=\oplus_{\lambda\,\in\,\textrm{spec}\,\mathcal H_0}W_\lambda$ also exists. Consider the state $M_1 |\chi \rangle \in \tilde V_N^d$ and decompose it as
$$M_1 | \chi \rangle = v_0 + \sum_{\lambda\,\in\,\textrm{spec }\mathcal H_0} v_\lambda$$
where $v_\lambda \in V_\lambda$ and where the sum omits the eigenvalue $\lambda=0$. To prove that there is a rank $2$ Jordan block, we show that $\mathcal H_0v_0 \neq 0$. This will also prove that $v_0 \neq 0$ and $|\chi \rangle \neq 0$.

The question of whether $\mathcal H_0v_0$ is non-zero can be reformulated by the use of the identities \eqref{eq:h0m0inter}:
$$\mathcal H_0(M_1|\chi\rangle)=(\mathcal H_0M_1+\mathcal H_1 M_0)|\chi \rangle = 
(M_1H_0+M_0H_1)|\chi\rangle=M_0H_1|\chi\rangle$$
where the fact that $|\chi\rangle\in\ker M_0$ was used for the first equality and that $|\chi\rangle$ is an eigenvector of $H_0$ with zero eigenvalue for the last. Therefore the component $\mathcal H_0v_0$ in $\mathcal H_0(M_1|\chi\rangle)=\mathcal H_0v_0+\sum' \mathcal H_0v_\lambda$ will be non-zero if $(M_0H_1|\chi\rangle)_{\lambda=0}$ is non-zero. (The subscript ``$\lambda=0$'' stands for the component of the vector in the subspace $V_{\lambda=0}$.)

Let $\{\beta_i\}$ be an orthonormal basis of $W_{d/2}$ made of eigenvectors of the hermitian $H_0$. (The inner product is the usual one.) The orthonormality of the basis implies that $\sum_i |\beta_i\rangle\langle \beta_i|$ is the identity. Moreover, since $M_0W_\lambda\subset V_\lambda$, the component of $M_0H_1|\chi\rangle$ in $V_{\lambda = 0}$ can only come from the action of $M_0$ on $W_{\lambda = 0}$ and 
\begin{equation}\label{eq:h0v0}
(M_0H_1|\chi\rangle)_{\lambda = 0}=\Big(\sum_i M_0|\beta_i\rangle\langle \beta_i|H_1|\chi\rangle\Big)_{\lambda = 0}=
\sum_{|\beta_i\rangle\in W_{\lambda = 0}}M_0|\beta_i\rangle\langle \beta_i|H_1|\chi\rangle.
\end{equation}
This same property ($M_0W_\lambda\subset V_\lambda$) also implies that the projection of any element of $\ker M_0$ on any of the eigenspaces $W_\lambda$ is itself an element of the kernel. The (orthonormal) basis of $W_{\lambda = 0}$ can then be chosen so that it splits into elements in $\ker M_0$ and elements in its orthogonal complement $(\ker M_0)^\perp$. Let $\mathcal B_0^\perp$ denote the latter subset. Then the last sum of \eqref{eq:h0v0} can be restricted to the elements of $\mathcal B_0^\perp$. Since $\dim W_{\lambda = 0}=
\dim \ker (M_0|_{W_{\lambda = 0}}) + \dim \im(M_0|_{W_{\lambda = 0}})$, the set $\{M_0|\beta_i\rangle\}_{|\beta_i\rangle\in\mathcal B_0^\perp}$ must be a basis of $M_0W_{\lambda = 0}$. Therefore $\mathcal H_0v_0$ is non-zero if and only if any of the components $\langle \beta_i|H_1|\chi\rangle$ of $(M_0H_1|\chi\rangle)_{\lambda = 0}$, with $|\beta_i\rangle\in\mathcal B_0^\perp$, is non-zero. An explicit basis $\mathcal B_0^\perp$ might be hard to compute. However $\ker M_0=\im I_0$ and $(\ker M_0)^\perp$ is therefore the left nullspace of $I_N^d$ characterized in section \ref{sec:kern}. Corollary \ref{sec:coro2} tells us that the state $\langle \nu_3|$ belongs to the left kernel of $I_N^d$. 
(Of course $|\nu_3\rangle$ is in $W_{\lambda=0}$, as observed at the beginning of paragraph \ref{sec:imker}, and can be chosen, after normalization, as one of the elements of $\mathcal B_0^\perp$.)
The final step is therefore to compute the matrix element  $\langle \nu_3 | H_1 |\chi \rangle$. This is done in appendix \ref{app:a}, and the result is
\begin{equation*} \langle \nu_3 | H_1 |\chi \rangle = - q_c^{P^2 + Pk -1} (q_c - q_c^{-1}) \frac{2P^2(N-2k)}{N(N-1)} \begin{pmatrix} N \\  k \end{pmatrix}\begin{pmatrix} N \\  P+k \end{pmatrix}\begin{pmatrix} N \\  2P+k \end{pmatrix},\end{equation*}
which is non-zero. The vector 
$(M_1|\chi\rangle)_{\lambda = 0}$ is thus a Jordan partner in a Jordan cell of size $2$ associated to the zero eigenvalue of $\mathcal H$. Using lemma 4.1 and proposition 4.10 of \cite{AMDSA}, the result can also be extended to the transfer matrix $T_N(\lambda,\nu)$ in the representation $\omega_d$.

\begin{Theoreme}\label{thm:intra} Hypotheses \ref{hyp:hypHourra} are assumed. Let $M_1$ be the second term in the expansion $(I_N^d)^{-1}(q,v_c)=(q-q_c)^{-1}M_0+M_1+\dots$ and 
$$|\chi\rangle=x_1|\nu_1\rangle-x_2|\nu_2\rangle \qquad
\textrm{with} \quad |\nu_1 \rangle = S^{+(k)}T^{-(P+k)}|0 \rangle, \qquad
|\nu_2 \rangle = S^{+(P+k)}T^{-(2P+k)}|0 \rangle$$
and $x_1 = \langle 0| S^{+(P)}|\nu_2\rangle$ and  $x_2 = \langle 0| S^{+(P)}|\nu_1\rangle$.
Then the component $v_0$ of $M_1|\chi\rangle$ lying in the generalized eigenspace $V_{\lambda=0}$ of $\mathcal H_0=\mathcal H(q=q_c)$ forms, with $\mathcal H_0v_0$, a non-trivial Jordan cell in the loop Hamiltonian $\mathcal H_0$. 
Under the same hypotheses the transfer 
matrices $\omega_d(T_N(\lambda,\nu))$, with $q=-e^{i\lambda}$, have Jordan cells for all but a finite set of values of $\nu$. 
\end{Theoreme}

It is useful to stress that no restrictions have been put on the parity of $N$. The simplest Jordan blocks obtained in this section appear for $P=2$. 
The (generalized) eigenspaces of $\mathcal H$ lie in the sector with $S^z=0$ for $N=4$ and with $S^z=\frac12$ for $N=5$. We close this section by giving explicitly the case of $(N,d) = (4,0)$ at $(q_c,v_c)=(i,1)$, which corresponds to $\beta=0$ and $\alpha=2$. These values satisfy hypotheses \ref{hyp:hypHourra} for $N=4$ and $d=0$ and were used in paragraph \ref{sec:observations} to provide examples of the expansion of $I_N^d$ and its inverse. The Hamiltonian $\mathcal H$ and its Jordan decomposition are
$$
\omega_0(\mathcal H)=\left(
\begin{matrix}
 2 \beta  & 2 & \alpha  & 0 & \alpha  & 2 \\
 1 & \beta  & 0 & 0 & 0 & 0 \\
 0 & 0 & \beta  & 1 & 0 & 0 \\
 0 & \alpha  & 2 & 2 \beta  & 2 & \alpha  \\
 0 & 0 & 0 & 1 & \beta  & 0 \\
 1 & 0 & 0 & 0 & 0 & \beta 
\end{matrix}
\right)\quad\textrm{and}\quad
\textrm{Jordan form of }\displaystyle{\omega_0 (\mathcal H)\Big|_{\substack{\beta = 0\\ \alpha=2}}}=
\left(
\begin{matrix}
  0 & 0 & 0 & 0 & 0 & 0 \\
 0 & 0 & 0 & 0 & 0 & 0 \\
 0 & 0 & 0 & 1 & 0 & 0 \\
 0 & 0 & 0 & 0 & 0 & 0 \\
 0 & 0 & 0 & 0 & -2 \sqrt{2} & 0 \\
 0 & 0 & 0 & 0 & 0 & 2 \sqrt{2} \\
\end{matrix}
\right).
$$

\begin{figure}[h!]
\begin{center}
\begin{pspicture}(0,-1)(12,11.5)

\rput(1,11){(1,1)}
\rput(0,10){(2,0)}
\rput(2,10){(2,2)}
\rput(1,9){(3,1)}
\rput(3,9){(3,3)}
\rput(0,8){\fbox{(4,0)}}
\rput(2,8){(4,2)}
\rput(4,8){(4,4)}
\rput(1,7){\fbox{(5,1)}}
\rput(3,7){(5,3)}
\rput(5,7){(5,5)}
\rput(0,6){\dashbox{0.05}(1.0,0.585)[cc]{(6,0)}}
\rput(2,6){\fbox{(6,2)}}
\rput(4,6){(6,4)}
\rput(6,6){(6,6)}
\rput(1,5){\dashbox{0.05}(1.0,0.585)[cc]{(7,1)}}
\rput(3,5){\fbox{(7,3)}}
\rput(5,5){(7,5)}
\rput(7,5){(7,7)}
\rput(0,4){\dashbox{0.05}(1.2,0.7)[cc]{\dashbox{0.05}(1.0,0.5)[cc]{(8,0)}}}
\rput(2,4){\dashbox{0.05}(1.0,0.585)[cc]{(8,2)}}
\rput(4,4){\fbox{(8,4)}}
\rput(6,4){(8,6)}
\rput(8,4){(8,8)}
\rput(1,3){\dashbox{0.05}(1.2,0.7)[cc]{\dashbox{0.05}(1.0,0.5)[cc]{(9,1)}}}
\rput(3,3){\dashbox{0.05}(1.0,0.585)[cc]{(9,3)}}
\rput(5,3){\fbox{(9,5)}}
\rput(7,3){(9,7)}
\rput(9,3){(9,9)}
\rput(0,2){\dashbox{0.05}(1.3,0.7)[cc]{\dashbox{0.05}(1.1,0.5)[cc]{(10,0)}}}
\rput(2,2){\dashbox{0.05}(1.3,0.7)[cc]{\dashbox{0.05}(1.1,0.5)[cc]{(10,2)}}}
\rput(4,2){\dashbox{0.05}(1.1,0.585)[cc]{(10,4)}}
\rput(6,2){\fbox{(10,6)}}
\rput(8,2){(10,8)}
\rput(10,2){(10,10)}
\rput(1,1){\dashbox{0.05}(1.3,0.7)[cc]{\dashbox{0.05}(1.1,0.5)[cc]{(11,1)}}}
\rput(3,1){\dashbox{0.05}(1.3,0.7)[cc]{\dashbox{0.05}(1.1,0.5)[cc]{(11,3)}}}
\rput(5,1){\dashbox{0.05}(1.1,0.585)[cc]{(11,5)}}
\rput(7,1){\fbox{(11,7)}}
\rput(9,1){(11,9)}
\rput(11,1){(11,11)}
\rput(0,0){\dashbox{0.05}(1.6,0.9)[cc]{\dashbox{0.05}(1.4,0.7)[cc]{\dashbox{0.05}(1.2,0.5)[cc]{(12,0)}}}}
\rput(2,0){\dashbox{0.05}(1.4,0.7)[cc]{\dashbox{0.05}(1.2,0.5)[cc]{(12,2)}}}
\rput(4,0){\dashbox{0.05}(1.4,0.7)[cc]{\dashbox{0.05}(1.2,0.5)[cc]{(12,4)}}}
\rput(6,0){\dashbox{0.05}(1.2,0.585)[cc]{(12,6)}}
\rput(8,0){\fbox{(12,8)}}
\rput(10,0){(12,10)}
\rput(12,0){(12,12)}
\psline{-}(0,-0.35)(0,-0.6)(8,-0.6)(8,-0.29)
\psline{-}(4,-0.35)(4,-0.45)(7.9,-0.45)
\psline{-}(8.1,-0.45)(12,-0.45)(12,-0.3)
\psline{-}(2,0.35)(2,0.45)(6,0.45)(6,0.29)
\psline{-}(6,0.45)(10,0.45)(10,0.3)
\psline{-}(0,1.65)(0,1.55)(8,1.55)(8,1.7)
\psline{-}(2,2.35)(2,2.45)(6,2.45)(6,2.29)
\psline{-}(6,2.45)(10,2.45)(10,2.3)
\psline{-}(0,3.65)(0,3.55)(8,3.55)(8,3.7)
\psline{-}(2,4.29)(2,4.45)(6,4.45)(6,4.3)
\psline{-}(2,6.29)(2,6.45)(6,6.45)(6,6.3)
\rput(0,-0.85){$\vdots$}
\rput(4,-0.85){$\vdots$}
\rput(8,-0.85){$\vdots$}
\rput(12,-0.85){$\vdots$}

\end{pspicture}
\end{center}
\caption{The Bratelli diagram for $ q = e^{i \pi /2}$. Jordan cells in $\omega_d$ occur for the pairs $(N,d)$ that are contained in a box. 
The construction of section \ref{sec:BJintra} is for boxes with solid segments, while boxes with dashed lines are for Jordan blocks found by our 
computer exploration. The number of multiple dashed boxes indicates the maximum number of Jordan partners in the largest Jordan cell of the given $\tilde V_N^d$. 
For example, the first occurrence of rank $4$ Jordan blocks within a sector appears for $(N,d) = (12,0)$. Jordan cells between sectors $d$ and $d'$ in the representation $\rho$ are indicated by solid segments connecting pairs $(N,d)$. For these, $\alpha = 2$ is assumed for $N$ even.}\label{fig:Bratelli} 
\end{figure}
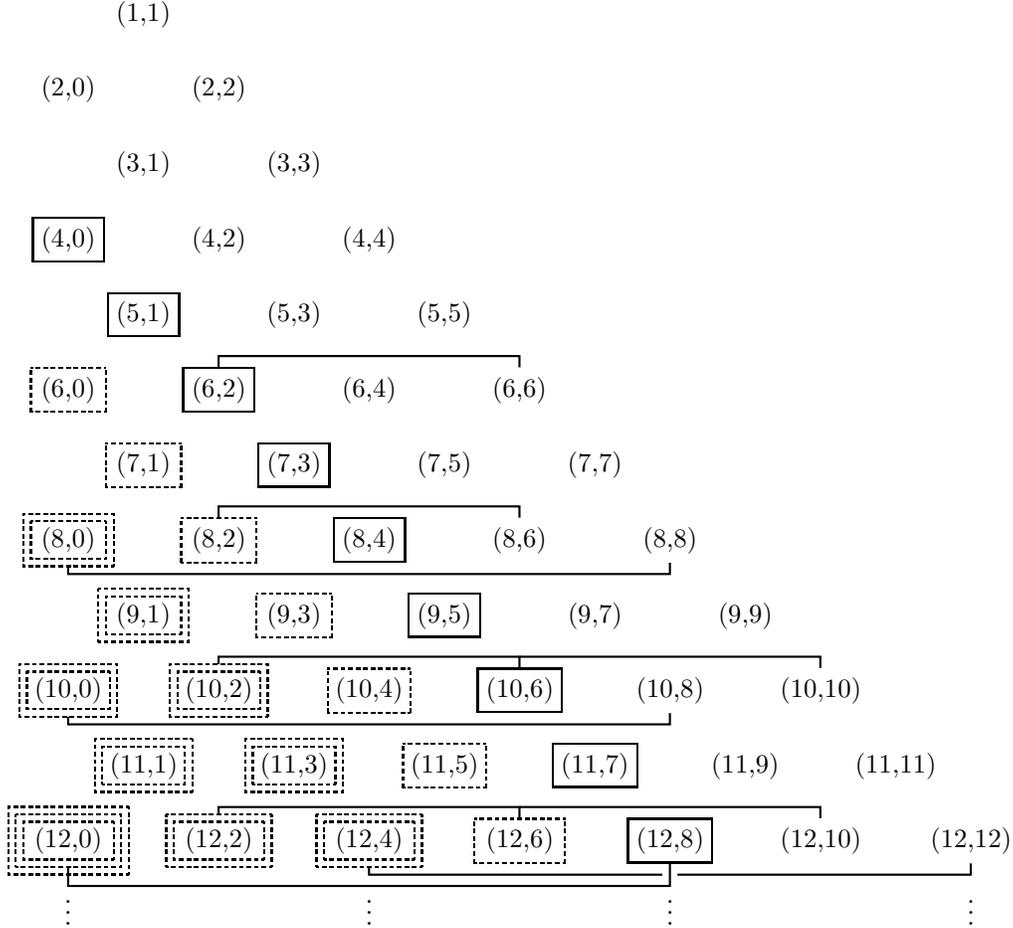

%
\section{Concluding remarks}
%

The Jordan structure of the transfer matrix $T_N$ for loop models on the cylinder was investigated for the two representations $\rho$ and $\omega_{d,v}$ defined in paragraph \ref{sec:EPTL}. A complete answer for the Jordan blocks between sectors of $\rho(F_N)$ and $\rho(T_N)$ is given in corollary \ref{coro:leCoroFN} and theorem \ref{thm:inter} and the various subcases are reminiscent of the characterization of Jordan blocks for the double-row transfer matrix arising for loop models on the strip as recalled in theorem \ref{thm:interstrip}. The structure of $\omega_d(\mathcal H)$ is partially revealed in theorem \ref{thm:intra}. The result 
relies on {\em (i)} the freedom introduced by the twist parameter 
of the representation $\omega_{d,v}$ in the condition $(qv^2)^N=1$ of lemmas \ref{sec:firstcomm} and \ref{sec:firstcomm2},
{\em (ii)} the existence in the representations $U_q(sl_2)\rightarrow \textrm{End }\cn$ of maps that commute with $H$ (paragraph \ref{sec:uqsl2}), 
and {\em (iii)} the intertwiner $\tilde i_N^d$ and its characterization (paragraph \ref{sec:linkAndSpin}), not only at the critical point $(q_c,v_c)$, but in a neighborhood of it (paragraph \ref{sec:observations}). 

The techniques used for the two representations $\rho$ and $\omega_d$ are quite different. But both have at their core a homomorphism that changes its algebraic nature at the critical point. For $\rho$, this homomorphism is the left multiplication by the central element $F_N$, which stops being diagonalizable at the critical $q_c$. For $\omega_d$ it is the intertwiner $\tilde i_N^d$ that becomes singular at $(q_c,v_c)$. These two techniques are obviously non-trivial and, to our knowledge, new. The first settles completely the existence of Jordan blocks, the second constructs them for an infinite family of models by concentrating on a single eigenvalue. Most importantly, both find Jordan blocks that, given $\Lambda$, exist for an infinite subsequence of $N$s. 
This stability of Jordan blocks is remarkable. From the seminal work of Pasquier and Saleur \cite{PasquierSaleur} to more recent ones where fusion of modules over the Temperley-Lieb algebra is related to that of modules over the Virasoro algebra \cite{GainutVasseur}, there are several indications that the continuum limit of a large family of lattice models, including those studied here, is described by a conformal field theory and that, in some sense, the modules over the first algebra carry in some limit a representation of the second one. (See \cite{gainutdinov1,gainutdinov2} for significant progress in this direction.) In this correspondence the Hamiltonian (of the loop or the XXZ models) should go, after proper rescaling, to the mode $L_0$ of the Virasoro algebra. Therefore, stable Jordan cells in $\mathcal H$ would translate into Jordan cells in $L_0$ and the limiting representations would be staggered modules over the Virasoro algebra. These modules are one of the signatures of logarithmic conformal field theories. 
One more similarity between the two techniques is worth underlining: They both rely on lenghty computations. But maybe these technicalities were to be expected due to the unusual nature of the question within representation theory, namely that of the Jordan structure of a single element of the algebra instead of the structure of the whole representation. Notwithstanding these, the techniques are promising and open a whole new set of questions.

In the case of the loop transfer matrix on a strip we limited our analysis to one  boundary condition, often referred to as the {\em open} boundary condition. What happens for other boundary conditions? Or, more generally, what happens when the boundary conditions are allowed to change along the strip? Algebraic tools, the so-called blob algebra and its representations, were introduced to probe this more difficult case a long time ago \cite{blobAlgebra}. (This tool is also called the Temperley-Lieb algebra of type $B$ \cite{Lehrer2}.) A recent work \cite{GJSV} using this larger algebra has identified modules of physical relevance that have Jordan cells of rank larger than $2$. The authors therefore conclude that more complicated Virasoro modules than those envisaged up to now are to be expected in the continuum limit. 

Figure \ref{fig:Bratelli} illustrates what has been accomplished and what is still open in the periodic case. It shows, for the value $P=2$ ($q=e^{i\pi/2}$, $\beta=0$), the sectors $\tilde V_N^d$ denoted by $(N,d)$ (with action $\omega_{d,v}$) where our rigorous arguments (solid boxes) and our computer explorations (dashed ones) found non-trivial cells. The construction of Jordan partners of theorem \ref{thm:intra} works on the diagonal $N-d = 4$ and under the hypotheses \ref{hyp:hypHourra}. These pairs $(N,d)$ appear in solid boxes, and our numerical data shows that they contain a unique Jordan block, the one constructed in section \ref{sec:jordan}. On lower diagonals, 
Jordan blocks associated with other eigenvalues of $\rho(\mathcal H)$ are found for $(qv^2)^N = q^{2k}$, $k = 0,1$. On the diagonals $N-d \ge 8$, we find Jordan cells of rank larger than $2$ and, for $(N,d) = (12,0)$, a Jordan cell with rank $4$ appears for the first time. The figure also shows the Jordan cells tying different sectors in the representation $\rho$, studied in section \ref{sec:BJentre}.

Improvements on the techniques introduced in section \ref{sec:BJintra} will be needed in order to explore this figure deeper. Proving the existence of Jordan cells with $\lambda \neq 0$ or with rank larger than $2$, found by our explorations, appears more involved. A first reason is that many eigenvalues of $\omega_d(\mathcal H)$ are degenerate and the identification of a vector similar to $|\chi\rangle$ used in paragraph \ref{sec:jordan} might be delicate. This difficulty cannot be easily addressed by representation-theoretic methods, or even by others like the Bethe ansatz.
A second reason is that more than one term in the factorization \eqref{eq:detINd} of the determinant of $I_N^d$ can be zero. Our explorations show that the exponent $\iota$ in \eqref{eq:defM0} is then greater than $1$, and our 
understanding of $M_0$ and $M_1$ remains incomplete. Nevertheless, our computer explorations indicate that, for $q$ and $v$ on the unit circle, any Jordan blocks, even in sectors $d < N-P$, appear only if hypotheses 4.1 (i) and (ii) are satisfied. The improvements just suggested aim at extending the picture uncovered in section \ref{sec:BJintra}. There are also questions about what these structures appearing on lattices become in the continuum limit. Do they match those obtained when the properties of the operator product expansion are analysed? Do the maps defined by $M_0$ and $M_1$ have a limit? Are these physically relevant? Even such simple questions are intriguing.
Clearly more work needs to be done before a global picture emerges.

\section*{Acknowledgements} 

We would like to thank J\o{}rgen Rasmusen and David Ridout for helpful discussions and encouragement. AMD holds a postdoctoral fellowship and YSA a grant of the Canadian Natural Sciences and Engineering Research Council. This support is gratefully acknowledged. 

\bigskip

\bigskip

%
%
\newpage
\noindent{\LARGE\bfseries Appendices}

%
%

\appendix

\section{States in the left kernel of $\boldsymbol{I_N^d}$}\label{app:0}

The corollaries \ref{sec:coro} and \ref{sec:coro2} of section \ref{sec:kern} give some states in the left-kernel of $I_N^d$. This paragraph provides the details leading to these corollaries. The following lemma is devoted to the computation of $S^{+(x)}\tilde i_N^d(w_{N^y}^N)$.

\begin{Proposition} \label{sec:ouach} For $y = (N-d)/2$ and $1\le x\le y\le N/2$,
\begin{equation}
S^{+(x)} \tilde i_N^d(w_{N^y}^N) = i^y q^{(y-x)/2} \frac{\langle d/2 +x \rangle!}{\langle d/2\rangle !} \displaystyle{\sum_{\substack{1 \le j_1 < j_2 < \dots < j_x \le y \\ J \equiv \{j_1, j_2, \dots, j_x\}}}} \Big( \displaystyle{\prod_{\substack{1\le r\le y\\ r\notin J}} G_{r,s(r,J)}} \Big) |0 \rangle
\label{eq:bigspformula}\end{equation}
where 
\begin{itemize}
\item  $\langle x \rangle! = \prod_{k=1}^x \langle k\rangle$ if $x>0$ and $\langle 0\rangle!=1$ (recall that $\langle x \rangle = v^N q^x - v^{-N}q^{-x}$);
\item $G_{j,k} = q^{-k} v^{2j-1} \sigma_j^- - q^{k-1} v^{-(2j-1)} \sigma_{N+1-j}^-$;
\item $s(r, J) = - |\{ j \in J | j>r  \}|$.
\end{itemize}
\end{Proposition}
\noindent{\scshape Proof\ \ } The proof is by induction. We start by noting that $\tilde i_N^d(w_{N^y}^N) = (\prod_{j=1}^y \tilde T_{N+1-j, N+j}) |0 \rangle \allowbreak=\allowbreak (i q^{\frac12})^y \times \allowbreak (\prod_{j=1}^y G_{j,0}) |0 \rangle$. If $x = 0$, $J$ is the empty set and the sum in equation \eqref{eq:bigspformula} is replaced by one, which is the correct answer. We now assume the result for $x$. 
Let $F_{i,j} = q^{-j} S^+_i + q^j S^+_{N+1-i}$. We can then write
\begin{equation*} S^{+(x+1)} \tilde i_N^d(w_{N^y}^N)  =  \frac{1}{[x+1]} \Big( \sum_{j=1}^y F_{j,0} \Big) S^{+(x)} \tilde i_N^d(w_{N^y}^N).
\end{equation*}
One can compute the following multiplication rules:
\begin{align} F_{i,j} G_{k,l} &=
\begin{cases}
G_{k,l} F_{i, j+1}, & i<k, \\ 
G_{k,l-1} F_{i, j}, & i>k, 
\end{cases} \label{eq:multrules}
\\ \qquad F_{i,j}G_{i,k} |0 \rangle &=  |0 \rangle ~q^{-1/2} \langle N/2 + 1 - j - k - i\rangle.
\label{eq:multrules2}
\end{align}
Then, 
\begin{align*}S^{+(x+1)} \tilde i_N^d(w_{N^y}^N)  &= \frac{i^y q^{(y-x)/2}}{[x+1]} \frac{\langle d/2 +x \rangle!}{\langle d/2 \rangle !} \sum_{j = 1}^y  \displaystyle{\sum_{\substack{1 \le j_1 < j_2 < \dots < j_x \le y \\ J \equiv \{j_1, j_2, \dots, j_x\}}}} \Big( \displaystyle{F_{j,0}\prod_{\substack{1\le r\le y\\ r\notin J}} G_{r,s(r, J)}} \Big) | 0 \rangle \\
&= \frac{i^y q^{(y-x)/2}}{[x+1]} \frac{\langle d/2 +x\rangle!}{\langle d/2\rangle !} \sum_{t = 1}^{x+1}  \displaystyle{\sum_{\substack{1 \le j_1 < \dots  < j_{t-1} < j < j_{t} < \dots  < j_x \le y \\ J \equiv \{j_1, j_2, \dots, j_x\}}}} \Big( \displaystyle{F_{j,0}\prod_{\substack{1\le r\le y\\ r\notin J}} G_{r,s(r, J)}} \Big) | 0 \rangle, 
\end{align*}
where in the last line it is implicit that $j_0 = 0$ and $j_{x+1} = y+1$. On the first line, if $j$ equals $j_i$ for some $i = 1, \dots, x$, then
$F_{j,0}$ acts directly on $| 0 \rangle$ and the result is zero. We now rename the variables
\begin{equation*}
k_i = \left\{ \begin{array}{l l} j_i, & i<t, \\ j, & i=t, \\ j_{i-1}, & i>t, \end{array}\right.
\end{equation*}
and get
\begin{equation*}S^{+(x+1)} \tilde i_N^d(w_{N^y}^N) = \frac{i^y q^{(y-x)/2}}{[x+1]} \frac{\langle d/2 +x\rangle!}{\langle d/2 \rangle !}  \displaystyle{\sum_{\substack{1 \le k_1 < k_2 < \dots < k_{x+1} \le y \\ K \equiv \{k_1, \dots, k_{x+1}\}}}} \sum_{t = 1}^{x+1} \Big( \displaystyle{F_{k_t, 0}\prod_{\substack{1\le r\le y \\ r \notin K \setminus \{k_t\}}}}  G_{r, s(r,J)}\Big) |0 \rangle.
\end{equation*} 
Using the multiplication rules (\ref{eq:multrules}), we simplify the summand:
\begin{align*}
\Big(F_{k_t, 0} \displaystyle{\prod_{\substack{1\le r\le y \\ r \notin K \setminus \{k_t\}}}}  G_{r, s(r,J)}\Big) |0 \rangle &=  \Big( \displaystyle{\prod_{\substack{1\le r\le k_t-1 \\ r \notin K }}} G_{r, s(r,J)-1}\Big)  \Big( \displaystyle{\prod_{\substack{k_t+1\le r\le y \\ r \notin K }}} G_{r, s(r,J)}\Big) F_{k_t, y-k_t-(x+1-t)} G_{k_t, s(k_t, J)}|0 \rangle \\
& = q^{-1/2} \langle N/2 +1 - y + 2 (x + 1 - t) \rangle  \Big( \displaystyle{\prod_{\substack{1\le r\le y \\ r \notin K }}} G_{r, s(r,K)} \Big) |0 \rangle
\end{align*} 
because $s(k_t, J) = -(x+1 -t)$. Finally,
\begin{equation*}S^{+(x+1)} \tilde i_N^d(w_{N^y}^N) = \frac{i^y q^{(y-x-1)/2}}{[x+1]} \frac{\langle d/2 +x \rangle!}{\langle d/2\rangle !}  \displaystyle{\sum_{\substack{1 \le k_1 < k_2 < \dots  < k_{x+1} \le y \\ K \equiv \{k_1, \dots, k_{x+1}\}}}} \Big( \displaystyle{\prod_{\substack{1\le r\le y \\ r \notin K }}} G_{r, s(r,K)} \Big) |0 \rangle \times \sum_{t=1}^{x+1} \langle N/2 +1 - y + 2 (x + 1 - t) \rangle
\end{equation*}
which yields the correct result when we use the identity $\sum_{t=1}^{x+1} \langle A + (x+2-2t) \rangle = [x+1] \langle A \rangle$ with $A = N/2 + x - y + 1 = d/2 + x + 1$. This completes the induction.
\hfill$\square$ 

\medskip

Note that in equation \eqref{eq:bigspformula}, the sum over subsets $J\subset \{1,2,\dots, y\}$ of $\prod_{1\le r\le y, r\notin J} G_{r,s(r,J)}  |0 \rangle$  gives a non-zero vector. 
Indeed $S^{+(x)}\tilde i_N^d(w_{N^y}^N)$ is in the subspace with $S^z=N/2-(y-x)$ 
and, because $G_{j,k}$ changes the spin states at positions $j\le y$ or $j\ge N+1-y$ only, the only contribution to 
the coefficient of the state $|--\dots -++\dots+\rangle$ starting with precisely $(y-x)$ minus signs comes from the subset $J=\{y-x+1, y-x+2, \dots, y\}$ and is non-zero. 
Corollary \ref{sec:coro} follows. 

The next result characterizes vectors in the left nullspace of $I_N^d$ for pairs $(q,v)$ satisfying some of the hypotheses of lemma \ref{sec:firstcomm2}.

\begin{Proposition}
Let $q$ and $v$ be such that $q^{N-2k}v^{2N}=1$. Then $S^{-( N-p-k)}\tilde i_N^d(w_{N^p}^N) = 0$ for every integer $p$ such that $0\le k< p\le N/2$.
\label{sec:nu3}
\end{Proposition}
\noindent{\scshape Proof\ \ }  
Using (\ref{eq:OSxO}) with $q^{N-2k}v^{2N}-1 = 0$, we have
\begin{equation*}
S^{-(N-p-k)}\tilde i_N^d(w_{N^p}^N) = \bar \Omega q^{(N-p-k)\sigma_1^z} S^{-(N-p-k)} \tilde i_N^d(w_{1^p}^N) = \dots =  \bar \Omega^p q^{(N-p-k) \sum_{j=1}^p\sigma_j^z} S^{-(N-p-k)} \tilde i_N^d(w_{p^p}^N)
\end{equation*}
where, we recall, $w_{j^p}^N$ denotes the link state with $p$ concentric half-arcs centered at $j$.
We now express the generator $S^-$ as $S^-_A + S^-_B$ where $S^\pm_A = \sum_{j=1}^{2p} S^\pm_j$ and $S^\pm_B = \sum_{j=2p+1}^{N} S^\pm_j$. With the commutation relation $S_A^\pm S_B^\pm=q^{\pm 2}S_B^\pm S_A^\pm$, a simple induction on $x$ gives
\begin{equation*}
S^{-(x)} = \sum_{j=0}^x q^{-j(x-j)} S^{-(x-j)}_B S^{-(j)}_A
\end{equation*}
where $S^{\pm(j)}_X = (S^{\pm}_X)^j / [j]!$ for both $X = A$ and $B$. Because $S_B^{-(N-2p + 1)} = S_A^{-(2p + 1)}=  0$, the lower and upper bounds of the sum can be changed respectively to $p-k$ and $2p$ if $x > N-2p$. Since $p-k$ is a positive integer, the index $j$ runs over an interval of positive integers. Clearly the result will follow if $S_A^{-(j)}\tilde i_N^d(w_{p^p}^N) = 0$ for $j>0$, an identity that we now set out to prove. Let
\begin{equation*}
F'_{i,j} = q^{-j} S^-_i + q^j S^-_{2p+1-i} \qquad \textrm{and} \qquad G'_{k,l} = q^{-l}v^{(2p+1-2k)} \sigma^-_{2p+1-k} - q^{l-1} v^{-(2p+1-2k)} \sigma^-_k.
\end{equation*}
The analog of \eqref{eq:multrules} now reads
\begin{align} F_{i,j}' G_{k,l}' &=
\begin{cases}
G_{k,l}' F_{i, j+1}', & i<k, \\ 
G_{k,l+1}' F_{i, j}', & i>k,
\end{cases} 
\end{align}
and that of \eqref{eq:multrules2} is
\begin{equation*}
F'_{i,j}G'_{i,k} |0 \rangle = \sigma^-_{i} \sigma^-_{2p+1-i} |0 \rangle \times q^{(N-1)/2} v^{2p-N} (q^{-(i+j+k)} - q^{i+j+k-2p}).
\end{equation*}
Then
\begin{align*}
S^-_A \tilde i_N^d (w_{p^p}^N) &=  (i q^{1/2})^p \sum_{i=1}^p F'_{i,0} \Big(\prod_{j=1}^p G'_{j,0} \Big) |0 \rangle \\
&=  (i q^{1/2})^p \sum_{i=1}^p \Big( \prod_{j=1}^{i-1} G'_{j,1} \Big)
\Big(\prod_{k=i+1}^{p} G'_{k,0} \Big) F'_{i,p-i} G'_{i,0}|0 \rangle = 0.
\end{align*}
Because this holds independently of $q$ and $v$, any renormalized power of $S^-_A$ also annihilates $\tilde i_N^d (w_{p^p}^N)$. \hfill$\square$ 

\medskip 

\noindent Setting $p=P$ in this proposition yields corollary \ref{sec:coro2}.

%
%

\section{Computations leading to theorem \ref{thm:intra}}\label{app:a}
\subsection[Computation of $\langle \nu_3 | H_1 |\chi \rangle$]{Computation of $\boldsymbol{\langle \nu_3 | H_1 |\chi \rangle}$}\label{app:a1}

The objective of this appendix is to compute the matrix element $ I = \langle \nu_3 | H_1 |\chi \rangle$ and show that it is non-zero. Throughout, we assume the hypotheses \ref{hyp:hypHourra} stated at the beginning of section \ref{sec:BJintra}. By writing $$H_1 = \sum_{i=0}^{N-1} \bar\Omega^i \left.\frac{d (e_{N-1})}{dq}\right|_{q = q_c} \bar\Omega^{-i},$$ 
and using the commutation of $\bar \Omega^{\pm 1}$ with $S^{+(k)}T^{-(P+k)}$, $S^{+(P+k)}T^{-(2P+k)}$ and $T^{+(N-P-k)}S^{-(N-k)}$, we get a much simpler expression for $I$, namely $$I = \left. N \langle \nu_3 | \frac{d (e_{N-1})}{dq} | \chi \rangle \right|_{q = q_c} \hspace{-0.2cm}=  \left. N\big(x_1 A(k) - x_2 A(P+k)\big) \right|_{q = q_c} \, \textrm{where} \quad A(x)= \langle \nu_3 | \frac{d (e_{N-1})}{dq} S^{+(x)}T^{-(P+x)} |0\rangle,$$
and $x_1, x_2$ are given in equation \eqref{eq:valeurx1x2}. From equation (\ref{eq:ebar}), $\frac{d (e_{N-1})}{dq} = q^{-2} f^{-+} - f^{+-}$ with $f^{s_1 s_2} = \sigma_{N-1}^{s_1}\sigma_{N-1}^{-s_1}\sigma_{N}^{s_2}\sigma_{N}^{-s_2}$ and $s_1, s_2 \in \{+,-\}$. By writing 
$$S^\pm = \underbrace{\sum_{j = 1}^{N-2}S^\pm_j}_{S^\pm_A} + \underbrace{S^\pm_{N-1} + S^\pm_{N}}_{S^\pm_B}, \qquad T^\pm = \underbrace{\sum_{j = 1}^{N-2}T^\pm_j}_{T^\pm_A} + \underbrace{T^\pm_{N-1} + T^\pm_{N}}_{T^\pm_B},$$  
we expand $S^{+(x)}$ and $T^{-(P+x)}$ with 
\begin{align}
S^{\pm(x)} &= S^{\pm (x)}_A + q^{\mp(x-1)} S^{\pm(x-1)}_A S^\pm_B + q^{\mp 2(x-2)}  S^{\pm{(x-2)}}_A S^{\pm(2)}_B, \label{eq:devST}\\
T^{\pm(P+x)} &= T^{\pm (P+x)}_A + q^{\pm(P+x-1)}  T^{\pm(P+x-1)}_A T^\pm_B + q^{\pm 2(P+x-2)}  T^{\pm{(P+x-2)}}_A T^{\pm(2)}_B, \nonumber
\end{align}
and write down the multiplication rules
\begin{equation*}
[f^{s_1 s_2}, S^+_A]  = [f^{s_1 s_2}, T^-_A] = 0, \quad
f^{+-} S^+_B = S^+_{N-1} f^{--}, \quad
f^{-+} S^+_B = S^+_{N} f^{--}, \quad
f^{+-} S^{+(2)}_B = f^{-+} S^{+(2)}_B =  0. 
\end{equation*}
This allows us to write
\begin{align*}
f^{+-} S^{+(x)} &= S^{+(x)}_A f^{+-} + q^{-(x-1)} S^{+(x-1)}_A S^+_{N-1}f^{--},\\
f^{-+} S^{+(x)} &= S^{+(x)}_A f^{-+} + q^{-(x-1)} S^{+(x-1)}_A S^+_{N}f^{--}.
\end{align*}
With more multiplication rules
\begin{equation*}
f^{+-} T^-_B = T^-_{N} f^{++}, \quad
f^{-+} T^-_B = T^-_{N-1} f^{++}, \quad
f^{+-} T^{-(2)}_B = f^{-+} T^{-(2)}_B =  0, \quad 
f^{--} T^-_B = T^-_{N-1} f^{+-} + T^-_{N} f^{-+}, \end{equation*}
\begin{equation*}
f^{++} | 0 \rangle = | 0 \rangle, \qquad
f^{+-} | 0 \rangle = f^{-+} | 0 \rangle = f^{--} | 0 \rangle = 0,
\end{equation*}
we find
\begin{align*}
f^{+-} S^{+(x)}T^{-(P+x)} | 0 \rangle &= q^{-(P+x-1)}S^{+(x)}_A T_A^{-(P+x-1)}T^-_N |0 \rangle+ q^{-x} S^{+(x-1)}_A 
T_A^{-(P+x-2)} T^-_N |0 \rangle,\\
f^{-+} S^{+(x)}T^{-(P+x)} | 0 \rangle &= q^{-(P+x-1)}S^{+(x)}_A T_A^{-(P+x-1)}T^-_{N-1} |0 \rangle+ q^{-x+2} S^{+(x-1)}_A 
T_A^{-(P+x-2)} T^-_{N-1} |0 \rangle,
\end{align*}
and finally,
\begin{align*}
\frac{d (e_{N-1})}{dq} S^{+(x)} T^{-(P+x)} | 0 \rangle = q^{-(P+x-1)}S^{+(x)}_A T_A^{-(P+x-1)}(q^{-2}T^-_{N-1}-T^-_N) |0 \rangle+ q^{-x} S^{+(x-1)}_A 
T_A^{-(P+x-2)} (T^-_{N-1}- T^-_N) |0 \rangle.
\end{align*}
We can now expand $\langle \nu_3 |$ by using equation (\ref{eq:devST}). This yields nine terms, of which only two survive when applied to $\frac{d (e_{N-1})}{dq} S^{+(x)} T^{-(P+x)} | 0 \rangle$:
$$ \langle \nu_3 | \rightarrow q^{P-y-1} \langle 0 | T^{+(P+y-1)}_A S^{-(y)}_A T^+_B + q^{2P-y-1} \langle 0 | T^{+(P+y-2)}_AS^{-(y-1)}_A T^{+(2)}_B S^-_B$$
where $y = N-P-k$. The amplitude $A(x)$ is then the sum of four terms
\begin{align}
A(x) &= q^{-y-x} \langle 0 | T^{+(P+y-1)}_A S^{-(y)}_A S^{+(x)}_A T_A^{-(P+x-1)} | 0 \rangle \times \langle 0 | T^+_B (q^{-2} T^-_{N-1} - T^-_N) |0 \rangle \nonumber \\
& + q^{P-y-x} \langle 0 | T^{+(P+y-2)}_A S^{-(y-1)}_A S^{+(x)}_A T_A^{-(P+x-1)} | 0 \rangle \times\langle 0 | T^{+(2)}_B S^-_B (q^{-2} T^-_{N-1} - T^-_N) |0 \rangle \label{eq:4lines}\\
& + q^{P-y-x-1} \langle 0 | T^{+(P+y-1)}_A S^{-(y)}_A S^{+(x-1)}_A T_A^{-(P+x-2)} | 0 \rangle \times\langle 0 | T^{+}_B (T^-_{N-1} - T^-_N) |0 \rangle \nonumber\\
& + q^{2P-y-x-1} \langle 0 | T^{+(P+y-2)}_A S^{-(y-1)}_A S^{+(x-1)}_A T_A^{-(P+x-2)} | 0 \rangle \times\langle 0 | T^{+(2)}_BS^-_B (T^-_{N-1} - T^-_N) |0 \rangle \nonumber,
\end{align}
where the following factors are easy to compute
\begin{align*}
\langle 0 | T^+_B (q^{-2} T^-_{N-1} - T^-_N) |0 \rangle & = -q^{N-3}(q^2 - q^{-2}), \\
\langle 0 | T^{+(2)}_B S^-_B (q^{-2} T^-_{N-1} - T^-_N) |0 \rangle  & = -q^{N-3}(q - q^{-1}), \\
\langle 0 | T^{+}_B (T^-_{N-1} - T^-_N) |0 \rangle & = -q^{N-2}(q - q^{-1}), \\
\langle 0 | T^{+(2)}_BS^-_B (T^-_{N-1} - T^-_N) |0 \rangle & = 0.
\end{align*}
The remaining are harder and appendix \ref{app:a2} gives their expression
\begin{align}
 \langle 0 | T^{+(N-k-1)}_A S^{-(N-P-k)}_A S^{+(x)}_A T_A^{-(P+x-1)} | 0 \rangle|_{q = q_c} &= 0 \qquad \textrm{for} \quad x = k, P+k, \nonumber \\
 \langle 0 | T^{+(N-k-2)}_A S^{-(N-P-k-1)}_A S^{+(k)}_A T_A^{-(P+k-1)} | 0 \rangle|_{q = q_c} & =  q_c^2 \times \Big(\left(\begin{smallmatrix}N-2 \\ k\end{smallmatrix}\right) \left(\begin{smallmatrix}N-2 \\ P+k -1 \end{smallmatrix}\right) - \left(\begin{smallmatrix}N-2 \\ k-1\end{smallmatrix}\right) \left(\begin{smallmatrix}N-2 \\ P+k \end{smallmatrix}\right)\Big), \label{eq:amplits} \\
 \langle 0 | T^{+(N-k-2)}_A S^{-(N-P-k-1)}_A S^{+(P+k)}_A T_A^{-(2P+k-1)} | 0 \rangle|_{q = q_c} & =  
q_c^{P^2+P+2} \times \Big(\left(\begin{smallmatrix}N-2 \\ k\end{smallmatrix}\right) \left(\begin{smallmatrix}N-2 \\ 2P+k -1 \end{smallmatrix}\right) - \left(\begin{smallmatrix}N-2 \\ k-1\end{smallmatrix}\right) \left(\begin{smallmatrix}N-2 \\ 2P+k \end{smallmatrix}\right)\Big).
\nonumber \label{eq:amplits} 
\end{align}
Only the second and third terms of equation (\ref{eq:4lines}) are non-zero and a short calculation yields
\begin{align*}
A(k)|_{q = q_c} & = -Pq_c^{-1}(q_c-q_c^{-1})\frac{ (N-P)(N-2k)+2k^2}{N^2(N-1)}\begin{pmatrix} N \\ P+k \end{pmatrix} \begin{pmatrix} N \\ k \end{pmatrix},\\
A(P+k)|_{q = q_c} & = -2Pq_c^{P^2-1}(q_c-q_c^{-1})\frac{ (N-2P)(N-2k)+2k^2}{N^2(N-1)}\begin{pmatrix} N \\ 2P+k \end{pmatrix} \begin{pmatrix} N \\ k \end{pmatrix}.
\end{align*}
The amplitude $I=\langle \nu_3|H_1|\chi\rangle$ is then
\begin{equation*}
I = - q_c^{P^2 + Pk -1} (q_c - q_c^{-1}) \frac{2P^2(N-2k)}{N(N-1)} \begin{pmatrix} N \\  k \end{pmatrix}\begin{pmatrix} N \\  P+k \end{pmatrix}\begin{pmatrix} N \\  2P+k \end{pmatrix}\end{equation*}
which is non-zero. 

%
%

\subsection[Specific $U_q(sl_2)$ amplitudes]{Specific $\boldsymbol{U_q(sl_2)}$ amplitudes}\label{app:a2}

In paragraph \ref{sec:jordan}, the proof of the existence of the Jordan generalized eigenvector for the loop hamiltonian is shown to follow from $\langle \nu_3 | H_1 |\chi \rangle$ being non zero, which in turn requires the calculation of some amplitudes of the renormalized generators of the $U_q(sl_2)$ algebra. We compute these expressions here. Hypotheses \ref{hyp:hypHourra} are assumed throughout. The evaluation will require various ingredients, among which the first is the following rewriting of equation (\ref{eq:bigcomm}):
\begin{equation}
[S^{+(m)}, S^{-(n)}] = \sum_{j=1}^{\textrm{min}(m,n)}S^{-(n-j)}S^{+(m-j)} \times \left\{\begin{array}{ll} (-1)^j\left[ \begin{matrix} j-1-(2S^z + m - n)   \\ j \end{matrix} \right] & \textrm{if} \, \, 2S^z + m - n <0,  \vspace{0.1cm}\\   \left[ \begin{matrix} 2S^z + m - n \\ j \end{matrix} \right]  & \textrm{if} \, \, 2S^z + m - n \ge 0, \end{array}\right.
\label{eq:bigcomm2}
\end{equation}
Another is the following lemma.
\begin{Lemme} Let $x,y$ be non-negative integers. Then the inverse of the matrix 
\begin{equation*} M({x,y}) = \left( \begin{matrix}
1 & -\left[ \begin{smallmatrix} x \\ 1\end{smallmatrix}\right] & \left[ \begin{smallmatrix} x+1 \\ 2\end{smallmatrix}\right] & -\left[ \begin{smallmatrix} x+2 \\ 3\end{smallmatrix}\right] & \dots & (-1)^y \left[ \begin{smallmatrix} x+y-1 \\ y\end{smallmatrix}\right] \\
0 & 1 & -\left[ \begin{smallmatrix} x \\ 1\end{smallmatrix}\right] & \left[ \begin{smallmatrix} x+1 \\ 2\end{smallmatrix}\right] & -\left[ \begin{smallmatrix} x+2 \\ 3\end{smallmatrix}\right] & \vdots \\
0 & 0 & 1 & -\left[ \begin{smallmatrix} x \\ 1\end{smallmatrix}\right] & \left[ \begin{smallmatrix} x+1 \\ 2\end{smallmatrix}\right] & -\left[ \begin{smallmatrix} x+2 \\ 3\end{smallmatrix}\right] \\
0 & 0 & 0 & 1 &-\left[ \begin{smallmatrix} x \\ 1\end{smallmatrix}\right] & \left[ \begin{smallmatrix} x+1 \\ 2\end{smallmatrix}\right] \\
0 & 0 & 0 & 0 & 1 & -\left[ \begin{smallmatrix} x \\ 1\end{smallmatrix}\right] \\
0 & 0 & 0 & 0 & 0 & 1
\end{matrix} \right) \end{equation*}
is given by 
\begin{equation}\hspace{-1.2cm}
M({x,y})^{-1} = \left( \begin{matrix}
1 & \left[ \begin{smallmatrix} x \\ 1\end{smallmatrix}\right] & \left[ \begin{smallmatrix} x \\ 2\end{smallmatrix}\right] & \left[ \begin{smallmatrix} x \\ 3\end{smallmatrix}\right] & \dots &  \left[ \begin{smallmatrix} x \\ y\end{smallmatrix}\right] \\
0 & 1 & \left[ \begin{smallmatrix} x \\ 1\end{smallmatrix}\right] & \left[ \begin{smallmatrix} x \\ 2\end{smallmatrix}\right] & \left[ \begin{smallmatrix} x \\ 3\end{smallmatrix}\right] & \vdots \\
0 & 0 & 1 & \left[ \begin{smallmatrix} x \\ 1\end{smallmatrix}\right] & \left[ \begin{smallmatrix} x \\ 2\end{smallmatrix}\right] & \left[ \begin{smallmatrix} x \\ 3\end{smallmatrix}\right] \\
0 & 0 & 0 & 1 &\left[ \begin{smallmatrix} x \\ 1\end{smallmatrix}\right] & \left[ \begin{smallmatrix} x \\ 2\end{smallmatrix}\right] \\
0 & 0 & 0 & 0 & 1 & \left[ \begin{smallmatrix} x \\ 1\end{smallmatrix}\right] \\
0 & 0 & 0 & 0 & 0 & 1
\end{matrix} \right).
\label{eq:invM}
\end{equation}
\label{sec:invlemme}
\end{Lemme}
\noindent{\scshape Proof\ \ } The result is trivial for $y = 0$ and $y = 1$. The matrix $M(x,y)$ is upper triangular and has the form 
$$M(x,y) = \left(\begin{array}{cccccc}
\multicolumn{5}{c}{\multirow{5}{*}{$M(x,y-1)$}}
&  (-1)^y \left[ \begin{smallmatrix} x+y-1 \\ y\end{smallmatrix}\right] \\
&&&&& \vdots \\
&&&&& -\left[ \begin{smallmatrix} x+2 \\ 3\end{smallmatrix}\right] \\
&&&&& \left[ \begin{smallmatrix} x+1 \\ 2\end{smallmatrix}\right] \\
&&&&& -\left[ \begin{smallmatrix} x \\ 1\end{smallmatrix}\right]\\
0 &0&0&0&0& 1 
 \end{array}\right)= \left(\begin{array}{cccccc}
 1 & -\left[ \begin{smallmatrix} x \\ 1\end{smallmatrix}\right] & \left[ \begin{smallmatrix} x+1 \\ 2\end{smallmatrix}\right] & -\left[ \begin{smallmatrix} x+2 \\ 3\end{smallmatrix}\right] & \dots & (-1)^y \left[ \begin{smallmatrix} x+y-1 \\ y\end{smallmatrix}\right] \\
0 &\multicolumn{5}{c}{\multirow{5}{*}{$M(x,y-1)$}} \\ 0 \\ 0 \\ 0 \\ 0
 \end{array}\right)
$$
and similarly for $M(x,y)^{-1}$.
We proceed by induction and suppose the expression (\ref{eq:invM}) is correct for $M(x, y-1)^{-1}$. To check it is also valid for $M(x,y)^{-1}$ we only need to prove that the element in the upper right corner of the product $M(x,y)M(x,y)^{-1}$ is indeed zero, that is
$$\sum_{i = 0}^y (-1)^i  \left[\begin{matrix} x \\ y-i\end{matrix}\right] \left[\begin{matrix} x + i - 1\\ i\end{matrix}\right]=0.$$
An intermediate step is showing that
\begin{equation}s_j=\sum_{i = 0}^j (-1)^i  \left[\begin{matrix} x \\ y-i\end{matrix}\right] \left[\begin{matrix} x + i - 1\\ i\end{matrix}\right]=(-1)^j  \left[\begin{matrix} x+j \\ y\end{matrix}\right]\left[\begin{matrix} y-1 \\ j\end{matrix}\right]\label{eq:sj}\end{equation}
which is trivially true for $j = 0$ and indeed gives zero for $j = y$. Equation (\ref{eq:sj}) is then proved recursively. 
\hfill$\square$\\

We now turn to the evaluation of the amplitudes of equation (\ref{eq:amplits}).

\begin{Proposition}\label{sec:amplitude0} Under the hypotheses \ref{hyp:hypHourra},
$$ \langle 0 | T^{+(N-k-1)}_A S^{-(N-P-k)}_A S^{+(x)}_A T_A^{-(P+x-1)} | 0 \rangle = 0, \qquad \textrm{for} \quad x = k \textrm{\ and\ } P+k.$$
\end{Proposition}
\noindent{\scshape Proof\ \ } Recall that the index $A$ on generators means that the defining sums run from $1$ to $N-2$, e.g.~$S^\pm_A=\sum_{j=1}^{N-2}S^\pm_j$. Therefore
\begin{equation}S_A^{\pm (x)} = \left. S^{\pm (x)} \right|_{N-2} \otimes v^{\pm x}q^{x \sigma^z /2} \otimes v^{\pm x}q^{x \sigma^z /2}, \quad \qquad T_A^{\pm (x)} = \left. T^{\pm (x)} \right|_{N-2} \otimes v^{\pm x}q^{-x \sigma^z /2} \otimes v^{\pm x}q^{-x \sigma^z /2}
\label{eq:Nm2}\end{equation}
where $\left. S^{\pm (x)} \right|_{N-2}$ and $\left. T^{\pm (x)} \right|_{N-2}$ are the matrices $S^{\pm (x)}$ and $T^{\pm (x)}$ in the representation $\tau$ with $N-2$ spins. Then, 
$$ \langle 0 | T^{+(N-k-1)}_A S^{-(N-P-k)}_A S^{+(x)}_A T_A^{-(P+x-1)} | 0 \rangle = q^{-2P+2}\langle 0' | T^{+(N-k-1)} S^{-(N-P-k)} S^{+(x)} T^{-(P+x-1)} | 0' \rangle$$
where the amplitudes on the right-hand side are evaluated in the representation with $N-2$ spins. Then the state $|0' \rangle$ with $N-2$ spins up satisfies $S^z |0' \rangle = (\frac{N}2 -1) |0' \rangle$ and $\langle 0'|S^{+(x)}T^{-(x)}|0' \rangle = \langle 0'|T^{+(x)}S^{-(x)}|0' \rangle=\left(\begin{smallmatrix} N-2 \\ x \end{smallmatrix} \right)$. (See \eqref{eq:plustot} and the following lines.)

We set $x = k$ and define 
$$f_j = \langle 0' | T^{+(N-k-1)} S^{-(N-P-k-j)} S^{+(k-j)} T^{-(P+k-1)} | 0' \rangle, \qquad j = 0, \dots, k.$$
The amplitude to be computed is then $f_0$.
By using equation (\ref{eq:bigcomm2}) to commute $S^{-(N-P-k-j)}$ and $S^{+(k-j)}$, we find 
\begin{equation*} f_j = t_j - \sum_{i = 1}^{k-j} (-1)^i \left[ \begin{matrix} P-1+i \\ i \end{matrix}\right] f_{i+j} \quad
\textrm{where} \quad t_j = \langle 0' | T^{+(N-k-1)} S^{+(k-j)} S^{-(N-P-k-j)} T^{-(P+k-1)} | 0' \rangle.\end{equation*}
Using Lemma \ref{sec:invlemme}, this can be reformulated as $t_j = \sum_{i = 0}^k M(P,k)_{j,i}f_i$ and inverted to give $f_i = \sum_{j = 0}^k M(P,k)^{-1}_{i,j}t_j$. The amplitude $f_0$ is then $\sum_{j=0}^k \left[ \begin{smallmatrix}P \\ j \end{smallmatrix}\right] t_j$. Because
$\left[\begin{smallmatrix}P \\ j \end{smallmatrix}\right]=0$ for $q^{2P} = 1$ and $0<j<P$, $f_0 = t_0$. But $t_0$ is also zero because, when applied to $S^{-(N-P-k)} T^{-(P+k-1)} | 0' \rangle$, $S^z$ evaluates to $-\frac{N}2$, which is smaller than the minimum value $-(\frac{N}2 -1)$ taken by $S^z$ in the representation with $N-2$ spins.
 
If instead we set $x = P+k$ and  define
$$\tilde f_j = \langle 0' | T^{+(N-k-1)} S^{-(N-P-k-j)} S^{+(P+k-j)} T^{-(2P+k-1)} | 0' \rangle, \qquad j = 0, \dots, P+k,$$
the same technique leads to
$$\tilde t_j = \sum_{i = 0}^{a} M(2P,a)_{j,i}\tilde f_i, \qquad \tilde f_i = \sum_{j = 0}^a M(2P,a)^{-1}_{i,j}\tilde t_j$$
$$\textrm{with} \qquad \tilde t_j = \langle 0' | T^{+(N-k-1)} S^{+(P+k-j)} S^{-(N-P-k-j)} T^{-(2P+k-1)} | 0' \rangle,$$
$a = \textrm{min}(P+k, N-P-k)$ and $P \le a<2P$. Then $\tilde f_0 = \sum_{j=0}^a \left[ \begin{smallmatrix}2P \\ j \end{smallmatrix}\right] \tilde t_j$. In the limit $q^{2P} = 1$, only the terms $j = 0$ and $j = P$ can potentially contribute, but $S^z$ exits its natural interval in both cases, so $\tilde t_0$ and $\tilde t_P$ are zero.\hfill$\square$\\ 
\begin{Proposition} Under the hypotheses \ref{hyp:hypHourra},
\begin{equation*}
 \langle 0 | T^{+(N-k-2)}_A S^{-(N-P-k-1)}_A S^{+(k)}_A T_A^{-(P+k-1)} | 0 \rangle  =  q_c^2 \times \Big(\left(\begin{smallmatrix}N-2 \\ k\end{smallmatrix}\right) \left(\begin{smallmatrix}N-2 \\ P+k -1 \end{smallmatrix}\right) - \left(\begin{smallmatrix}N-2 \\ k-1\end{smallmatrix}\right) \left(\begin{smallmatrix}N-2 \\ P+k \end{smallmatrix}\right)\Big).
\end{equation*}
\end{Proposition}
\noindent{\scshape Proof\ \ } We remove the subscripts $A$ using equation (\ref{eq:Nm2}) 
$$ \langle 0 | T^{+(N-k-2)}_A S^{-(N-P-k-1)}_A S^{+(k)}_A T_A^{-(P+k-1)} | 0 \rangle = q^{-2P+2}\langle 0' | T^{+(N-k-2)} S^{-(N-P-k-1)} S^{+(k)} T^{-(P+k-1)} | 0' \rangle$$
and then define
$$\begin{array}{l} g_j = \langle 0' | T^{+(N-k-2)} S^{-(N-P-k-j-1)} S^{+(k-j)} T^{-(P+k-1)} | 0' \rangle, \\  u_j = \langle 0' | T^{+(N-k-2)} S^{+(k-j)} S^{-(N-P-k-j-1)} T^{-(P+k-1)} | 0' \rangle, \end{array} \qquad \textrm{for} \quad j = 0, \dots, k.$$
The strategy used in proposition \ref{sec:amplitude0} yields the relation $u_j = \sum_{i = 0}^k M(P-1,k)_{j,i}g_i$, which is inverted to $g_i = \sum_{j = 0}^k M(P-1,k)^{-1}_{i,j}u_j$. In particular,  $g_0 = \sum_{j=0}^k \left[ \begin{smallmatrix}P-1 \\ j \end{smallmatrix}\right] u_j =  \sum_{j=0}^k (-q_c^P)^j u_j$ where $q^{2P} = 1$ has been used for the last equality. This is more complicated than before, and writing three more sets of recursion relations will be necessary. The first concerns the $u_j$s. Let $r = 0, \dots, k$. From equations (\ref{eq:SandTcommute}) and (\ref{eq:bigcomm2}),
\begin{align}
u_{k-r} &=  \langle 0' | S^{+(r)} [T^{+(N-k-2)}, T^{-(P+k-1)}] S^{-(N-P-2k+r-1)}  | 0' \rangle \nonumber\\
	& = \sum_{j=1}^{P+k-1} \left[\begin{matrix} P+2k -2r-1 \\ j \end{matrix} \right]\langle 0' | S^{+(r)} T^{-(P+k-1-j)} T^{+(N-k-2-j)} S^{-(N-P-2k+r-1)}  | 0' \rangle \nonumber\\
	& = \sum_{i=0}^{P+k-2} \left[\begin{matrix} P+2k -2r-1 \\ P+k-1-i \end{matrix} \right]\langle 0' | S^{+(r)} T^{-(i)} T^{+(N-2k-P-1+i)} S^{-(N-P-2k+r-1)}  | 0' \rangle \nonumber\\	
	& = \sum_{i=0}^{r} \left[\begin{matrix} P+2k -2r-1 \\ k-2r+i \end{matrix} \right]\langle 0' | S^{+(r)} T^{-(i)} T^{+(Y+i)} S^{-(Y+r)}  | 0' \rangle =  \sum_{i=0}^{r} \left[\begin{matrix} P+2k -2r-1 \\ k-2r+i \end{matrix} \right] a_{r,i}	 \label{eq:ukr}
\end{align}
where $a_{r,i} = \langle 0' | S^{+(r)} T^{-(i)} T^{+(Y+i)} S^{-(Y+r)}  | 0' \rangle,\, Y = N-P-2k-1\ge 0$ 
and the subscript $i$ must therefore remain smaller or equal to $r$.
If $i = r$, $a_{r,r}= \left(\begin{smallmatrix} N-2 \\ r\end{smallmatrix}\right)\left(\begin{smallmatrix} N-2 \\ Y+r\end{smallmatrix}\right)$. For a pair $(r,i)$ with $r>i$,
\begin{align*} a_{r,i} &= \langle 0' | S^{+(r)} T^{+(Y+i)} T^{-(i)} S^{-(Y+r)}  | 0' \rangle - \langle 0' | S^{+(r)} [T^{+(Y+i)}, T^{-(i)}] S^{-(Y+r)}  | 0' \rangle \\
& =  \langle 0' |  T^{+(Y+i)} [S^{+(r)}, S^{-(Y+r)}]T^{-(i)}   | 0' \rangle - \sum_{j=1}^i \left[\begin{matrix} P+2k -2r-1 \\ j \end{matrix} \right] \langle 0' | S^{+(r)} T^{-(i-j)} T^{+(Y+i-j)}  S^{-(Y+r)}  | 0' \rangle \\
& =  \sum_{j = 1}^r \left[\begin{matrix} P+2k -2i-1 \\ j \end{matrix} \right]\langle 0' |  T^{+(Y+i)} S^{-(Y+r-j)} S^{+(r-j)} T^{-(i)}   | 0' \rangle - \sum_{j=1}^i \left[\begin{matrix} P+2k -2r-1 \\ j \end{matrix} \right] a_{r,i-j} \\
& = \sum_{s = 0}^i \left[\begin{matrix} P+2k -2i-1 \\ r-s \end{matrix} \right] b_{i,s} - \sum_{j'=0}^{i-1} \left[\begin{matrix} P+2k -2r-1 \\ i-j' \end{matrix} \right] a_{r,j'}
\end{align*}
where $b_{i,s} = \langle 0' |  T^{+(Y+i)} S^{-(Y+s)} S^{+(s)} T^{-(i)}   | 0' \rangle = a_{i,s}^*$ 
where ``$*$'' denotes complex conjugation. 
We obtain
\begin{equation}
(r,i): \quad  \sum_{j=0}^{i} \left[\begin{matrix} P+2k -2r-1 \\ i-j \end{matrix} \right] a_{r,j}= \sum_{s = 0}^i \left[\begin{matrix} P+2k -2i-1 \\ r-s \end{matrix} \right] a_{i,s}^{*}.
\label{eq:ri}\end{equation}
The $(r,i)$ equation allows one to write $a_{r,i}$ in terms of $a_{r',i'}$ and $a^*_{r'',i''}$ where $r'$ and $r''$ are smaller than $r$. The complex conjugate of equation (\ref{eq:ri}) allows us to do the same for $a_{r,i}^*$. Because the matrices $S^{\pm(m)}$ and $T^{\pm(m)}$ can be made independent of $v$ by the transformation $\mathcal O = v^{\sum_{j=1}^N j \sigma^z_j}$, the amplitudes we calculate do not depend on $v$. 
Moreover the coefficients in the recursion \eqref{eq:ri} are real for $q$ on the unit circle and so are the coefficients $a_{r,r}$, which are the only ones needed to construct recursively all $a_{r,i}$s. 
It follows that $a_{r,i}$ is real and equal to $a^*_{r,i}$. 

It only remains to understand how simplifications occuring at $q^{2P} =1$ allow for the computation of $g_0$. In what follows we show that the recursion \eqref{eq:ri} leads, in this limit, to 
\begin{equation}  a_{k-1,0} = q_c^P u_k  \qquad \textrm{and} \qquad a_{k-1, k-j} + q_c^P a_{k-1, k-j-1}= q_c^P u_j \quad \textrm{for} \quad j = 2, \dots, k-1. \label{eq:lastrec}\end{equation}
For $k=0$ or $1$, the relevant $a_{r,i}$s are $a_{0,0}$ and $a_{1,1}$ which are known and $a_{1,0}$ that can be obtained readily from \eqref{eq:ri}. The index $k$ above can therefore be assumed to be larger or equal to $2$. 
This new recursion will then complete the proof as
\begin{align*} 
g_0 &= u_0 - q_c^{P} u_1 + \sum_{i = 2}^k (-q_c^P)^i u_i, \quad \textrm{where} \quad \begin{array}{l} u_0 = a_{k,k} = \left(\begin{smallmatrix} N-2 \\ k\end{smallmatrix}\right)\left(\begin{smallmatrix} N-2 \\ Y+k\end{smallmatrix}\right) \\ u_1 = \left[\begin{smallmatrix}P+1\\1 \end{smallmatrix}\right] a_{k-1,k-1} + a_{k-1,k-2} = q_c^P \left(\begin{smallmatrix} N-2 \\ k-1\end{smallmatrix}\right)\left(\begin{smallmatrix} N-2 \\ Y+k-1\end{smallmatrix}\right) + a_{k-1,k-2}\end{array} \\
&\textrm{and} \quad a_{k-1,k-2} = q_c^P (u_2 - a_{k-1,k-3}) = q_c^P (u_2 - q_c^P u_3) + a_{k-1,k-4} = \dots = q_c^P \sum_{i=2}^k (-q_c^P)^i u_i
\end{align*}
which gives $g_0 = \left(\begin{smallmatrix} N-2 \\ k\end{smallmatrix}\right)\left(\begin{smallmatrix} N-2 \\ Y+k\end{smallmatrix}\right) - \left(\begin{smallmatrix} N-2 \\ k-1\end{smallmatrix}\right)\left(\begin{smallmatrix} N-2 \\ Y+k-1\end{smallmatrix}\right) $ as claimed. 

We now prove equation (\ref{eq:lastrec}). The relation 
\begin{equation}
\lim_{q \rightarrow q_c} \left[\begin{matrix}P+a \\ k\end{matrix}\right] = \left\{\begin{array}{ll} 0 & a<k<P, \\ q_c^{Pk}\left[\begin{matrix}a \\ k\end{matrix}\right] & k \le a, \end{array}\right.
\label{eq:q2Psimp}
\end{equation}
will be used.
For the first part of (\ref{eq:lastrec}), setting $(r,i) = (k-1,0)$ in equation (\ref{eq:ri}) and using equation (\ref{eq:q2Psimp}) gives 
$$a_{k-1,0} = \left[ \begin{matrix} P+2k-1 \\ k-1\end{matrix}\right] a_{0,0}= q_c^{P(k-1)}\left[ \begin{matrix} 2k-1 \\ k-1\end{matrix}\right] a_{0,0}$$ while equation (\ref{eq:ukr}) gives $$u_k = \left[ \begin{matrix} P+2k-1 \\ k\end{matrix}\right] a_{0,0} = q_c^{Pk}\left[ \begin{matrix} 2k-1 \\ k\end{matrix}\right] a_{0,0}= q_c^{P} a_{k-1,0}$$ as expected. For the second part of (\ref{eq:lastrec}), by setting $(r,i) = (k-1,k-j)$, we find, on one hand,
$$\sum_{i=0}^{k-j} \left[\begin{matrix} P+1 \\ k-j-i \end{matrix} \right] a_{k-1,i}= \sum_{s = 0}^{k-j} \left[\begin{matrix} P+2j-1 \\ k-1-s \end{matrix} \right] a_{k-j,s}.$$
In the left-hand side, the last two terms are the only to survive in the limit $q^{2P}=1$ and the result is $a_{k-1,k-j} + [P+1] a_{k-1,k-j-1} = a_{k-1,k-j} + q_c^P a_{k-1,k-j-1}$. We can rewrite the right-hand side as
\begin{align*}
\sum_{s = 0}^{k-j} \left[\begin{matrix} P+2j-1 \\ k-1-s \end{matrix} \right] a_{k-j,s} = 
\sum_{t=j-1}^{\min(k-1,2j-1)} q_c^{P t}\left[\begin{matrix} 2j-1 \\ t \end{matrix} \right] a_{k-j,k-1-t}.
\end{align*}
On the other hand, from (\ref{eq:ukr}), we have 
\begin{align*}
u_j &=  \sum_{i=0}^{k-j} \left[\begin{matrix} P+2j-1 \\ -k+2j+i \end{matrix} \right] a_{k-j,i} 
= \sum_{s=j-1}^{\min(k-1, 2j-1)} \left[\begin{matrix} P+2j-1 \\ 2j-1-s \end{matrix} \right] a_{k-j,k-1-s} \\ 
& 
= q_c^{P} \sum_{s=j-1}^{\min(k-1, 2j-1)} q^{Ps}\left[\begin{matrix} 2j-1 \\ s \end{matrix} \right] a_{k-j,k-1-s} = q_c^P(a_{k-1,k-j} + q_c^P a_{k-1,k-j-1})
\end{align*}
which completes the proof.
\hfill$\square$\\

\begin{Proposition}  Under the hypotheses \ref{hyp:hypHourra}, 
\begin{equation*}
 \langle 0 | T^{+(N-k-2)}_A S^{-(N-P-k-1)}_A S^{+(P+k)}_A T_A^{-(2P+k-1)} | 0 \rangle  =  {q}_c^{P^2+P+2} \times \Big(\left(\begin{smallmatrix}N-2 \\ k\end{smallmatrix}\right) \left(\begin{smallmatrix}N-2 \\ 2P+k -1 \end{smallmatrix}\right) - \left(\begin{smallmatrix}N-2 \\ k-1\end{smallmatrix}\right) \left(\begin{smallmatrix}N-2 \\ 2P+k \end{smallmatrix}\right)\Big) .
\end{equation*}
\end{Proposition}
\noindent{\scshape Proof\ \ }
The argument is similar to the one of the previous proposition. A highlight of the steps should suffice. The amplitude to be computed differs from $\langle 0' | T^{+(N-k-2)} S^{-(N-P-k-1)} S^{+(P+k)} T^{-(2P+k-1)} | 0' \rangle$ only by a factor of $q_c^{-2P+2} = q_c^2$. Defining 
$$\begin{array}{l}\tilde g_j = \langle 0' | T^{+(N-k-2)} S^{-(N-P-k-j-1)} S^{+(P+k-j)} T^{-(2P+k-1)} | 0' \rangle, \\  \tilde u_j = \langle 0' | T^{+(N-k-2)} S^{+(P+k-j)} S^{-(N-P-k-j-1)} T^{-(2P+k-1)} | 0' \rangle, \end{array} \qquad \textrm{for} \quad j = 0, \dots, a,$$
with $a = \min (P+k, N-P-k-1)$ and 
$P-1\le a<2P$, we write down four sets of recursion relations, the first being 
\begin{equation}\tilde u_j = \sum_{i = 0}^a M(2P-1,a)_{j,i}\tilde g_i, \qquad \tilde g_i = \sum_{j = 0}^a M(2P-1,a)^{-1}_{i,j}\tilde u_j. \label{eq:firstset}\end{equation}
A considerable simplification stems from the fact that $\tilde u_j =0$ for $j = 0, \dots, P-1$, because at some point, $S^z$ exits the interval $[\frac{N}2-1, -(\frac{N}2-1)]$. 
For $q^{2P}=1$, equation (\ref{eq:firstset}) gives $\tilde g_0$ as simply $\sum_{i = P}^a \tilde u_i \left[\begin{smallmatrix} 2P-1\\ i \end{smallmatrix} \right] = 
\sum_{i = P}^a (-1)^{i+1}\tilde u_i$. 
The second and third sets of recursion 
relations are found to be: 
\begin{equation}
\tilde u_{a-r} = \sum_{i = 0}^r  \left[ \begin{matrix} 2a - 2r -1 \\ a-P+i-2r \end{matrix} \right] \tilde a_{r,i}, \qquad r = 0, \dots, a-1
\label{eq:uar}\end{equation} 
\begin{equation*} \textrm{with} \quad \tilde a_{r,i} = 
\langle 0 | S^{+(r)} T^{-(i)} T^{+(Z+i)}S^{-(Z+r)} |0 \rangle,\qquad Z = \begin{cases} N-2P-2k-1,  &\textrm{if} \,\, a = P+k,  \\ -(N-2P-2k-1),  &\textrm{if} \,\, a = N-P-k-1, \end{cases}
\end{equation*}
\begin{equation}
(r,i): \quad  \sum_{j=0}^{i} \left[\begin{matrix} 2a -2r-1 \\ i-j \end{matrix} \right] \tilde a_{r,j}= \sum_{s = 0}^i \left[\begin{matrix} 2a -2i-1 \\ r-s \end{matrix} \right] \tilde a_{i,s} \qquad (r>i)
\label{eq:ri3}\end{equation}
and, obviously, $\tilde a_{r,r}= \left(\begin{smallmatrix} N-2 \\ r\end{smallmatrix}\right)\left(\begin{smallmatrix} N-2 \\ Z+r\end{smallmatrix}\right)$. 
(We note that the form of \eqref{eq:ri3} results from the reality of the $\tilde a_{i,s}$, which itself stems from the same equation \eqref{eq:ri3} with $\tilde a_{i,s}$ replaced with $\tilde a_{i,s}^*$.) Again, equations (\ref{eq:uar}) and (\ref{eq:ri3}), evaluated at $q^{2P} = 1$, lead to the final recursion
\begin{equation}  \tilde a_{a-P-1,0} = \tilde u_{a} \qquad \textrm{and} \qquad \tilde a_{a-P-1, a-P-j} + \tilde a_{a-P-1, a-P-j-1}= \tilde u_{P+j}  \quad \textrm{for} \quad j = 2, \dots, a-P-1 \label{eq:lastrec2}\end{equation}
whose solution gives $\tilde g_0 = (-1)^{P+1} \Big(\left(\begin{smallmatrix}N-2 \\ k\end{smallmatrix}\right) \left(\begin{smallmatrix}N-2 \\ 2P+k -1 \end{smallmatrix}\right) - \left(\begin{smallmatrix}N-2 \\ k-1\end{smallmatrix}\right) \left(\begin{smallmatrix}N-2 \\ 2P+k \end{smallmatrix}\right)\Big)$ for both values of $a$. Finally, if $P$ is the smallest integer such that $q_c^{2P} = 1$, we can replace $(-1)^{P+1}$ by $q_c^{P^2 + P}$.
\hfill$\square$\\ 

%
 
\end{document}